\documentclass[final,5p,times,twocolumn]{elsarticle}

\journal{TexExchange}
    \journal{Journal of \LaTeX\ Templates}
    
    \usepackage{tgtermes}
    \usepackage{nicematrix}
    \usepackage{tikz}
    \usepackage{multirow}
    \usepackage{hyperref}
    
    \usepackage{changepage}

    \usepackage{siunitx}

    \usepackage{hhline}  
    \usepackage{float}
    \usepackage{multirow}
    \usepackage{graphicx,booktabs}
    \usepackage{booktabs,tabularx}
    \usepackage{xfrac}
    \usepackage{titlesec}
    \titleformat{\paragraph}[runin]{\normalfont\bfseries}{\theparagraph}{1em}{}
    \usepackage{wrapfig,booktabs}
    \usepackage{mwe}
    \usepackage{adjustbox}
    \usepackage{pdfpages}  
    \usepackage{caption}
    \usepackage{subcaption}
    \usepackage{makecell}
    \usepackage{rotating}

    \usepackage{algorithm}
    \usepackage{algpseudocode}
    \usepackage{amsmath,array, amssymb}
    \usepackage{xcolor,soul}
    \usepackage{tcolorbox}
    \usepackage{longtable}
    \usepackage{ragged2e}

    \usepackage[indent=1cm]{parskip}
    \usepackage{tabularray}

    \usepackage{makecell}
    \usepackage{setspace}
    \usepackage{float}
    \usepackage[singlelinecheck=false]{caption}
\hypersetup{colorlinks,breaklinks,
            urlcolor=Maroon,
            linkcolor=Maroon}
    \newcommand{\etal}{\textit{et al}.}
    \def\correction#1{%
    \abovedisplayshortskip=#1\baselineskip\relax\belowdisplayshortskip=#1\baselineskip\relax%
    \abovedisplayskip=#1\baselineskip\relax\belowdisplayskip=#1\baselineskip\relax}

    \fboxsep=\tabcolsep\relax
    \fboxrule=\arrayrulewidth\relax

    \newcolumntype{A}[2]{%
    >{\minipage{\dimexpr#1\linewidth-2\tabcolsep-#2\arrayrulewidth\relax}\vspace\tabcolsep}%
    c<{\vspace\tabcolsep\endminipage}}

    \usepackage{siunitx}

    \usepackage{tikz}
    \usepackage{color}
    \definecolor{scalebgcolor}{rgb}{0.08,0.52,0.80}     
\biboptions{numbers,sort&compress}
\begin{document}

\begin{frontmatter}
\title{\Large{A BRAVE Alloy Design Campaign\\
(Bayesian Risk-aware Alloy discovery and Exploration)}}



\author{Mrinalini Mulukutla$^{a}$\corref{mycorrespondingauthor}}
\author{Danial Khatamsaz$^{a}$}
\author{Trevor Hastings$^{a}$}
\author{Wenle Xu $^{a}$}
\author{Daniel Salas$^{a}$}
\author{Joydeep Kundu$^{a}$}
\author{Vasanth C. Shunmugasamy$^{a}$}
\author{Daniel Lewis $^{a}$}
\author{Jacob Hempel $^{a}$}
\author{Clinton Strosser $^{a}$}
\author{Alexandra Salinas $^{b}$}
\author{Brent Vela$^{a}$}
\author{Nicolas Flores $^{a}$}
\author{Sina Hossein Zadeh$^{a}$}
\author{Ali Rachidi$^{d}$}
\author{David Elbert$^{d}$}
\author{Brady Butler $^{c}$}
\author{James Paramore $^{e}$}
\author{Dimitris Lagoudas $^{a}$}
\author{Justin Wilkerson $^{b}$}
\author{George Pharr $^{a}$}
\author{Douglas Allaire $^{b}$}
\author{Vahid Attari $^{a}$}
\author{Ibrahim Karaman $^{a}$}
\author{Ankit Srivastava $^{a}$}
\author{Raymundo Arr\'{o}yave$^{a,b}$}

\address{$^a$Department of Materials Science and Engineering, Texas A\&M University, College Station, TX 77843, USA}
\address{$^b$Mechanical Engineering Department, Texas A\&M University, College Station, TX, USA 77840}
\address{$^c$DEVCOM Army Research Laboratory South at Texas A\&M University, College Station, TX 77843-3003, USA}
\address{$^d$Hopkins Extreme Materials Institute, Johns Hopkins University, Baltimore, Maryland, 21218, USA}
\address{$^e$Bush Combat Development Complex, Texas A\&M University System, 3479 TAMU, College Station, TX 77843-3479, USA}

\cortext[mycorrespondingauthor]{Corresponding author email:\textrm{mrinalini.mulukutla@tamu.edu}}

\begin{abstract}

In constrained alloy optimization, the compositions with the highest performance potential often reside at the boundary of phase stability---where the risk of experimental failure is also highest. This work demonstrates this principle through a risk-aware Bayesian optimization campaign on single-phase FCC high-entropy alloys in the Al--V--Cr--Mn--Fe--Co--Ni--Cu system. A learned feasibility classifier, integrated directly into the multi-objective acquisition function, probabilistically penalizes candidates likely to produce failed experiments while preserving access to high-performing boundary compositions. From approximately 27{,}000 CALPHAD-screened candidates, 48 alloys were synthesized over three closed-loop iterations targeting five objectives (yield strength, UTS/YS ratio, strain at UTS, dynamic-to-quasi-static hardness ratio, and simulated depth of penetration), exploring 0.12\% of the feasible space. Two compositional regimes emerged: a V-rich, Ni-rich high-strength regime (UTS up to ${\sim}1480$~MPa at 50\% elongation) and a Mn-containing high-ductility regime (UTS/YS up to 4.20 at $>$50\% elongation). Among feasible alloys, vanadium simultaneously drives yield strength ($r = 0.84$) and sigma-phase formation ($r = 0.54$ with infeasibility); at V = 24~at.\%, the three strongest alloys and three sigma failures share the same compositional point. Additionally, the strongest performing alloys cluster in a narrow region of compositional space (V $\geq$ 20 at.\%, Ni $\geq$ 36 at.\%), representing ${\sim}100$ of $27{,}074$ feasible candidates---a probability of $P \approx 6.5 \times 10^{-6}$ under random sampling. This dual role---consistent with the KKT prediction that constrained optima lie on active constraint boundaries---required feasibility-aware acquisition to access; hard filtering would have excluded this region entirely.

\end{abstract}

\end{frontmatter}

\section{Introduction}\label{sec:Intro}


Structural alloys for extreme environments---from ballistic armor to gas turbine blades---underpin performance in defense, aerospace, and energy systems~\cite{pollock2016alloy, reed2006superalloys, arroyave2022perspective}. Although significant advances have been made in computational screening, high-throughput synthesis, and machine learning, the alloy development cycle remains slow and resource-intensive~\cite{liu2022highthroughput}: typically, alloy development programs proceed through iterative synthesis--processing--characterization--testing loops, and each cycle demands reproducibility (i.e., multiple evaluations) across heats, processing routes, and testing protocols. This problem is compounded by the fact that the composition--process--microstructure--performance (PSPP) space over which alloy campaigns operate is vast, and only a minute fraction has been explored directly~\cite{miracle2017critical}. In high-entropy alloy (HEA) systems, where five or more principal elements define a combinatorially large design space, exhaustive exploration is impractical even with advanced experimental methods~\cite{zhang2014microstructures, cantor2004microstructural}.


Within this broad landscape, FCC HEAs present a rich but not sufficiently mapped composition--property space. Strength and ductility are governed by competing deformation mechanisms---dislocation glide, twinning, and local chemical ordering---whose relative contributions shift with composition in ways that binary or ternary subsystem data cannot predict. Interaction effects among four or more solutes (i.e., beyond pairwise or ternary interactions) are absent from lower-order assessments, leaving large regions of the compositional space without reliable property estimates. The strength--ductility trade-off, often treated as intrinsic, may instead reflect constrained design trajectories; multi-component spaces can contain regions where twinning-induced plasticity (TWIP), transformation-induced plasticity (TRIP), or both enable simultaneous improvements~\cite{gludovatz2014fracture, li2016metastable, otto2013influences}. Thermodynamic constraints compound the challenge: CALPHAD databases show significant prediction errors at FCC stability boundaries in higher-order systems, and compositions that appear promising from property models frequently form brittle intermetallics. Mapping these composition--structure--property relationships thus requires an exploration strategy that generates both optimized compositions and the physical insight needed to interpret them.


To date, several distinct frameworks have been applied to this problem. Combinatorial and high-throughput (HTP) approaches, for example, can generate candidate compositions at scale, but open-loop strategies tend to produce a significant fraction of infeasible compositions---alloys with undesired phase constitution---making such campaigns inherently resource-inefficient~\cite{potyrailo2011combinatorial, rajan2008combinatorial}. More targeted campaigns, guided by Integrated Computational Materials Engineering (ICME) models or domain expertise, reduce infeasibility risk but require substantial computational investment and often struggle to integrate multi-scale simulations at acceptable computational cost~\cite{national2008integrated, panchal2013key, arroyave2019systems}. Machine learning (ML) approaches complement these frameworks by enabling rapid screening, yet they frequently defer experimental validation to later stages, particularly for bulk structural alloys where synthesis and testing costs are significant~\cite{yang2022machine}. Despite the progress made, the dominant bottleneck across all these approaches is resource allocation: deciding which chemistries to interrogate under a finite experimental budget~\cite{arroyave2022perspective}.


In this context, Bayesian optimization (BO) has rapidly emerged as a principled approach to optimal resource allocation in materials discovery, framing the problem as sequential decision-making under uncertainty~\cite{shahriari2016taking, frazier2018tutorial}. In a BO framework, probabilistic surrogate models (typically Gaussian processes~\cite{Rasmussen:2005:GPM:1162254}) quantify epistemic uncertainty across the design space, and acquisition functions balance (a)~exploration of poorly characterized regions with (b)~exploitation of promising candidates~\cite{snoek2012practical, arroyave2022perspective}. Multi-objective formulations extend this framework to optimize competing performance metrics simultaneously, typically by maximizing the expected hypervolume improvement (EHVI) of the Pareto front~\cite{emmerich2006single, zhao2018fast, mamun2025accelerated, hanaoka2022comparison}. Furthermore, metallurgical knowledge (thermodynamic stability limits, processing windows, and microstructural descriptors) can be encoded directly into surrogate structure and constraint models, improving data efficiency while preserving physical interpretability~\cite{khatamsaz2022multiobjective, vela2023data, paramore2025twoshot}.


Over the past decade, closed-loop experimental campaigns that couple BO with iterative alloy synthesis have validated this approach in practice. Xue et al.~\cite{xue2016accelerated} pioneered adaptive alloy design by identifying low-hysteresis NiTi-based shape memory alloys through nine feedback loops exploring fewer than 40 compositions from a space of 800{,}000 candidates. Rao et al.~\cite{rao2022machine} applied iterative ML-guided design to discover high-entropy Invar alloys with near-zero thermal expansion. More recently, Sohail et al.~\cite{sohail2025machine} used domain-knowledge-informed active learning to design FeNiCoAlTa alloys achieving 1.8~GPa yield strength with 25\% elongation. Additional closed-loop campaigns have targeted lead-free solders~\cite{wei2025discovering}, magnesium alloys~\cite{ghorbani2024active}, high-temperature shape memory alloys~\cite{tian2024noise}, and Al--Si alloys~\cite{cai2025process}, with several incorporating multi-objective acquisition strategies~\cite{wei2025discovering, li2022strength}. These studies demonstrate that BO-guided campaigns can identify competitive alloy compositions while exploring less than 1\% of the feasible design space.

\begin{figure}[H]
    \centering
    \includegraphics[width=1\columnwidth]{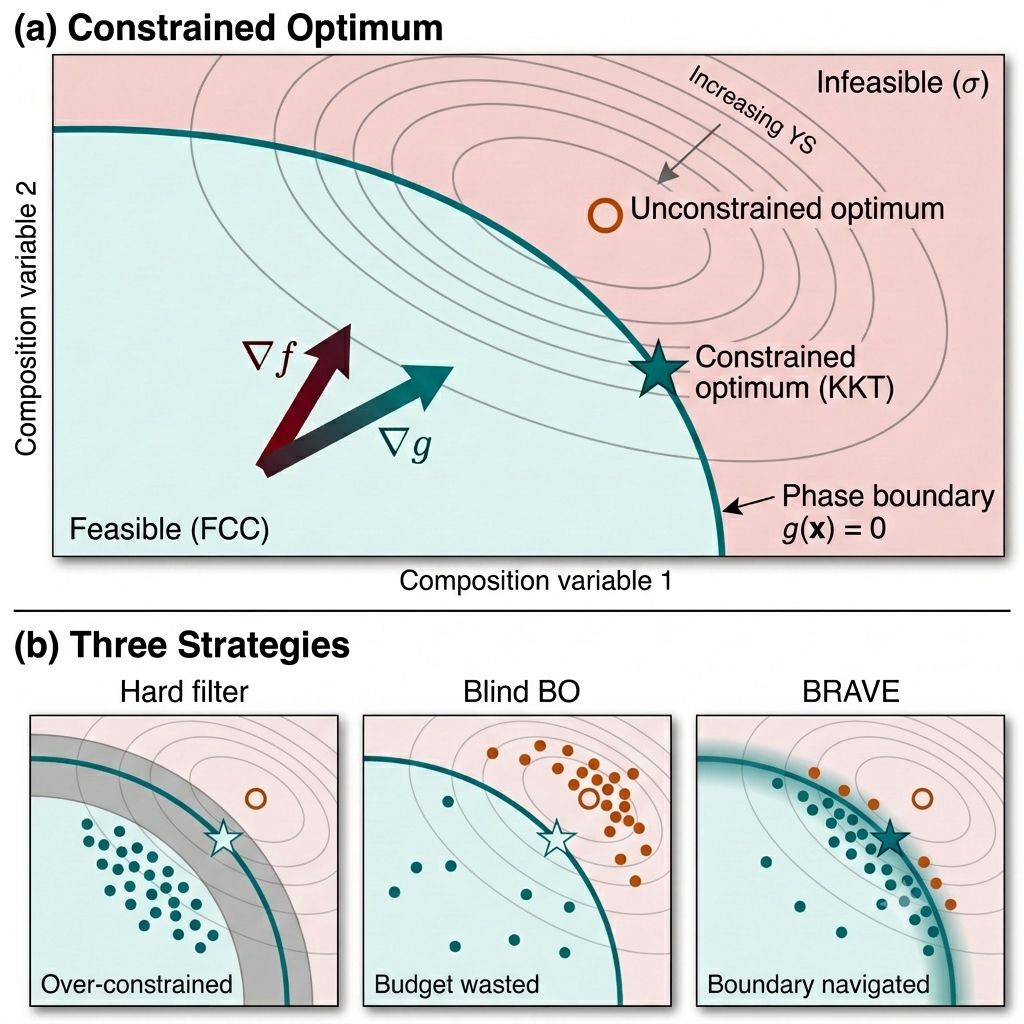}
    \caption{Constrained optimization in composition space. (a)~When the objective (e.g., yield strength) increases toward the infeasible region (sigma phase), the KKT conditions place the constrained optimum (star) on the phase stability boundary where $\nabla f$ and $\nabla g$ are aligned; the unconstrained optimum (open circle) lies in the infeasible region. (b)~Three acquisition strategies distribute experimental budget differently given the same total number of samples. Hard filtering over-constrains the search by excluding a buffer zone around the boundary, missing the KKT optimum entirely. Feasibility-blind BO concentrates samples near the unconstrained optimum, wasting budget on infeasible compositions. BRAVE navigates the boundary by concentrating samples in the feasible region near the phase stability limit, accepting a small number of infeasible outcomes as the cost of boundary exploration.}
    \label{fig:kkt_schematic}
\end{figure}

However, a practical constraint remains unaddressed in these campaigns: many candidate designs are infeasible and yield no measurable objective values, wasting experimental resources without contributing to surrogate model refinement. We note that in alloy discovery, failure does not indicate poor performance (e.g., low yield strength) but rather that experimental time and resources are expended without obtaining usable measurements---because the alloy cannot be synthesized, processed, or tested under the prescribed conditions. This distinction is consequential: a low-performing alloy still provides training data for the surrogate model, whereas a failed experiment provides none. The cost of each experimental iteration---several thousands of dollars in materials, instrument time, and personnel---means that even a small fraction of failed experiments can substantially reduce campaign efficiency. Yet standard BO frameworks implicitly assume that every queried candidate returns measurable objectives, offering no mechanism to account for experimental viability during candidate selection. \autoref{fig:kkt_schematic} illustrates why this matters: in constrained optimization, the best solutions tend to reside on the feasibility boundary (a consequence of the Karush--Kuhn--Tucker optimality conditions~\cite{nocedal2006numerical}), precisely where the risk of failure is highest. A framework that excludes boundary compositions sacrifices the best candidates; one that ignores feasibility wastes budget on infeasible ones.

Constraint-aware acquisition strategies have begun to address this gap across several domains. Hickman et al.~\cite{hickman2025anubis} introduced feasibility-aware acquisition functions that reduce failed experiments in perovskite and drug synthesis. Low et al.~\cite{low2024evolution} developed evolution-guided constrained optimization for self-driving chemistry laboratories. Chen et al.~\cite{chen2026materials} coupled constrained expected improvement with representation learning for computational catalyst and transparent conductor design. Tamura et al.~\cite{tamura2023nims} integrated feasibility constraints into autonomous robotic materials exploration. In alloy systems specifically, constraint-aware formulations have been proposed~\cite{khatamsaz2022multiobjective, khatamsaz2023bayesian}, and physics-informed feasibility classifiers have been developed~\cite{hardcastle2025physics}, but no experimental alloy campaign has yet integrated a learned feasibility model directly into the multi-objective acquisition function within an iterative workflow.

\begin{figure}[H]
    \centering
    \includegraphics[width=1\columnwidth]{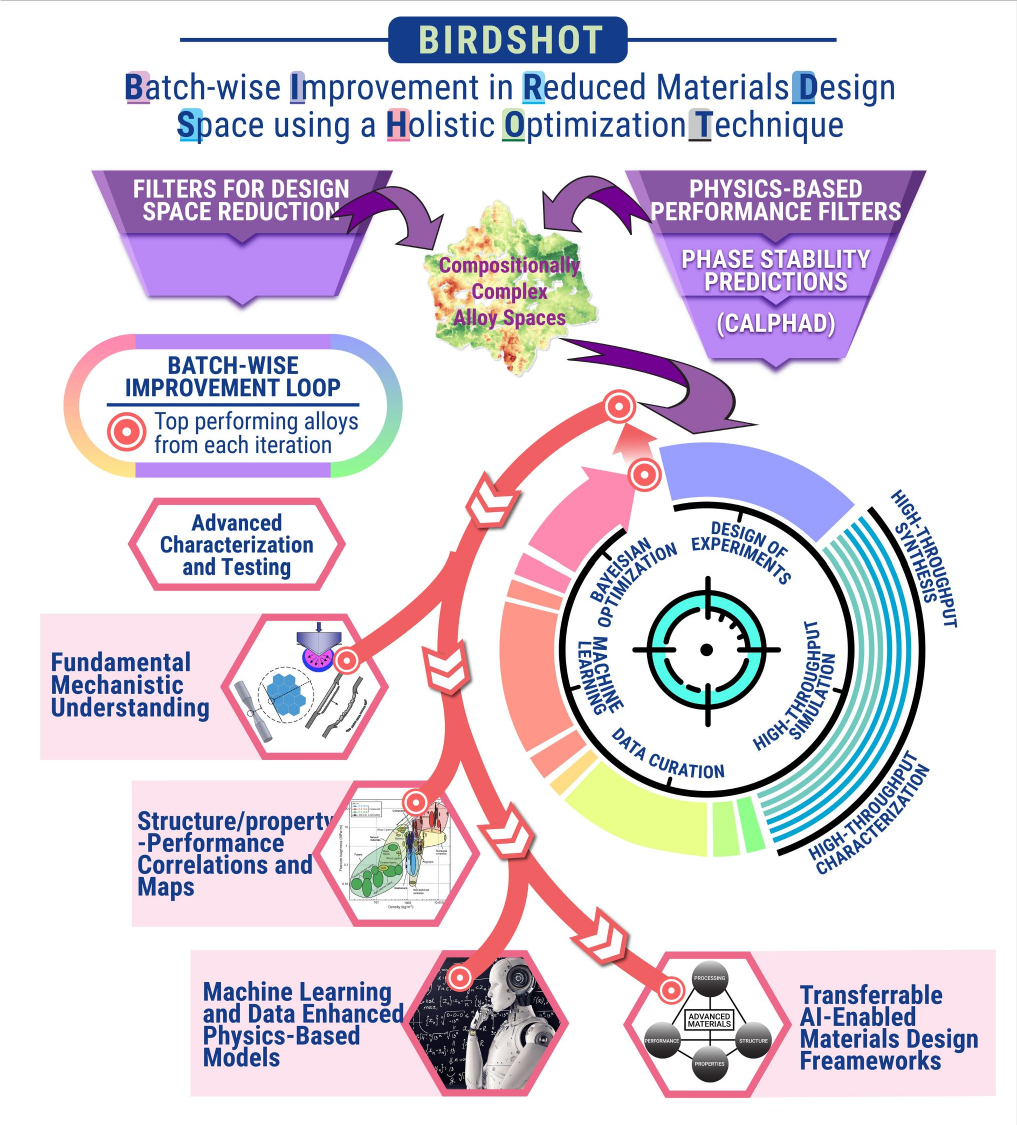}
    \caption{Overview of the BIRDSHOT framework. Physics-based filters (CALPHAD phase stability, property constraints) reduce the design space before optimization begins. The iterative loop cycles through design of experiments, high-throughput simulation, high-throughput characterization, data curation, and machine learning, with Bayesian optimization closing the loop at each iteration. Reproduced from:~\cite{hastings2025accelerated}.}
    \label{fig:birdshot_framework}
\end{figure}

The BIRDSHOT (Batch-wise Improvement in Reduced Design Space using a Holistic Optimization Technique) framework was developed to integrate computational screening, high-throughput experimentation, and batch Bayesian optimization within an iterative design--make--test--learn workflow~\cite{mulukutla2024illustrating, hastings2025accelerated} (\autoref{fig:birdshot_framework}).

In a prior campaign, we applied BIRDSHOT to the six-component Al--V--Cr--Fe--Co--Ni FCC HEA system, targeting three mechanical objectives (UTS/YS ratio, hardness, and strain rate sensitivity) under four constraints (single-phase FCC at 700$^\circ$C, density $<$ 8.5~g/cm$^3$, thermal conductivity $>$ 5~W/(K$\cdot$m), and SFE regime classification)~\cite{hastings2025accelerated}. Five iterative design--make--test--learn loops, each synthesizing 8 alloys via vacuum arc melting, identified Pareto-optimal compositions by exploring only 0.15\% of the feasible design space. That campaign established the viability of iterative, multi-objective Bayesian optimization for structural alloy development. Although successful in identifying Pareto-optimal compositions, it also exposed limitations. CALPHAD phase stability predictions were imperfect: several alloys predicted to be single-phase FCC showed sigma or BCC secondary phases. The framework had no mechanism to account for these failed experiments, which consumed resources without contributing to model refinement. A detailed summary of this earlier work is provided in the Supplemental Information.


In the present work, we report a second discovery campaign that extends the BIRDSHOT framework in three directions, addressing the limitations identified above. The campaign also exposes a tension inherent to this alloy system: vanadium, the dominant strengthening solute, simultaneously promotes sigma-phase formation, forcing the best alloys onto the FCC phase stability boundary where the risk of experimental failure is highest.


First, we introduce a risk-aware acquisition strategy---Bayesian Risk-Aware Alloy Discovery and Exploration (BRAVE)---that penalizes candidates likely to produce failed experiments (i.e., alloys that form secondary phases and yield no usable objective data). Alloys predicted to form single-phase FCC structures but found to contain secondary phases are classified as failures: they introduce processing challenges and do not provide a consistent basis for comparison within the intended single-phase design space. To mitigate this risk, we extend the EHVI acquisition function with a learned Gaussian process classifier that estimates the probability of experimental viability for each candidate composition. High-risk candidates receive a probabilistic penalty during acquisition scoring, reducing their selection priority. We note that \emph{the penalty is soft}: a candidate with sufficiently high predicted objective values can still be selected despite elevated risk, preserving controlled exploration in uncertain but potentially high-performing regions.


Second, we expand the design space from the previous 6-element system to an 8-element system, Al--V--Cr--Mn--Fe--Co--Ni--Cu. The inclusion of Mn and Cu introduces additional degrees of freedom for tuning ductility and phase stability, guided by prior experimental insights and domain knowledge. Mn is expected to reduce the stacking fault energy in FCC alloys, potentially promoting mechanical twinning and enhancing strain-hardening capacity. Cu expands the accessible composition space but has positive mixing enthalpies with Fe, Co, and Cr, driving FCC phase separation into Cu-rich and Cu-lean regions and creating thermodynamic boundaries that test the limits of current CALPHAD predictions. Grid sampling at 4~at.\% resolution yields approximately 3.4 million candidate compositions, of which $\sim$27{,}000 (i.e., less than 1\%) are predicted to form single-phase FCC structures after CALPHAD screening. To enable tractable exploration of this feasible space, we use a diversity-aware sampling strategy based on k-medoids clustering~\cite{paramore2025twoshot} across 37 chemically distinct subsystems. This prioritizes compositionally informative candidates over redundant clusters, improving early-stage surrogate learning through better coverage of the design space.


Third, five objectives are evaluated through a coordinated high-throughput characterization workflow: yield strength (YS), UTS/YS ratio, strain at UTS, dynamic-to-quasi-static hardness ratio ($H_\text{dyn}/H_\text{qs}$), and depth of penetration (DoP). These objectives serve as complementary probes of deformation behavior across strain rates, spanning (a)~quasi-static tensile response, (b)~rate-dependent behavior measured by nanoindentation and split Hopkinson pressure bar (SHPB) testing, and (c)~simulated impact resistance computed via finite-element modeling. Among these, DoP is derived from ballistic impact simulations parameterized by Cowper--Symonds constitutive models calibrated to the quasi-static tensile and dynamic SHPB data acquired within each iteration. This creates a simulation-in-the-loop architecture in which experimental measurements from one stage of characterization feed directly into a computational objective evaluated within the same design cycle, coupling experiment and simulation within a single design iteration.

\begin{figure*}[t]
    \centering
    \includegraphics[width=1\textwidth]{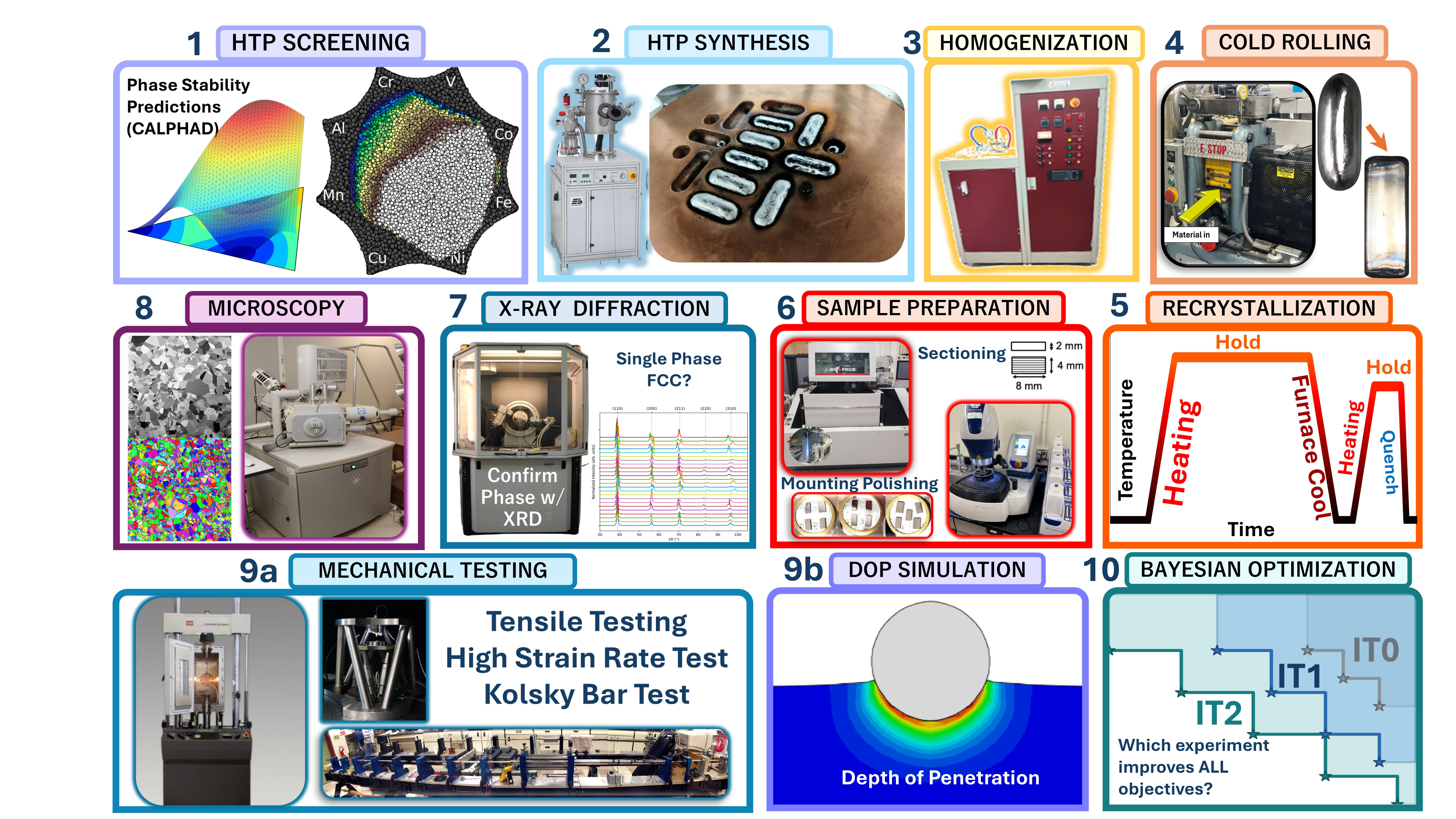}
    \caption{BIRDSHOT workflow from computational design through synthesis, characterization, and simulation. The iterative loop feeds experimental measurements and phase verification results back into the Bayesian optimization framework, which updates surrogate models and the feasibility classifier before selecting the next batch of candidate alloys.}
    \label{fig:full_framework}
\end{figure*}

The complete BIRDSHOT framework---encompassing computational design, synthesis, characterization, and simulation---is summarized in \autoref{fig:full_framework}. By addressing thermodynamic constraints, manufacturability, and experimental throughput within a closed-loop workflow, this campaign demonstrates risk-aware, multi-objective optimization through three iterative experimental cycles. The remainder of this paper presents the design methodology, experimental results across three closed-loop iterations, and the implications for feasibility-aware optimization in high-dimensional compositional spaces.

\section{Methods}\label{sec:Material and methods}

\subsection{Alloy Design Methodology}

\subsubsection{Design Space Definition and Compositional Constraints}

The design space is an 8-element HEA system, Al--V--Cr--Mn--Fe--Co--Ni--Cu, extending the 6-element Al--V--Cr--Fe--Co--Ni system investigated in Campaign~1~\cite{hastings2025accelerated} by the addition of Mn and Cu (rationale discussed in Section~1). The system includes all subsystems from 5 to 8 elements. Unlike Campaign~1, which imposed four constraints (single-phase FCC, density, thermal conductivity, and SFE regime), this study enforces only phase stability (i.e., single-phase FCC across the designated temperature range), adopting an open-ended strategy that avoids prematurely excluding regions of the composition space.

The initial compositional space is discretized at a resolution of 4 atomic percent, resulting in a pool of 3,365,848 unique compositions. To ensure manufacturability and focus on regions with potential for high performance, expert recommendations were used to impose upper limits on elemental fractions: Al $\leq 0.24$, V $\leq 0.24$, Cr $\leq 0.50$, Mn $\leq 0.40$, Fe $\leq 0.50$, Co $\leq 0.50$, Ni $\leq 0.60$, Cu $\leq 0.24$. These compositional bounds reduce the design space to 1,504,938 candidate alloys, of which 1,490,129 were successfully evaluated using CALPHAD due to solver limitations encountered on the high-performance computing cluster.


\subsubsection{Candidate Screening via CALculation of PHAse Diagram (CALPHAD) Predictions}

To enforce the FCC phase stability constraint, we screened the reduced compositional space using CALPHAD simulations. Thermo-Calc demonstrated effective single-phase predictions for the Al-V-Cr-Fe-Co-Ni system in Campaign~1~\cite{hastings2025accelerated}. Using the TCHEA6 database, we retained compositions with FCC phase fraction $\geq 0.99$ at 700$^\circ$C, reducing $\sim$3.36 million candidates to a feasible space of 26,621 alloys (approximately 0.8\% of the original design space). Visualizations of this downselection are included in the \textbf{SI}.

Phase prediction discrepancies in the first optimization iteration (BBB)---alloys predicted as single-phase FCC that experimentally formed sigma---motivated a reanalysis using the updated TCHEA7 database, released during this study. TCHEA7 enabled prediction of multiple phase regions (FCC, BCC, HCP, SIGMA, and liquid) across different temperature regimes. We reevaluated the design space at both 700$^\circ$C and 950$^\circ$C, advancing only alloys that retained FCC under both conditions. This dual-temperature screening yielded a final feasible set of approximately 27,240 candidate alloys.

\subsubsection{Sampling and Candidate Selection}

Given the large number of candidate alloys resulting from the initial filtering and screening (over 26,000), an efficient sampling strategy was critical to balance experimental feasibility with adequate coverage of the design space. For iteration BBA (zeroth iteration), sampling was performed over the initial TCHEA6-screened feasible space of 26,621 alloys (the TCHEA7 re-screening to 27,240 alloys occurred after iteration BBB). We implemented a hierarchical sampling strategy using k-medoids clustering. First, the entire feasible space was segmented based on chemical composition into 37 distinct subsystems, defined by the specific combination of elements present (for example, alloys containing all eight elements formed one subsystem, while seven-element alloys were grouped by which element is absent). Details of subsystem grouping are provided in \textbf{SI}.

Subsequently, we applied k-medoids clustering to the feasible space to extract a subset of 1000 representative alloys. In this clustering process, we enforced a constraint that at least one candidate alloy was selected from each of the 37 subsystems, with the number of selections per subsystem proportional to the number of feasible candidates available within that system. This approach guaranteed broad chemical coverage and preserved the diversity of the design space.

From this subset of 1000 alloys, a further down-selection identified 16 candidate alloys for synthesis. The downselection maximized representation across all 37 subsystems, ensuring that the final batch covered as many distinct chemistries as possible and provided a well-distributed initial training dataset for the Bayesian optimization loop.

\subsection{Synthesis}

The computationally designed alloys were synthesized using high-purity elemental ($>$ 99.9 wt.\%) precursors to produce $\sim$35~g dense, crack-free ingots using a Vacuum Arc Melting (VAM) (Edmund B\"{u}hler AM 200) system. The elements were initially melted in Cu crucibles under a high-purity Ar atmosphere. Each ingot was flipped and re-melted 10 times to promote overall uniform chemical homogeneity. Mass loss was monitored throughout processing, and ingots showing losses greater than 0.5\% were reprocessed. Post-solidification, the ingots were homogenized between 950$^\circ$C to 1150$^\circ$C temperature ranges for 24~h in a Centorr LF Series Model~22 high-temperature furnace. The homogenization heat treatment temperatures were selected based on the predicted solidus temperature. Prior to heating, the furnace chamber was purged with Ar gas multiple times to achieve a vacuum of $\sim$5~$\times$~10$^{-5}$~torr. The heat treatment was conducted under a low-pressure Ar atmosphere ($<$10$^{-2}$~torr), followed by furnace cooling.

The homogenized ingots were subsequently cold rolled to a $\sim$60\% reduction in thickness. For recrystallization, the rolled specimens were wrapped in stainless-steel foil with Ti sponge sacrificial getter to minimize oxidation. Recrystallization was performed at 950\textdegree C for 30~min, followed by water quenching to suppress secondary-phase precipitation during cooling.

\begin{figure}[t]
    \centering
    \includegraphics[width=0.9\columnwidth]{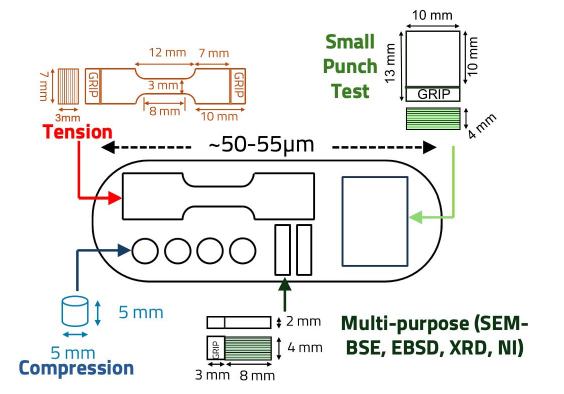}
    \caption{Representative schematics of ingots post-cold rolling with corresponding approximate sample dimensions for distinct characterization tasks: SEM/EBSD, EDS, XRD, and mechanical (Vickers, Nanoindentation) analyses, quasistatic tensile deformation, and Kolsky-bar compression dynamic mechanical response tests.}
    \label{fig:Sample Cutting Protocol}
\end{figure}

Test sample profiles were precision-cut using wire-Electrical Discharge Machining (EDM). Approximately 4 miniaturized flat tensile specimens, modeled following modified SS-3 designs, were prepared for quasistatic tension testing. Each tensile specimen had a total length of $\sim$26~mm, a gauge length of $\sim$8~mm, and a typical cross-sectional area of 3~$\times$~1~mm$^{2}$, with the loading axis aligned parallel to the rolling direction. Prior to testing, the tensile samples were ground using 120~grit SiC abrasive papers to ensure grip section parallelism, and pinholes were drilled using carbide drill bits. Additional specimens were prepared for microstructural and mechanical characterization, as follows: two specimens, each $\sim$2~mm thick with an exposed face area of 8.4~mm$^{2}$ normal to the long rolling direction, were designated for nanoindentation. Another two samples, 1.2~mm thick with a face area of 14.4~mm$^{2}$ (also normal to the rolling direction), were allocated for scanning electron microscopy (SEM), electron backscatter diffraction (EBSD), energy-dispersive X-ray spectroscopy (EDS), X-ray diffraction (XRD), and Vickers hardness (HV) measurements. A schematic of the specimen geometries is shown in \autoref{fig:Sample Cutting Protocol}.

We note that manganese contents exceeding 20 atomic percent introduced significant challenges during arc melting. Mn has a relatively low boiling point and high vapor pressure compared to the other constituent elements, making it volatile under the localized temperatures generated during arc melting. This volatility leads to Mn evaporation, resulting in deviations from the nominal composition and causing microstructural inhomogeneities and porosity in the cast ingots. Mn loss disrupts the stability of the intended single-phase FCC structure and promotes the formation of intermetallic compounds or segregated phases.

\subsection{Verification}

\subsubsection{Metallography}

Microstructural characterization and nanoindentation samples were prepared by mechanical polishing using a Buehler AutoMet 250 autopolisher with diamond suspensions down to 1~$\mu$m, followed by vibratory polishing in 0.04~$\mu$m colloidal silica for EBSD analysis. A high-throughput sample preparation workflow was developed over the course of the campaign, evolving from individual mounting (8 samples per batch) to large-format aluminum platens supporting 16 or more samples simultaneously, with laser marking for traceability. Details of this workflow evolution are provided in the \textbf{SI}.

\subsubsection{Chemical, Microstructural, and Phase Analyses}

\begin{figure}[t]
    \centering
    \includegraphics[width=0.9\columnwidth]{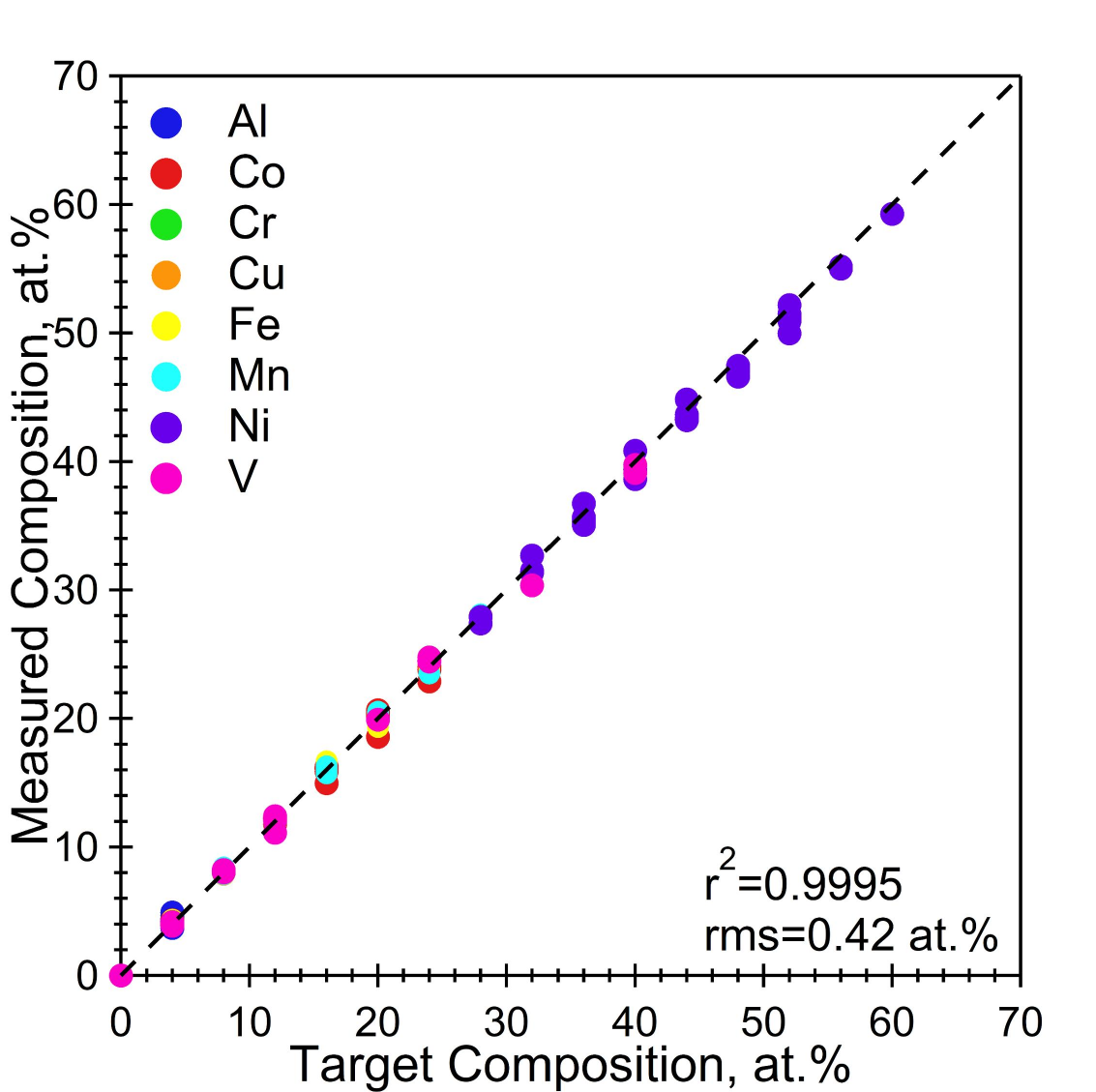}
    \caption{Comparison between target and realized chemistry for the samples generated in this study.}
    \label{fig:EDS}
\end{figure}

Compositional analysis was performed using a Tescan FERA-3 Model GMH Focused Ion Beam Microscope equipped with an Oxford energy-dispersive X-ray spectroscopy (EDS) detector, operated at an accelerating voltage of 20 kV. EDS point measurements were obtained from three spatially distinct regions of each processed ingot, with five acquisition points per region and an integrated time of $\sim$50~s per point. The overall composition of each ingot was determined as the arithmetic mean of the 15 measurements. Inter-regional variability was evaluated to assess chemical homogeneity. Representative EDS results from three independent synthesis iterations are presented in \autoref{fig:EDS}. The mean deviation between measured and nominal compositions was $\sim$0.42~at.\%, with 99\% of individual elemental measurements deviating by less than 1~at.\%, confirming compositional uniformity across all synthesis batches.

Microstructural characterization of the undeformed samples was performed by backscattered electron (BSE) imaging at 15~kV using a Phenom XL scanning electron microscope (SEM) with a working distance of $\sim$6~mm. Electron backscatter diffraction (EBSD) measurements were carried out using a FERA SEM equipped with an Oxford Symmetry CMOS-based EBSD detector, operated at an accelerating voltage of 20~kV, a beam current of 18~nA, and a specimen tilt angle of 70$^\circ$. Crystallographic orientation (IPF) maps were acquired on surfaces normal to the cold-rolling direction over an area of approximately 1~$\times$~1~mm$^{2}$, using a step size of $\sim$2~$\mu$m.

Room-temperature X-ray diffraction (XRD) measurements were conducted using a Bruker D8 Discover diffractometer with Cu K-$\alpha$ radiation ($\lambda$ = 1.5406~\AA) and a 1~mm aperture collimator. Diffraction patterns were collected using a Vantec 500 area detector over a 2$\theta$ range of 30--80$^\circ$. Phase identification and diffraction pattern analysis were performed using the MAUD software package.

\subsection{Characterization}

The optimization targets five objectives that together span the quasi-static--to--dynamic response space: (a)~yield strength (YS), (b)~the UTS/YS ratio (a proxy for strain hardening capacity and resistance to strain localization), and (c)~engineering strain at UTS (ductility reserve before peak load), all three extracted from uniaxial tensile tests; (d)~the dynamic-to-quasi-static hardness ratio ($H_\text{dyn}/H_\text{qs}$), obtained from nanoindentation and SHPB measurements, which quantifies how much the alloy strengthens under impact loading; and (e)~depth of penetration (DoP), computed from finite-element ballistic simulations calibrated to the tensile and SHPB data acquired within the same iteration---the only objective derived from simulation rather than direct measurement. Campaign~1 used hardness as an objective; here we replace it with the tensile-derived metrics, which connect more directly to bulk deformation behavior and are less sensitive to local microstructural heterogeneity.

\subsubsection{Quasistatic Tensile Deformation}

Uniaxial tensile tests were conducted at ambient temperature using an MTS 810 servo-hydraulic testing system equipped with an 11.12~kN (2500~lbf) load cell, under a constant quasi-static strain rate of 5~$\times$~10$^{-4}$~s$^{-1}$. Axial strain was measured using an MTS clip-on extensometer mounted directly on the specimen gauge section. An initial force-controlled loading sequence was performed to verify specimen alignment, modulus consistency, and linearity of the elastic response. At least 2 tests were conducted for each condition to ensure reproducibility. Tensile tests were terminated upon complete fracture. True stress--strain ($\sigma$--$\varepsilon$) curves were calculated, from which the ultimate tensile strength (UTS), 0.2\% offset yield strength (YS), and uniform elongation ($\varepsilon$) were determined.

\subsubsection{Nanoindentation}



Each sample was indented using a KLA Instruments iMicro nanoindenter fitted with a Berkovich diamond tip. Hardness ($H$) and reduced modulus ($E_r$) were determined following the Oliver and Pharr method \cite{Oliver1992}. Conventional nanoindentation hardness values were obtained from strain-rate jump tests using a constant $\dot{P}\!/P$ protocol, with hardness reported at an indentation depth of \SI{2000}{\nano\meter}. A minimum of 10 strain-rate jump indents was performed for each alloy. Indents were typically spaced \SI{100}{\micro\meter} apart to prevent overlap between neighboring plastic zones.
Strain-rate jump tests were conducted to assess rate-dependent mechanical behavior across the candidate alloys. These measurements were used to estimate indentation strain-rate sensitivity. The strain-rate sensitivities obtained from nanoindentation were cross-validated against SHPB results and were found to be in close agreement, typically within one standard deviation.
Projected contact areas were corrected for pile-up for each individual indent. The projected contact area was calculated from the calibrated tip area function, $A(h_c)$, determined from fused silica calibration \cite{Oliver1992}, and evaluated at the contact depth, $h_c$. The contact depth was calculated from the measured indentation depth, h, using an experimentally measured pile-up ratio,
\[
\eta = \frac{h_c}{h}
\]

The pile-up ratio was measured after each indentation using optical profilometry. First, the nominal indentation area, $A_N$, was determined from the region below the original sample surface. This area scales with $h^2$. The actual residual contact area, $A_R$, was then segmented using a gradient-based method that identifies the Berkovich impression boundary; this area scales with $h_c^2$. Assuming self-similar Berkovich geometry, the pile-up ratio was calculated as
\[
\eta = \frac{h_c}{h} \approx \sqrt{\frac{A_R}{A_N}}
\]

An independent value of $\eta$ was measured for every indent and applied to the corresponding hardness and modulus calculation. This correction was applied to strain-rate jump, quasistatic, and dynamic hardness measurements. The pile-up correction typically reduced the measured hardness by approximately 5 to 15\%, depending on alloy composition and testing condition. Custom analysis scripts were developed to extract the instrument data and automate projected contact area quantification; additional details of the image-analysis workflow are provided in the SI.

\subsubsection{High Strain Rate Nanoindentation}

High strain rate nanoindentation was performed on a separate custom instrument adapted for indentation impact testing at strain rates exceeding $10^{3}\ \mathrm{s^{-1}}$. The instrument includes a hexapod specimen stage and an InForce 1000 actuator customized to measure displacements at high acquisition rates using a Michelson interferometer; full details of this setup are provided by Hackett \etal\ (2023). All tests were performed with a Berkovich three-sided pyramidal diamond tip.

The high strain rate instrument was used to measure the dynamic-to-quasistatic hardness ratio, $H_{\mathrm{DYN}}/H_{\mathrm{QS}}$, which was incorporated into the Bayesian optimization loop. Dynamic hardness, $H_{\mathrm{DYN}}$, was measured from indentation impact tests with an effective $\dot{P}/P = 2 \times 10^{3}\ \mathrm{s^{-1}}$. Quasistatic hardness, $H_{\mathrm{QS}}$, was measured on the same instrument using a constant $\dot{P}/P = 2 \times 10^{-1}\ \mathrm{s^{-1}}$. A minimum of 10 dynamic impact indents and 10 quasistatic indents was performed for each alloy, with indents typically spaced $100\ \mu\mathrm{m}$ apart.

Dynamic tests used an impact force of $20\ \mathrm{mN}$, followed by reloading at the same indentation location to $1\ \mathrm{N}$ at $\dot{P}/P = 2 \times 10^{-1}\ \mathrm{s^{-1}}$. The reloading step was included to compare the dynamic-to-quasistatic hardness ratio obtained from separate tests with the ratio obtained from a single indentation location; those results will be presented in a separate study. Independent optical profilometry measurements were used to determine $\eta$ for each dynamic and quasistatic indent, and the corresponding pile-up correction was applied to each hardness value before calculating $H_{\mathrm{DYN}}/H_{\mathrm{QS}}$.

Custom analysis scripts were developed to extract data from the instrument's encrypted binary output and automated projected contact area quantification (see \textbf{SI} for details). High strain rate hardness data was acquired for all candidate alloy compositions and incorporated into the Bayesian optimization loop.

\subsubsection{Split-Hopkinson Pressure Bar (SHPB) Testing}

To characterize the high strain rate mechanical behavior of the high-entropy alloys (HEAs) developed in this program, Split-Hopkinson Pressure Bar (SHPB) testing was performed on all candidate alloys. For each composition, a minimum of two specimens were tested, and approximately half of the alloys were evaluated at multiple strain rates, namely 1500\,s$^{-1}$ or both 1500\,s$^{-1}$ and 3000\,s$^{-1}$.

The SHPB experiments used a pulse-shaping technique to achieve and maintain a uniform strain rate throughout the major portion of the specimen deformation. A Ni-30Mo alloy was used as the pulse shaper due to its strain hardening behavior, which closely resembles that of the tested HEAs. The SHPB setup captures three essential waveforms: the incident pulse, transmitted pulse, and reflected pulse.

\begin{figure}[t]
    \centering
    \includegraphics[width=0.9\columnwidth]{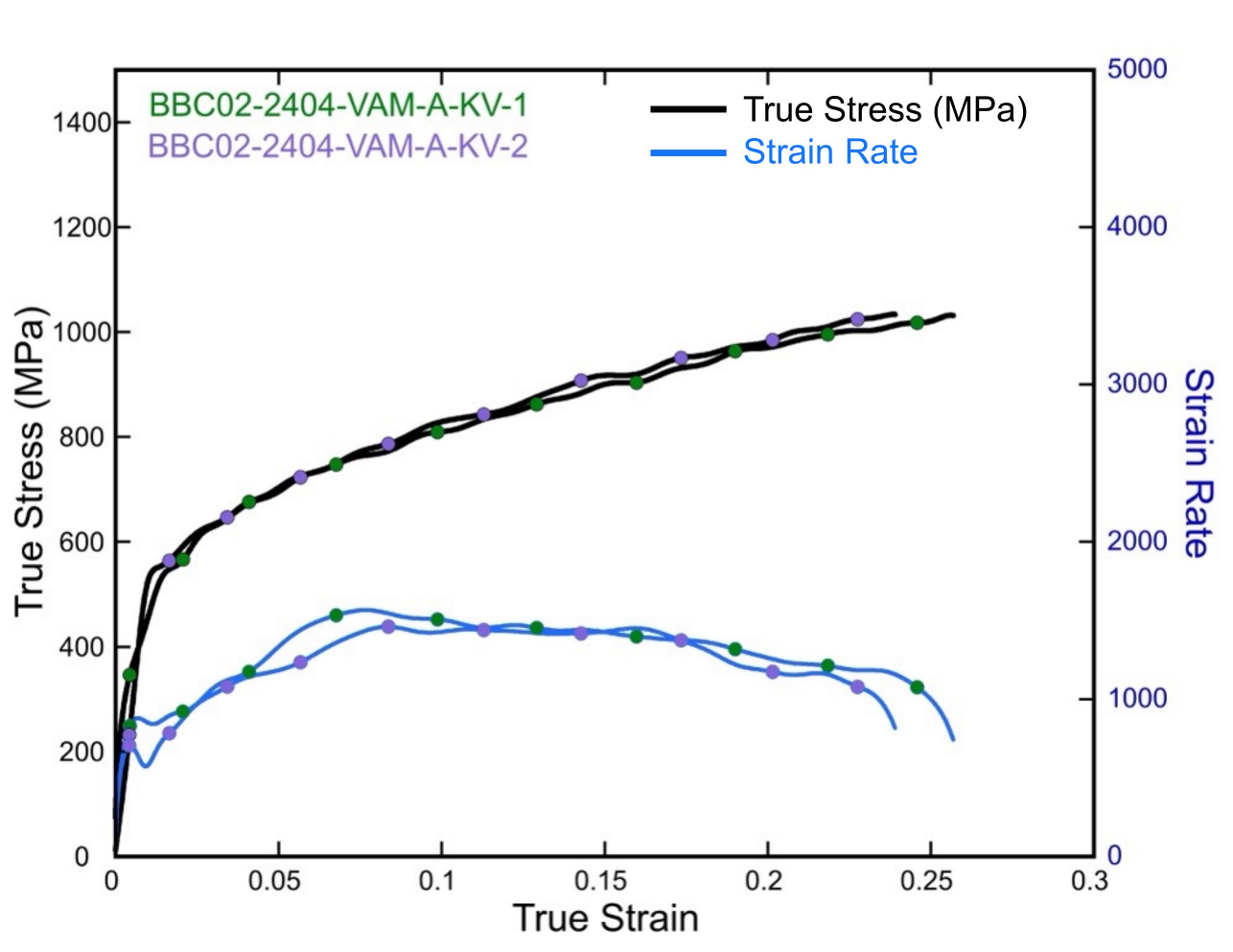}
    \caption{Dynamic true stress–strain responses (black) and corresponding strain rate as a function of strain (blue) for two HEA samples tested at 1500\,s$^{-1}$.}
    \label{fig:shpb_signals}
\end{figure}

These signals are time-synchronized to construct the dynamic stress–strain response of the material. \autoref{fig:shpb_signals} shows representative results from two SHPB tests performed at a nominal strain rate of 1500\,s$^{-1}$. The black curves represent the true stress–strain response, while the blue curves illustrate the instantaneous strain rate as a function of strain. These measurements provide the dynamic response data required for the multi-objective optimization framework.

\subsection{Depth of Penetration Simulation}

\emph{Viscoplasticity Constitutive Model Implementation:} We modeled the ballistic impact response using the Cowper-Symonds (CS) viscoplasticity framework to capture strain rate sensitivity under dynamic loading. The CS model provides a description of the flow stress evolution as a function of plastic strain rate through a multiplicative enhancement factor applied to the quasi-static flow strength. The constitutive relationship is expressed as:

\begin{equation}
\sigma_f = \sigma_{qs}^f \left(1 + \left(\frac{\dot{\varepsilon}_p}{D}\right)^{1/M}\right)
\end{equation}

where $\sigma_{qs}^f$ represents the quasi-static flow strength that evolves with accumulated plastic strain due to work hardening mechanisms, $\dot{\varepsilon}_p$ denotes the effective plastic strain rate, $D = 10^6 \text{ s}^{-1}$ serves as a material-dependent constant multiplier, and $M$ constitutes the strain rate sensitivity exponent within the CS formulation. This formulation captures dynamic strengthening at elevated loading rates, where dislocation motion becomes increasingly constrained.

In the Cowper--Symonds form, $D$ and $M$ are mathematically coupled: for a given pair of quasi-static and dynamic flow-stress measurements, a continuum of $(D, M)$ combinations can reproduce the data with comparable fidelity. With one SHPB curve and one quasi-static curve per alloy, joint identification of both parameters is underdetermined. We resolve this by fixing $D = 10^{6}\ \text{s}^{-1}$ uniformly across all alloys and calibrating $M$ per alloy by RMSE minimization against the experimental QS+SHPB data (Eq.~\ref{eq:rmse}. This convention isolates the alloy-to-alloy variation in a single rate-sensitivity parameter, allows direct comparison of $M$ across the composition set on a common basis.

\emph{Rate Sensitivity Parameter Identification:} The strain rate sensitivity parameter $M$ for each high entropy alloy composition was determined through a systematic calibration procedure based on experimental data obtained from quasistatic uniaxial tension tests and dynamic SHPB compression experiments conducted at strain rates ranging from $10^{-3} \text{ s}^{-1}$ to $10^4 \text{ s}^{-1}$. The parameter identification process used a nonlinear least-squares optimization algorithm to minimize the root mean squared error (RMSE) between the CS model predictions and experimental true stress measurements:

\begin{equation}
\text{RMSE} = \sqrt{\frac{1}{N}\sum_{i=1}^{N}\left(\sigma_{pred,i}^f - \sigma_{obs,i}^f\right)^2}
\label{eq:rmse}
\end{equation}

where $N$ represents the total number of discrete data points in the experimental stress-strain response, $\sigma_{pred,i}^f$ denotes the model-predicted flow stress at the $i$-th data point, and $\sigma_{obs,i}^f$ corresponds to the experimentally observed true stress value. The optimization procedure was implemented under isothermal conditions to isolate the strain rate effects from thermal softening mechanisms. The calibrated rate sensitivity exponents for the investigated high entropy alloy compositions ranged between 2 and 7, consistent with the expected strain rate sensitivity behavior of face-centered cubic polycrystalline materials.

\emph{Finite Element Ballistic Impact Simulation:} We implemented the calibrated material parameters into a user-defined material subroutine (UMAT) within the ABAQUS/Explicit finite element framework to simulate the ballistic impact response. The computational model used a two-dimensional axisymmetric formulation to simulate the normal impact of a rigid spherical projectile onto a cylindrical target of the candidate alloy. The target geometry consisted of a cylinder with radius $r_t = 34$ mm and thickness $t = 17$ mm, while the projectile was modeled as a rigid sphere with diameter $d_p = 10$ mm impacting at a velocity of $V_0 = 500$ m/s.

The finite element discretization used explicit, linear, axisymmetric stress elements (CAX4R) with a characteristic mesh size of 0.8 mm, resulting in a computational domain of 42 $\times$ 21 elements in the radial and axial directions, respectively. The target material properties included an elastic modulus $E_t = 200$ GPa, Poisson's ratio $\nu_t = 0.3$, and density $\rho_t = 7.85 \times 10^{-9}$ ton/mm³. Mesh distortion controls were implemented using default ABAQUS settings. The elastic constants and density are held fixed across all the alloys. This reflects both the available characterization data and a deliberate scoping choice. The quasi-static tension and SHPB experiments resolve only the plastic flow response of each alloy; elastic moduli and densities were not independently measured for individual compositions, so representative steel-class values ($E_t = 200$ GPa, $\rho_t = 7.85 \times 10^{-9}$ ton/mm$^3$) were used uniformly. For Ni-stabilized FCC HEAs representative of the type studied here, Young's modulus and Poisson's ratio are expected to vary by less than $\sim$10\% across compositions and to contribute negligibly to penetration response, which is dominated by plastic dissipation at 500 m/s.

\emph{Depth of Penetration Quantification:} The ballistic performance of each alloy composition was quantified through the depth of penetration (DoP), defined as the maximum axial displacement of the projectile centroid into the target at the instant when the projectile velocity approaches zero. DoP serves as the primary response variable for correlating material properties with impact performance. All alloys across the three optimization iterations were evaluated under identical loading conditions using their individually calibrated viscoplastic constitutive models.

\subsection{Computational Design Framework}

The computational component of BRAVE linked thermodynamic screening, physics-informed surrogate modeling, feasibility classification, multi-objective acquisition, and diversity-aware batch selection into one loop. In each iteration, experimentally verified FCC alloys provided objective data for surrogate updates, while infeasible alloys contributed only their phase labels to the feasibility model. A more detailed end-to-end workflow diagram is provided in the \textbf{SI}.

\subsubsection{Surrogate Models and Objective Representation}

We used Gaussian process surrogate models because they provide both property predictions and uncertainty estimates, which are needed for Bayesian optimization. The surrogates were initialized with informative priors rather than zero-mean priors, using the Varvenne--Luque--Curtin solid solution strengthening model together with prior campaign data to provide a physics-grounded starting point before substantial new data were collected. Detailed mathematical expressions for the GP posterior and prior construction are provided in the Appendix.

The optimization targeted the five objectives defined in Section~2.3. Four objectives (YS, UTS/YS, strain at UTS, and $H_\text{dyn}/H_\text{qs}$) were obtained experimentally, while DoP was computed from constitutive calibration and FEM simulation using data acquired in the same iteration. Because EHVI is implemented in minimization form, the four desirable-to-maximize objectives were sign-inverted before surrogate modeling and hypervolume calculation, while DoP remained in its original minimization form.

\subsubsection{Acquisition, Fusion, and Batch Selection}

Candidate ranking was based on Expected Hypervolume Improvement (EHVI), which favors alloys expected to improve the current Pareto front while still exploring uncertain regions. Where multiple computational estimates contributed to the same target, their predictions were fused through a reification-based multi-source model so that the acquisition function operated on a single posterior distribution for each objective. To reduce sensitivity to any one kernel hyperparameter setting in the low-data regime, we used an ensemble of Gaussian processes with different length-scale assumptions and then applied $k$-medoids clustering to select a diverse experimental batch. Mathematical details for EHVI, reification, and the batch ensemble strategy are provided in the Appendix.

\subsubsection{Constraint-Aware Acquisition with Learned Feasibility}

Phase stability constraints were enforced initially through CALPHAD screening, but deterministic rejection alone is too rigid near the FCC phase boundary. We therefore trained a Gaussian process classifier (GPC) on experimentally verified phase outcomes to estimate a continuous feasibility probability, $P_\text{feas}(\mathbf{x})$, for each candidate composition. This allowed the optimizer to distinguish compositions that were confidently infeasible from those that were merely uncertain.

To assess how robust CALPHAD-predicted FCC stability is to small local composition changes, we performed a perturbation analysis on candidate alloys that satisfied FCC $\geq 0.99$. Of the 1,154,207 perturbed compositions evaluated, 903,326 (78.3\%) retained FCC $\geq 0.99$. We therefore interpreted a feasibility probability near 0.8 as corresponding to a composition that is locally robust to small perturbations, which motivated the acquisition threshold $P_\text{feas}(\mathbf{x}) \geq 0.8$. Details are provided in the \textbf{SI}.

Candidates that passed the initial CALPHAD filter were assigned $P_\text{feas} = 1$. Candidates rejected by CALPHAD but predicted by the GPC to satisfy $P_\text{feas}(\mathbf{x}) \geq 0.8$ were reintroduced into the search space, but their acquisition value was penalized according to feasibility risk:
\begin{equation}
    \textrm{EHVI}(\mathbf{x})_{\textrm{constraint-aware}} = \textrm{EHVI}(\mathbf{x}) \times P_{\textrm{feas}}(\mathbf{x})
\end{equation}
This feasibility-weighted acquisition suppressed high-risk candidates whose apparent promise was driven mainly by uncertainty, while still allowing near-boundary alloys to be selected when their predicted performance was sufficiently strong. As new alloys were synthesized and phase-verified, the GPC was retrained so that the feasibility boundary improved over successive iterations. Manufacturing failures were not modeled by this classifier; it addressed phase-outcome feasibility only.

\subsubsection{Summary of the BRAVE Campaign Algorithm}

\begin{figure*}[t]
    \centering
    \includegraphics[width=0.9\textwidth]{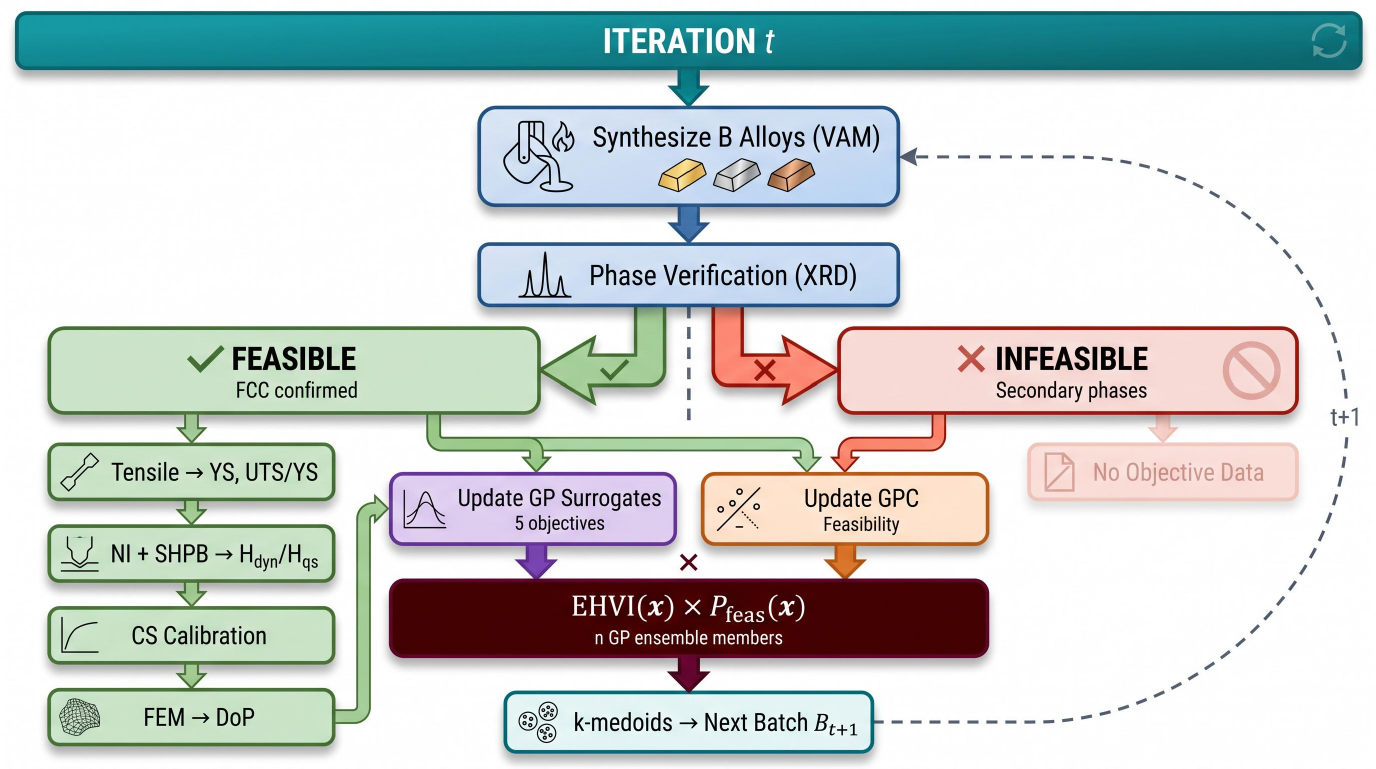}
    \caption{Data flow architecture for one iteration of the BRAVE campaign. Phase verification (XRD) classifies each synthesized alloy as feasible or infeasible, creating two parallel update paths: feasible alloys provide objective measurements for GP surrogate training, while all phase outcomes (feasible and infeasible) update the GPC feasibility classifier. The paths converge at the acquisition step, where EHVI is scaled by the feasibility probability $P_\text{feas}(\mathbf{x})$ to penalize high-risk candidates. k-medoids clustering selects the next batch from the top-ranked candidates.}
    \label{fig:brave_architecture}
\end{figure*}

\autoref{fig:brave_architecture} illustrates the data flow within a single iteration of the BRAVE campaign. The bifurcation at phase verification was the defining structural feature: feasible alloys proceeded through full characterization and fed objective data into the GP surrogates, while infeasible alloys contributed only their phase outcome to the GPC feasibility classifier. Both paths converged at the acquisition step, where EHVI was weighted by the learned feasibility probability. Algorithm~\ref{alg:brave} formalizes this workflow.

Algorithm~\ref{alg:brave} consolidates the full BRAVE campaign workflow, from design space initialization through iterative closed-loop optimization. Infeasible alloys (line~9) contribute to the feasibility classifier but not to surrogate model training, capturing the asymmetry between failed and successful experiments that motivates the risk-aware acquisition strategy. The DoP objective (lines~16--18) is computed within each iteration from experimentally calibrated constitutive models, creating the simulation-in-the-loop architecture described in Section~1.

\begin{algorithm}[!htbp]
\caption{BRAVE: Bayesian Risk-Aware Alloy Discovery and Exploration}\label{alg:brave}
\begin{algorithmic}[1]
\Statex \textbf{Input:} 8-element composition grid, CALPHAD database, batch size $B = 16$
\Statex
\Statex \textit{// Initialization}
\State Apply compositional bounds $\rightarrow$ 1,504,938 candidates
\State CALPHAD screen (TCHEA6, FCC $\geq 0.99$ at 700$^\circ$C) $\rightarrow$ 26,621 feasible
\State Partition feasible space into 37 chemical subsystems
\State Retain subsystem labels for feasible candidates
\State $k$-medoids clustering $\rightarrow$ 1,000 representative alloys
\State Diversity-aware down-selection $\rightarrow$ $B_0 = 16$ initial candidates
\Statex
\Statex \textit{// Mid-campaign update (from iteration BBB onward)}
\State Augment feasibility screening using TCHEA7 (700$^\circ$C and 950$^\circ$C)
\State Update feasible design space
\Statex
\Statex \textit{// Iterative closed-loop optimization}
\For{iteration $t = 0, 1, 2$}
    \State Synthesize $B_t$ alloys via vacuum arc melting
    \State Verify phase purity (XRD): classify each alloy as \textit{feasible} or \textit{infeasible}
    \State Update GPC feasibility classifier with new phase outcomes
    \For{each \textit{feasible} alloy}
        \State Characterize: tensile $\rightarrow$ YS, UTS/YS, strain at UTS
        \State Characterize: NI + HSRNI + SHPB $\rightarrow$ $H_\text{dyn}/H_\text{qs}$
        \State Calibrate Cowper--Symonds model from tensile + SHPB data
        \State Simulate depth of penetration (DoP) via FEM with calibrated model
    \EndFor
    \State Update GP surrogates with new objective measurements (feasible alloys only)
    \State Fuse multi-source predictions via reification
    \For{each of $n$ random length-scale configurations}
        \State Build GP ensemble member
        \State Compute $\textrm{EHVI}(\mathbf{x}) \times P_\text{feas}(\mathbf{x})$ for all remaining candidates
    \EndFor
    \State $k$-medoids select $B_{t+1}$ from top-ranked candidates
\EndFor
\Statex
\Statex \textbf{Output:} Pareto-optimal alloys, trained surrogates, feasibility classifier
\end{algorithmic}
\end{algorithm}

The campaign ran for three iterations (48 alloys total). This budget was set by experimental throughput---vacuum arc melting, thermomechanical processing, and multi-technique characterization limit the number of alloys that can be completed per cycle---rather than by a convergence criterion.

\subsection{Data Management and Overall Workflow}

Data management in this study follows a method-centric architecture with IGSN (International Generic Sample Number) registration for all physical samples. Details of the data architecture, folder structure, IGSN metadata schema, FAIR compliance, and a more detailed end-to-end workflow diagram are provided in the \textbf{SI}.

\section{Results and Discussion}\label{sec:Results}

\subsection{Phase Verification and Feasibility Outcomes}

\begin{figure}[htbp]
    \centering
    \includegraphics[width=0.95\columnwidth]{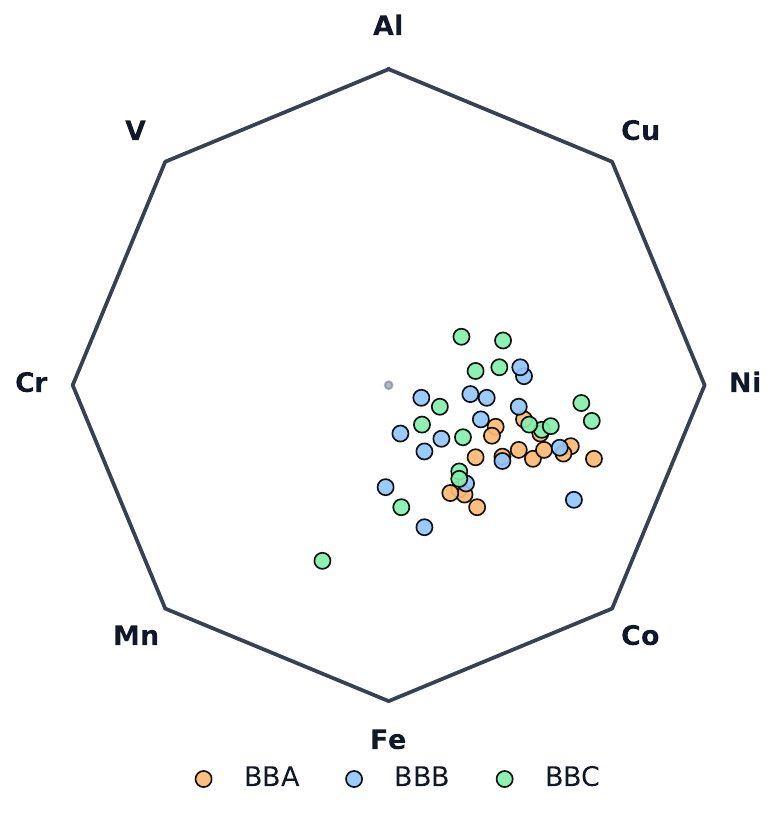}
    \caption{Affine projection of the target chemistries of all 48 alloys synthesized in this study onto a regular octagon representing the 8-element composition space (Al--V--Cr--Mn--Fe--Co--Ni--Cu). Each point is the affine-coordinate projection of one alloy composition and is colored by design iteration.} 
    \label{fig:alloy_affine_projection}
\end{figure}

The 16 candidate alloys selected for iteration BBA were chosen by diversity-aware k-medoids sampling across the 37 chemical subsystems (Section~2.1.3). \autoref{fig:alloy_affine_projection} shows the target chemistries of all 48 alloys synthesized in this work, projected from the 8-element composition simplex onto a regular octagon by affine coordinates and colored by iteration. This view emphasizes how the campaign progressively populated distinct regions of the accessible composition space across BBA, BBB, and BBC. A more detailed visualization of the iteration-0 batch selection is provided in the \textbf{SI}. All candidates across the three iterations were synthesized via vacuum arc melting, cold rolled, and recrystallized as described in Section~2.2. Phase purity was assessed by XRD for every alloy prior to mechanical testing.

\begin{figure*}[t]
    \centering
    \includegraphics[width=0.95\textwidth]{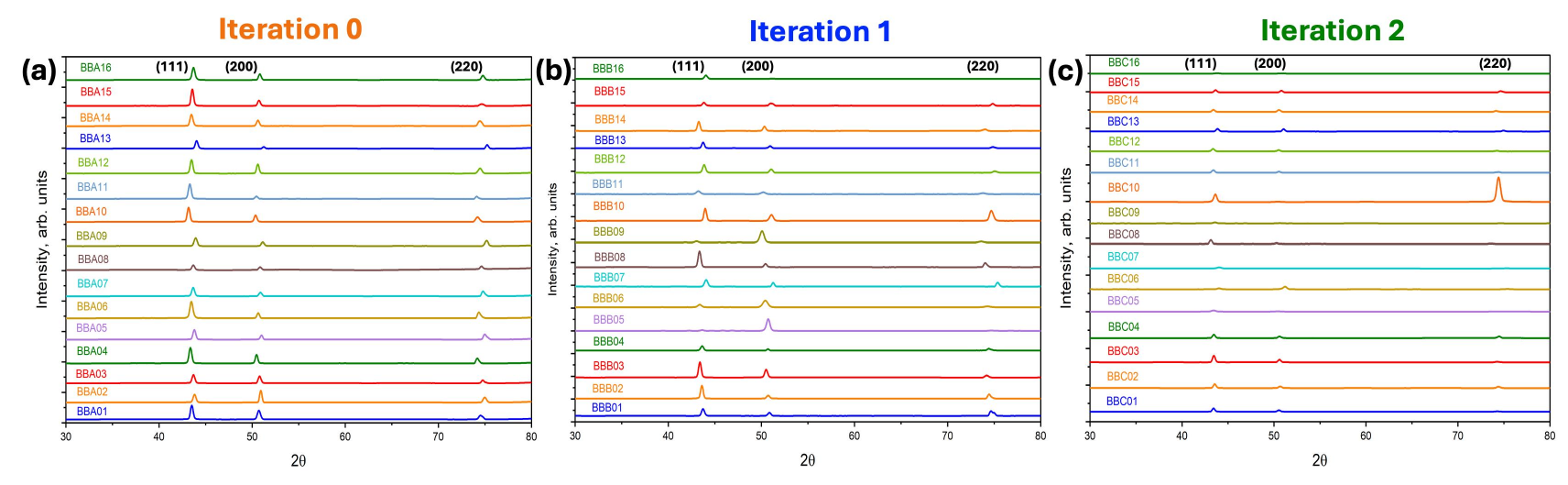}
    \caption{X-ray diffraction (XRD) patterns for all alloys across 3 iterations (a--c), showing FCC spectral peaks at (111), (200), and (220) planar reflections. Alloys with secondary sigma phase are indicated.}
    \label{fig:Phase_Analysis}
\end{figure*}

\autoref{fig:Phase_Analysis}(a--c) presents the diffraction patterns across all three iterations: the majority of alloys show single-phase FCC reflections at (111), (200), and (220), consistent with the CALPHAD-based design targets. Secondary intermetallic sigma ($P4_{2}/mnm$) phase was detected in a subset of alloys from iterations BBB and BBC.

Sigma phase appeared preferentially in alloys with combined V + Cr content $\geq$ 20~at.\%, although several alloys with high V + Cr (BBA09, BBB01, BBB03, BBC02, BBC04) remained single-phase FCC. These exceptions had Ni content of 36~at.\% or higher, consistent with increased valence electron concentration (VEC) destabilizing the sigma phase. We note that this compositional dependence---sigma formation correlating with V + Cr and suppressed by Ni---directly informed the GPC feasibility classifier training data.

The sigma-phase failures in BBB triggered a transition from the TCHEA6 to TCHEA7 database and the addition of a 950$^\circ$C FCC stability constraint (Section~2.1.2). Comparing TCHEA7 predictions with experimental XRD results for iterations BBA and BBB revealed five false positives (secondary phases predicted but not observed) and two false negatives (FCC predicted but sigma observed) at 700$^\circ$C, and one false positive and four false negatives at 950$^\circ$C. False negatives are the more consequential error: they lead to failed experiments (i.e., budget spent without usable objective data). These prediction discrepancies quantify the CALPHAD model uncertainty that the GPC feasibility classifier is designed to absorb.

\begin{figure*}[t]
    \centering
    \includegraphics[width=0.7\textwidth]{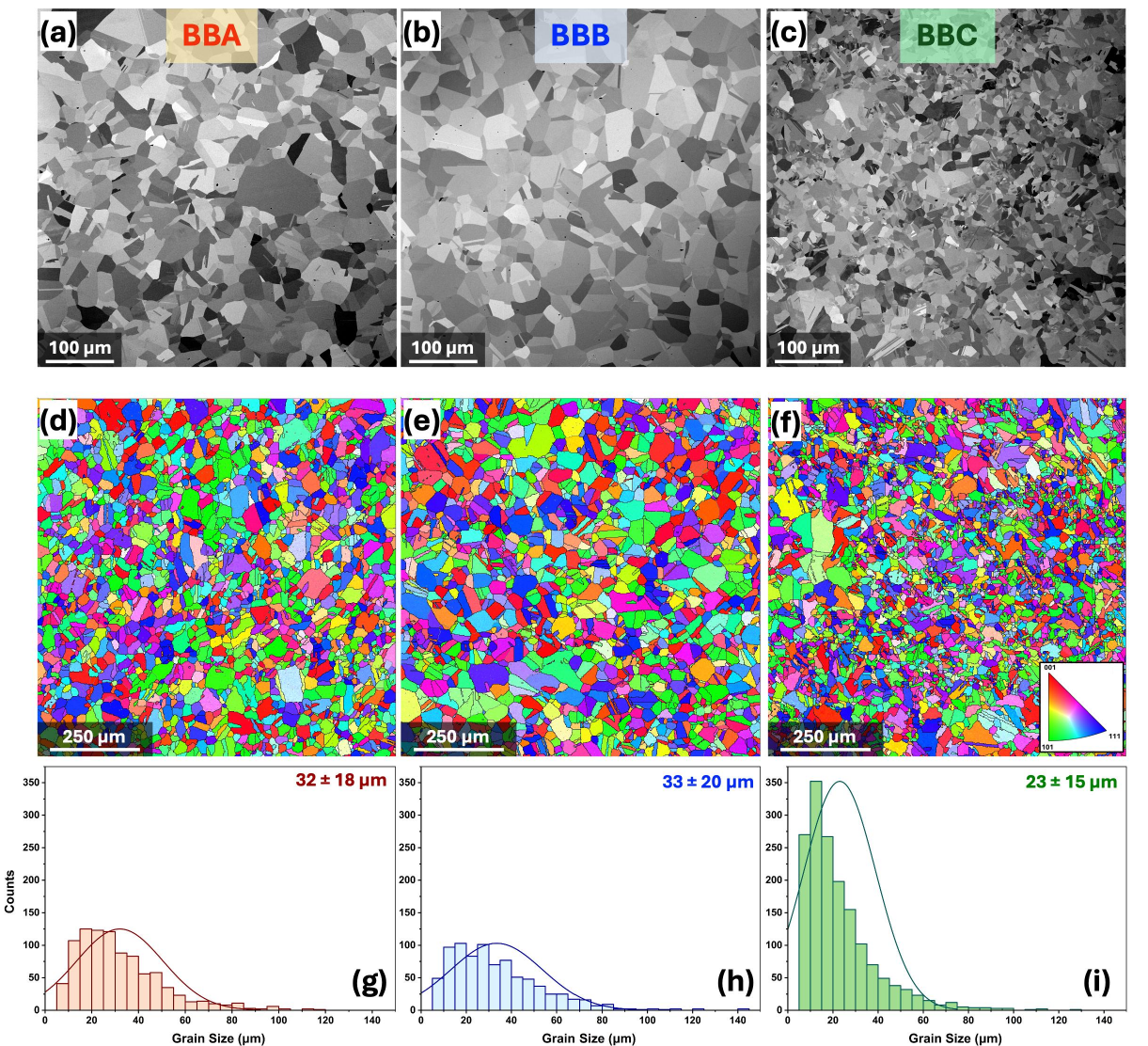}
    \caption{Microstructure of representative FCC alloy compositions across 3 iterations. (a--c) SEM-BSE micrographs showing fully recrystallized equiaxed FCC grains with annealing twins. (d--f) Corresponding EBSD IPF maps confirming random crystallographic texture. (g--i) Grain size histograms with descriptive statistics. Mean grain sizes range from $\sim$10~$\mu$m to 35~$\mu$m.}
    \label{fig:BSE_SEM}
\end{figure*}

\autoref{fig:BSE_SEM}(a--c) presents BSE-SEM images of three representative alloy compositions (one per iteration), showing fully recrystallized, equiaxed FCC grain structures with annealing twins. No significant porosity or macrosegregation was observed. The corresponding EBSD inverse pole figure (IPF) maps (\autoref{fig:BSE_SEM}(d--f)) confirm the absence of textural anisotropy, indicating that recrystallization reset the deformation-induced orientation bias from cold rolling. 
Grain size distributions (\autoref{fig:BSE_SEM}(g--i)) show average grain sizes between 10~$\mu$m and 35~$\mu$m, with compositionally complex alloys (7--8 elements) showing finer grains and narrower distributions, consistent with sluggish diffusion in concentrated solid solutions.

Alloys confirmed as single-phase FCC proceeded to full mechanical characterization, and we used their objective measurements to train the GP surrogates. Alloys containing sigma phase were classified as infeasible for the purposes of Bayesian optimization: their phase outcome was used to retrain the GPC feasibility classifier, but their objective measurements (when obtainable) were not included in surrogate model training. Several sigma-bearing alloys (BBC03, BBB13, BBC08, BBC14) retained sufficient FCC matrix to yield tensile specimens and were mechanically tested for scientific characterization, but their properties were excluded from the BO loop. Alloys that failed during processing (BBC05--07, BBC09--10, BBC16, BBA04) produced no specimens and contributed only their infeasible label to the GPC. \autoref{table_bba}, \autoref{table_bbb}, and \autoref{table_bbc} list the target and measured compositions (EDS) alongside XRD phase verification for all three iterations. Candidates that violated project constraints (manufacturing failures or secondary phases) are highlighted in red.

\begin{table*}[t]
\caption{Composition and phase analysis of BBA (Iteration 0) alloys}
\label{table_bba}
\centering
\scriptsize
\resizebox{\textwidth}{!}{
\begin{tblr}{
    hlines,
    hline{3,21} = {2pt},
    rowsep=1pt,
    colspec = {
        |X[1.2, c]
        |[2pt]X[0.7, c]|X[0.7, c]|X[0.7, c]|X[0.7, c]|X[0.7, c]|X[0.7, c]|X[0.7, c]|X[0.7, c]
        |[2pt]X[0.7, c]|X[0.7, c]|X[0.7, c]|X[0.7, c]|X[0.7, c]|X[0.7, c]|X[0.7, c]|X[0.7, c]
        |[2pt]X[2.0, c]
        |X[2.0, c]|
        },
    colsep=0pt,
    }
    \SetCell[r=2,c=1]{c}\textbf{\makecell{Alloy\\Name}}
    & \SetCell[c=8]{c} \textbf{Target Atomic \%}
    &&&&&&&& \SetCell[c=8]{c} \textbf{EDS Measured Atomic \%}
    &&&&&&&& \SetCell[r=2,c=1]{c}\textbf{XRD Phase} \\
    & Al & Co & Cr & Cu & Fe & Mn & Ni & V
    & Al & Co & Cr & Cu & Fe & Mn & Ni & V \\
    BBA01 & 4 & 8 & 4 & 4 & 16 & 12 & 48 & 4 & 4.1 & 8.2 & 4.2 & 4.2 & 16.1 & 12.2 & 47.0 & 4.1 & FCC \\
    BBA02 & 4 & 16 & \SetCell{gray!20}0 & 4 & 12 & 8 & 52 & 4 & 4.2 & 15.0 & \SetCell{gray!20}0.0 & 4.2 & 12.2 & 8.2 & 52.2 & 4.1 & FCC \\
    BBA03 & 4 & 12 & 8 & 4 & 16 & 8 & 48 & \SetCell{gray!20}0 & 4.3 & 12.1 & 8.2 & 4.2 & 16.1 & 8.2 & 47.0 & \SetCell{gray!20}0.0 & FCC \\
    \SetCell{red!20}\textbf{BBA04} & \SetCell{gray!20}0 & 12 & 8 & 4 & 16 & 20 & 36 & 4 & $\cdot$ & $\cdot$ & $\cdot$ & $\cdot$ & $\cdot$ & $\cdot$ & $\cdot$ & $\cdot$ & \SetCell{red!20}\textbf{---} \\
    BBA05 & 4 & 12 & 8 & \SetCell{gray!20}0 & 8 & 8 & 56 & 4 & 4.1 & 12.2 & 8.2 & \SetCell{gray!20}0.0 & 8.2 & 8.0 & 55.2 & 4.1 & FCC \\
    BBA06 & 4 & \SetCell{gray!20}0 & 8 & 4 & 24 & 12 & 44 & 4 & 4.2 & \SetCell{gray!20}0.0 & 8.3 & 4.1 & 24.0 & 11.9 & 43.4 & 4.1 & FCC \\
    BBA07 & 4 & 12 & \SetCell{gray!20}0 & \SetCell{gray!20}0 & 16 & 8 & 52 & 8 & 5.0 & 12.3 & \SetCell{gray!20}0.0 & \SetCell{gray!20}0.0 & 16.2 & 7.9 & 50.5 & 8.1 & FCC \\
    BBA08 & \SetCell{gray!20}0 & 24 & 8 & \SetCell{gray!20}0 & 12 & 16 & 32 & 8 & \SetCell{gray!20}0.0 & 22.9 & 8.1 & \SetCell{gray!20}0.0 & 12.2 & 16.0 & 32.7 & 8.1 & FCC \\
    BBA09 & 4 & 16 & 8 & \SetCell{gray!20}0 & 12 & \SetCell{gray!20}0 & 48 & 12 & 4.9 & 16.1 & 8.2 & \SetCell{gray!20}0.0 & 12.0 & \SetCell{gray!20}0.0 & 46.6 & 12.1 & FCC \\
    BBA10 & 4 & \SetCell{gray!20}0 & 4 & \SetCell{gray!20}0 & 20 & 8 & 52 & 12 & 4.4 & \SetCell{gray!20}0.0 & 4.2 & \SetCell{gray!20}0.0 & 20.2 & 8.0 & 51.0 & 12.2 & FCC \\
    BBA11 & \SetCell{gray!20}0 & 16 & \SetCell{gray!20}0 & 4 & 20 & 20 & 28 & 12 & \SetCell{gray!20}0.0 & 16.1 & \SetCell{gray!20}0.0 & 4.0 & 20.1 & 20.3 & 27.4 & 12.0 & FCC \\
    BBA12 & \SetCell{gray!20}0 & \SetCell{gray!20}0 & 8 & 4 & 16 & 12 & 52 & 8 & \SetCell{gray!20}0.0 & \SetCell{gray!20}0.0 & 8.3 & 4.1 & 16.1 & 12.3 & 51.2 & 8.0 & FCC \\
    BBA13 & 4 & 20 & 4 & 4 & 16 & \SetCell{gray!20}0 & 52 & \SetCell{gray!20}0 & 4.0 & 20.4 & 4.1 & 4.2 & 16.3 & \SetCell{gray!20}0.0 & 51.0 & \SetCell{gray!20}0.0 & FCC \\
    BBA14 & \SetCell{gray!20}0 & 16 & 8 & 4 & 16 & 20 & 36 & \SetCell{gray!20}0 & \SetCell{gray!20}0.0 & 16.1 & 8.2 & 4.0 & 16.0 & 20.3 & 35.3 & \SetCell{gray!20}0.0 & FCC \\
    BBA15 & \SetCell{gray!20}0 & 20 & 8 & 4 & \SetCell{gray!20}0 & 20 & 44 & 4 & \SetCell{gray!20}0.0 & 19.9 & 8.2 & 4.2 & \SetCell{gray!20}0.0 & 20.5 & 43.2 & 4.0 & FCC \\
    BBA16 & 4 & 12 & \SetCell{gray!20}0 & 12 & 20 & 8 & 44 & \SetCell{gray!20}0 & 4.1 & 12.3 & \SetCell{gray!20}0.0 & 11.8 & 20.4 & 8.1 & 43.2 & \SetCell{gray!20}0.0 & FCC \\
\end{tblr}
}
\end{table*}

\begin{table*}[t]
\caption{Composition and phase analysis of BBB (Iteration 1) alloys}
\label{table_bbb}
\centering
\scriptsize
\resizebox{\textwidth}{!}{
\begin{tblr}{
    hlines,
    hline{3,17} = {2pt},
    rowsep=1pt,
    colspec = {
        |X[1.2, c]
        |[2pt]X[0.7, c]|X[0.7, c]|X[0.7, c]|X[0.7, c]|X[0.7, c]|X[0.7, c]|X[0.7, c]|X[0.7, c]
        |[2pt]X[0.7, c]|X[0.7, c]|X[0.7, c]|X[0.7, c]|X[0.7, c]|X[0.7, c]|X[0.7, c]|X[0.7, c]
        |[2pt]X[2.0, c]
        |X[2.0, c]|
        },
    colsep=0pt,
    }
    \SetCell[r=2,c=1]{c}\textbf{\makecell{Alloy\\Name}}
    & \SetCell[c=8]{c} \textbf{Target Atomic \%}
    &&&&&&&& \SetCell[c=8]{c} \textbf{EDS Measured Atomic \%}
    &&&&&&&& \SetCell[r=2,c=1]{c}\textbf{XRD Phase} \\
    & Al & Co & Cr & Cu & Fe & Mn & Ni & V
    & Al & Co & Cr & Cu & Fe & Mn & Ni & V \\
    BBB01 & \SetCell{gray!20}0 & 24 & 4 & \SetCell{gray!20}0 & 8 & 4 & 36 & 24 & \SetCell{gray!20}0.0 & 20.5 & 4.2 & \SetCell{gray!20}0.0 & 8.1 & 4.0 & 38.9 & 24.4 & FCC \\
    BBB02 & 4 & 32 & 8 & 4 & 4 & 20 & 24 & 4 & 4.2 & 30.4 & 8.2 & 4.1 & 4.1 & 20.1 & 24.7 & 4.1 & FCC \\
    BBB03 & 4 & \SetCell{gray!20}0 & 12 & 4 & 4 & 16 & 52 & 8 & 4.1 & $\cdot$ & 12.3 & 4.1 & 4.1 & 16.1 & 51.2 & 8.2 & FCC \\
    BBB04 & 4 & 12 & 4 & \SetCell{gray!20}0 & 28 & \SetCell{gray!20}0 & 40 & 12 & 4.0 & 11.1 & 4.1 & \SetCell{gray!20}0.0 & 28.0 & \SetCell{gray!20}0.0 & 40.8 & 12.0 & FCC \\
    \SetCell{red!20}\textbf{BBB05} & \SetCell{gray!20}0 & 44 & 12 & \SetCell{gray!20}0 & 4 & 4 & 12 & 24 & $\cdot$ & $\cdot$ & $\cdot$ & $\cdot$ & $\cdot$ & $\cdot$ & $\cdot$ & $\cdot$ & \SetCell{red!20}\textbf{$\mathbf{FCC+\sigma}$} \\
    \SetCell{red!20}\textbf{BBB06} & \SetCell{gray!20}0 & 16 & 20 & \SetCell{gray!20}0 & 4 & 4 & 36 & 20 & $\cdot$ & $\cdot$ & $\cdot$ & $\cdot$ & $\cdot$ & $\cdot$ & $\cdot$ & $\cdot$ & \SetCell{red!20}\textbf{$\mathbf{FCC+\sigma}$} \\
    BBB07 & 4 & 40 & \SetCell{gray!20}0 & \SetCell{gray!20}0 & 12 & 4 & 36 & 4 & 4.0 & 39.1 & \SetCell{gray!20}0.0 & \SetCell{gray!20}0.0 & 12.3 & 3.9 & 36.7 & 4.0 & FCC \\
    BBB08 & 4 & 8 & 12 & 4 & 4 & 24 & 40 & 4 & 3.8 & 8.0 & 12.3 & 4.2 & 4.1 & 24.0 & 39.4 & 4.1 & FCC \\
    \SetCell{red!20}\textbf{BBB09} & \SetCell{gray!20}0 & 4 & 8 & \SetCell{gray!20}0 & 4 & 28 & 40 & 16 & $\cdot$ & $\cdot$ & $\cdot$ & $\cdot$ & $\cdot$ & $\cdot$ & $\cdot$ & $\cdot$ & \SetCell{red!20}\textbf{$\mathbf{FCC+\sigma}$} \\
    BBB10 & 4 & \SetCell{gray!20}0 & 4 & 4 & 8 & 16 & 52 & 12 & $\cdot$ & $\cdot$ & $\cdot$ & $\cdot$ & $\cdot$ & $\cdot$ & $\cdot$ & $\cdot$ & FCC \\
    \SetCell{red!20}\textbf{BBB11} & \SetCell{gray!20}0 & 24 & 4 & \SetCell{gray!20}0 & 4 & 32 & 20 & 16 & $\cdot$ & $\cdot$ & $\cdot$ & $\cdot$ & $\cdot$ & $\cdot$ & $\cdot$ & $\cdot$ & \SetCell{red!20}\textbf{$\mathbf{FCC+\sigma}$} \\
    BBB12 & \SetCell{gray!20}0 & 20 & 4 & \SetCell{gray!20}0 & 4 & 4 & 48 & 20 & \SetCell{gray!20}0.0 & 20.2 & 4.2 & \SetCell{gray!20}0.0 & 4.1 & 4.0 & 47.1 & 20.4 & FCC \\
    \SetCell{red!20}\textbf{BBB13} & \SetCell{gray!20}0 & 40 & 4 & \SetCell{gray!20}0 & 28 & 4 & 4 & 20 & 0.0 & 39.7 & 4.2 & 0.0 & 27.9 & 3.9 & 4.2 & 20.1 & \SetCell{red!20}\textbf{$\mathbf{FCC+\sigma}$} \\
    BBB14 & 4 & \SetCell{gray!20}0 & 12 & 16 & 4 & 12 & 52 & \SetCell{gray!20}0 & 4.4 & \SetCell{gray!20}0.0 & 12.1 & 15.7 & 4.1 & 12.4 & 51.3 & \SetCell{gray!20}0.0 & FCC \\
    BBB15 & 4 & \SetCell{gray!20}0 & 8 & 24 & 4 & 16 & 44 & \SetCell{gray!20}0 & 4.3 & \SetCell{gray!20}0.0 & 8.3 & 23.5 & 4.1 & 16.2 & 43.7 & \SetCell{gray!20}0.0 & FCC \\
    BBB16 & 4 & 36 & \SetCell{gray!20}0 & \SetCell{gray!20}0 & 4 & 4 & 40 & 12 & $\cdot$ & $\cdot$ & $\cdot$ & $\cdot$ & $\cdot$ & $\cdot$ & $\cdot$ & $\cdot$ & FCC \\
\end{tblr}
}
\end{table*}

\begin{table*}[t]
\caption{Composition and phase analysis of BBC (Iteration 2) alloys}
\label{table_bbc}
\centering
\scriptsize
\resizebox{\textwidth}{!}{
\begin{tblr}{
    hlines,
    hline{3,17} = {2pt},
    rowsep=1pt,
    colspec = {
        |X[1.2, c]
        |[2pt]X[0.7, c]|X[0.7, c]|X[0.7, c]|X[0.7, c]|X[0.7, c]|X[0.7, c]|X[0.7, c]|X[0.7, c]
        |[2pt]X[0.7, c]|X[0.7, c]|X[0.7, c]|X[0.7, c]|X[0.7, c]|X[0.7, c]|X[0.7, c]|X[0.7, c]
        |[2pt]X[2.0, c]
        |X[2.0, c]|
        },
    colsep=0pt,
    }
    \SetCell[r=2,c=1]{c}\textbf{\makecell{Alloy\\Name}}
    & \SetCell[c=8]{c} \textbf{Target Atomic \%}
    &&&&&&&& \SetCell[c=8]{c} \textbf{EDS Measured Atomic \%}
    &&&&&&&& \SetCell[r=2,c=1]{c}\textbf{XRD Phase} \\
    & Al & Co & Cr & Cu & Fe & Mn & Ni & V
    & Al & Co & Cr & Cu & Fe & Mn & Ni & V \\
    BBC01 & 4 & 12 & 4 & 4 & 20 & 16 & 32 & 8 & 4.0 & 12.2 & 4.2 & 4.0 & 20.1 & 16.0 & 31.4 & 8.2 & FCC \\
    BBC02 & 4 & 8 & \SetCell{gray!20}0 & \SetCell{gray!20}0 & 4 & 8 & 52 & 24 & 4.0 & 8.2 & \SetCell{gray!20}0.0 & \SetCell{gray!20}0.0 & 4.1 & 8.0 & 51.2 & 24.4 & FCC \\
    \SetCell{red!20}\textbf{BBC03} & 4 & 4 & 12 & \SetCell{gray!20}0 & \SetCell{gray!20}0 & 4 & 52 & 24 & 4.0 & 4.1 & 12.2 & \SetCell{gray!20}0.0 & \SetCell{gray!20}0.0 & 3.9 & 51.3 & 24.4 & \SetCell{red!20}\textbf{$\mathbf{FCC+\sigma}$} \\
    BBC04 & 4 & \SetCell{gray!20}0 & 4 & \SetCell{gray!20}0 & 4 & 4 & 60 & 24 & 4.0 & \SetCell{gray!20}0.0 & 4.2 & \SetCell{gray!20}0.0 & 4.1 & 4.0 & 59.3 & 24.5 & FCC \\
    \SetCell{red!20}\textbf{BBC05} & \SetCell{gray!20}0 & 24 & 8 & \SetCell{gray!20}0 & 16 & 40 & 4 & 8 & $\cdot$ & $\cdot$ & $\cdot$ & $\cdot$ & $\cdot$ & $\cdot$ & $\cdot$ & $\cdot$ & \SetCell{red!20}\textbf{---} \\
    \SetCell{red!20}\textbf{BBC06} & 4 & 28 & 4 & \SetCell{gray!20}0 & 4 & 4 & 44 & 12 & $\cdot$ & $\cdot$ & $\cdot$ & $\cdot$ & $\cdot$ & $\cdot$ & $\cdot$ & $\cdot$ & \SetCell{red!20}\textbf{---} \\
    \SetCell{red!20}\textbf{BBC07} & 4 & 32 & 4 & \SetCell{gray!20}0 & \SetCell{gray!20}0 & 4 & 44 & 12 & $\cdot$ & $\cdot$ & $\cdot$ & $\cdot$ & $\cdot$ & $\cdot$ & $\cdot$ & $\cdot$ & \SetCell{red!20}\textbf{---} \\
    \SetCell{red!20}\textbf{BBC08} & \SetCell{gray!20}0 & 8 & \SetCell{gray!20}0 & 4 & 4 & 28 & 36 & 20 & $\cdot$ & 8.1 & $\cdot$ & 4.1 & 4.2 & 28.1 & 35.7 & 19.9 & \SetCell{red!20}\textbf{$\mathbf{FCC+\sigma}$} \\
    \SetCell{red!20}\textbf{BBC09} & \SetCell{gray!20}0 & 36 & 4 & \SetCell{gray!20}0 & 16 & 16 & 8 & 20 & $\cdot$ & $\cdot$ & $\cdot$ & $\cdot$ & $\cdot$ & $\cdot$ & $\cdot$ & $\cdot$ & \SetCell{red!20}\textbf{---} \\
    \SetCell{red!20}\textbf{BBC10} & 4 & 4 & 16 & \SetCell{gray!20}0 & 8 & \SetCell{gray!20}0 & 52 & 16 & $\cdot$ & $\cdot$ & $\cdot$ & $\cdot$ & $\cdot$ & $\cdot$ & $\cdot$ & $\cdot$ & \SetCell{red!20}\textbf{---} \\
    BBC11 & 4 & 12 & 4 & 4 & 12 & 16 & 36 & 12 & 4.3 & 12.2 & 4.2 & 3.9 & 12.2 & 15.9 & 35.1 & 12.3 & FCC \\
    BBC12 & 4 & 4 & 4 & 8 & 28 & 16 & 32 & 4 & 4.1 & 4.3 & 4.2 & 7.8 & 28.0 & 16.1 & 31.3 & 4.1 & FCC \\
    BBC13 & 4 & 28 & \SetCell{gray!20}0 & 20 & 4 & 8 & 36 & \SetCell{gray!20}0 & 4.3 & 28.0 & $\cdot$ & 19.9 & 4.1 & 8.4 & 35.4 & $\cdot$ & FCC \\
    \SetCell{red!20}\textbf{BBC14} & \SetCell{gray!20}0 & 12 & 4 & \SetCell{gray!20}0 & 4 & 16 & 40 & 24 & $\cdot$ & 12.1 & 4.1 & $\cdot$ & 4.3 & 16.1 & 39.0 & 24.5 & \SetCell{red!20}\textbf{$\mathbf{FCC+\sigma}$} \\
    BBC15 & 4 & 16 & \SetCell{gray!20}0 & 20 & 4 & 12 & 44 & \SetCell{gray!20}0 & 4.4 & 16.2 & $\cdot$ & 19.7 & 4.2 & 12.3 & 43.3 & $\cdot$ & FCC \\
    \SetCell{red!20}\textbf{BBC16} & 4 & 28 & 4 & \SetCell{gray!20}0 & 8 & \SetCell{gray!20}0 & 40 & 16 & $\cdot$ & $\cdot$ & $\cdot$ & $\cdot$ & $\cdot$ & $\cdot$ & $\cdot$ & $\cdot$ & \SetCell{red!20}\textbf{---} \\
\end{tblr}
}
\end{table*}

\subsection{Mechanical Properties Across Iterations}

\subsubsection{Tensile Response}

\begin{figure}[H]
\centering
\includegraphics[width=0.9\columnwidth]{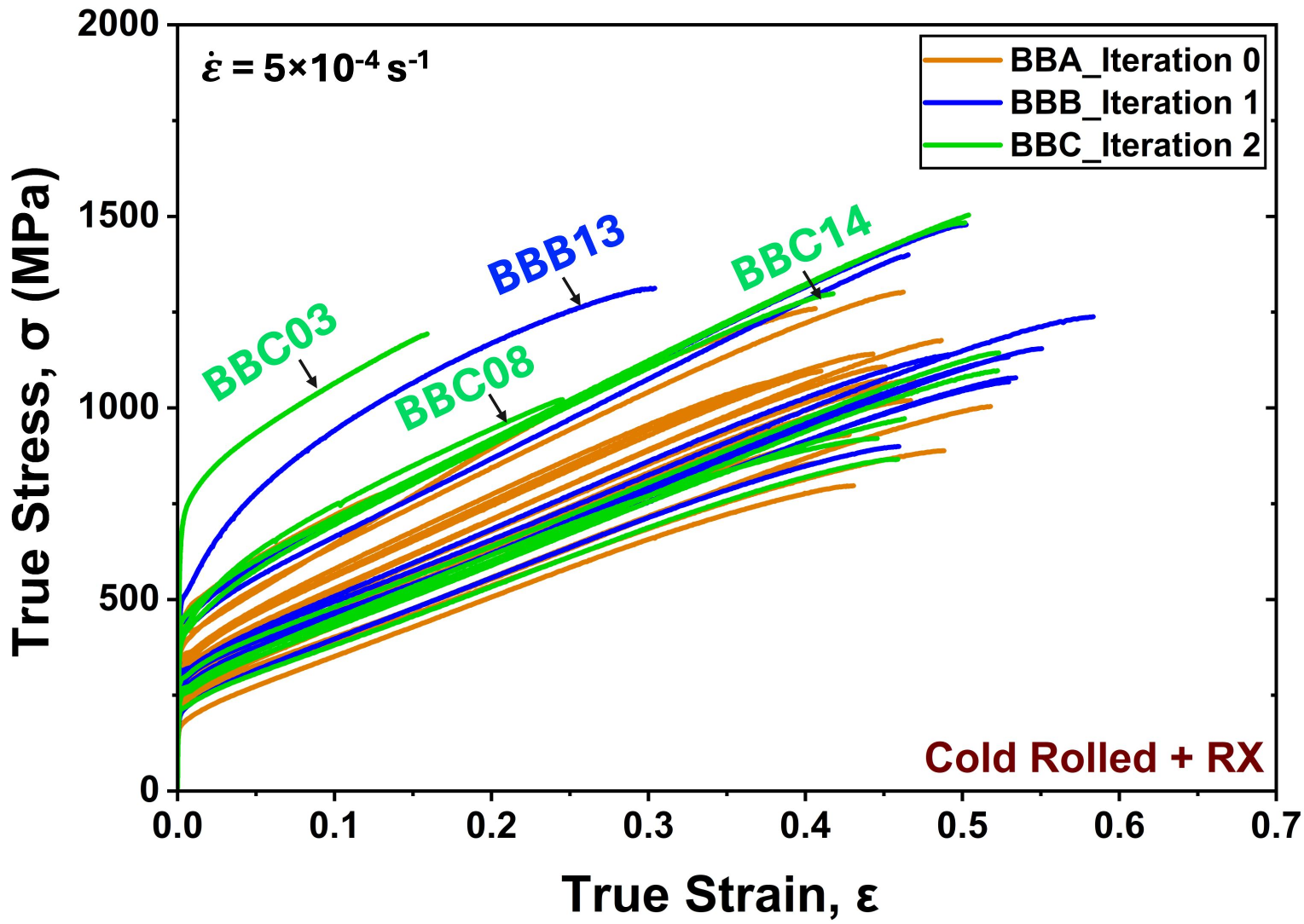}
\caption{Consolidated true stress--strain curves from ambient temperature quasistatic uniaxial tensile tests for all tested FCC alloys across 3 iterations. Alloys containing sigma phase are indicated by black arrows. RX refers to recrystallized alloys.}
\label{fig:true_stress_strain}
\end{figure}

\autoref{fig:true_stress_strain} presents the true stress--strain curves for all tested FCC alloys across the three iterations. True fracture stress increased progressively across successive iterations, with the majority of alloys surpassing 1~GPa in true tensile strength by iteration BBB. Ductility remained comparable across iterations, with the top-performing alloys sustaining strain at UTS in the range of 45--50\%. Enhanced strain hardening across iterations delayed the onset of strain localization, manifesting as high UTS/YS ratios. Compositions with UTS/YS $>$ 3.5 demonstrated superior strain hardening capacity, contributing to delayed failure under quasi-static loading. For context, Campaign~1 achieved UTS/YS ratios ranging from 1.38 to 2.93~\cite{hastings2025accelerated}. In Campaign~2, every successfully tested single-phase FCC alloy exceeded Campaign~1's maximum: the minimum UTS/YS was 3.07 (BBA07), the median was 3.79, and 9 alloys exceeded 4.0. 
We note that this wholesale improvement over Campaign~1 is not attributable to the optimizer alone. The GP surrogates in Campaign~2 were initialized with informative priors trained on Campaign~1 experimental data and the Varvenne--Luque--Curtin solid solution strengthening model (Section~2.6), giving the optimizer a physics-grounded starting point rather than a zero-mean prior. The design space was shaped by Campaign~1's experience: we expanded from 6 to 8 elements based on the observation that the 6-element space lacked sufficient compositional freedom to simultaneously maximize strength and ductility, and we replaced hardness with tensile-derived objectives (YS, UTS/YS, strain at UTS) based on the finding that hardness alone is an incomplete proxy for bulk mechanical response. The UTS/YS shift from 1.38--2.93 (Campaign~1) to 3.07--4.53 (Campaign~2) thus reflects the compounding effect of prior-informed surrogates, a better-chosen design space, and more directly relevant optimization objectives. Tabulated values of YS, UTS, and strain to failure for all tested alloys are provided in the \textbf{SI}.

\subsubsection{Dynamic Hardness and Rate Sensitivity}

\begin{figure*}[t]
    \centering
    \includegraphics[width=\textwidth]{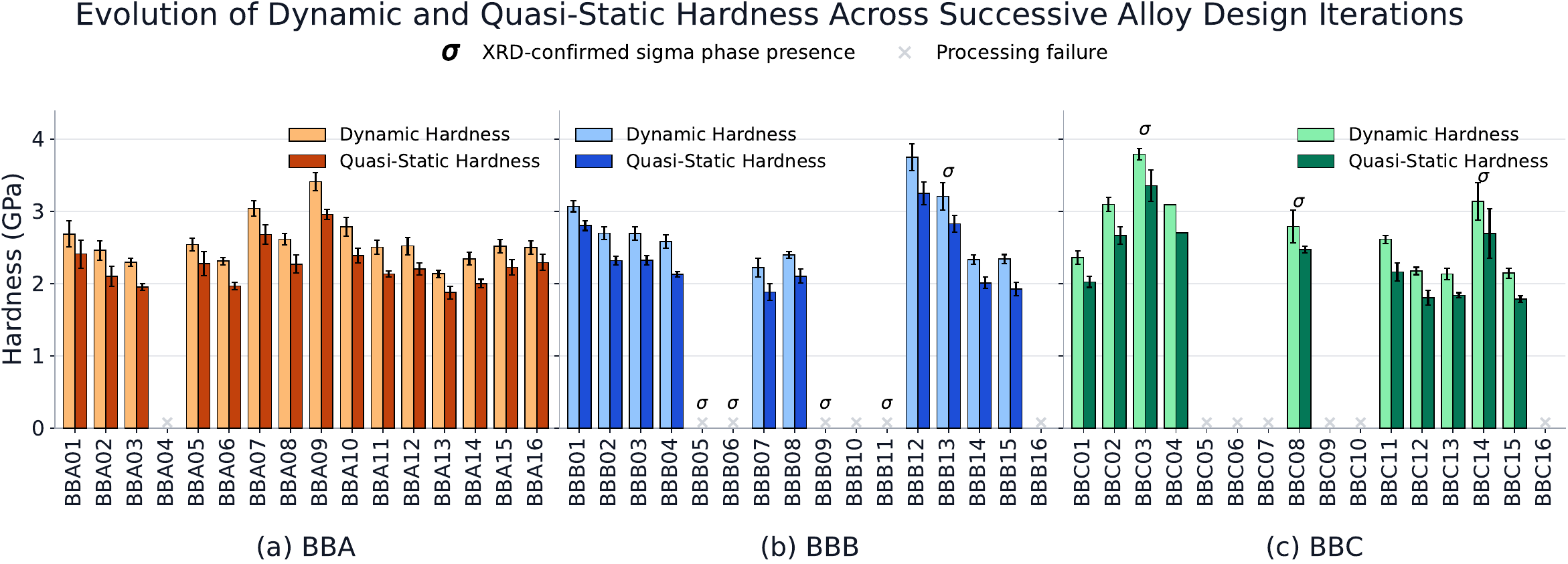}
    \caption{Dynamic and quasistatic hardness values for all candidate alloys across iterations (a) BBA, (b) BBB, and (c) BBC. Bar colors are iteration-specific within each panel. Gray \texttimes{} markers denote processing failures with no hardness data, and $\sigma$ markers indicate alloys with XRD-confirmed sigma phase presence.}
    \label{fig:HSRNI1}
\end{figure*}

Dynamic hardness consistently exceeded quasi-static hardness across all tested compositions (\autoref{fig:HSRNI1}), consistent with the higher strain rates associated with impact testing ($\sim 10^3$\,s$^{-1}$) relative to quasi-static indentation ($\sim 10^0$\,s$^{-1}$). Among single-phase FCC alloys, BBB12 showed the highest hardness. BBC03, which contained secondary sigma phase, showed the overall highest hardness.

\begin{figure*}[t]
    \centering
    \includegraphics[width=\textwidth]{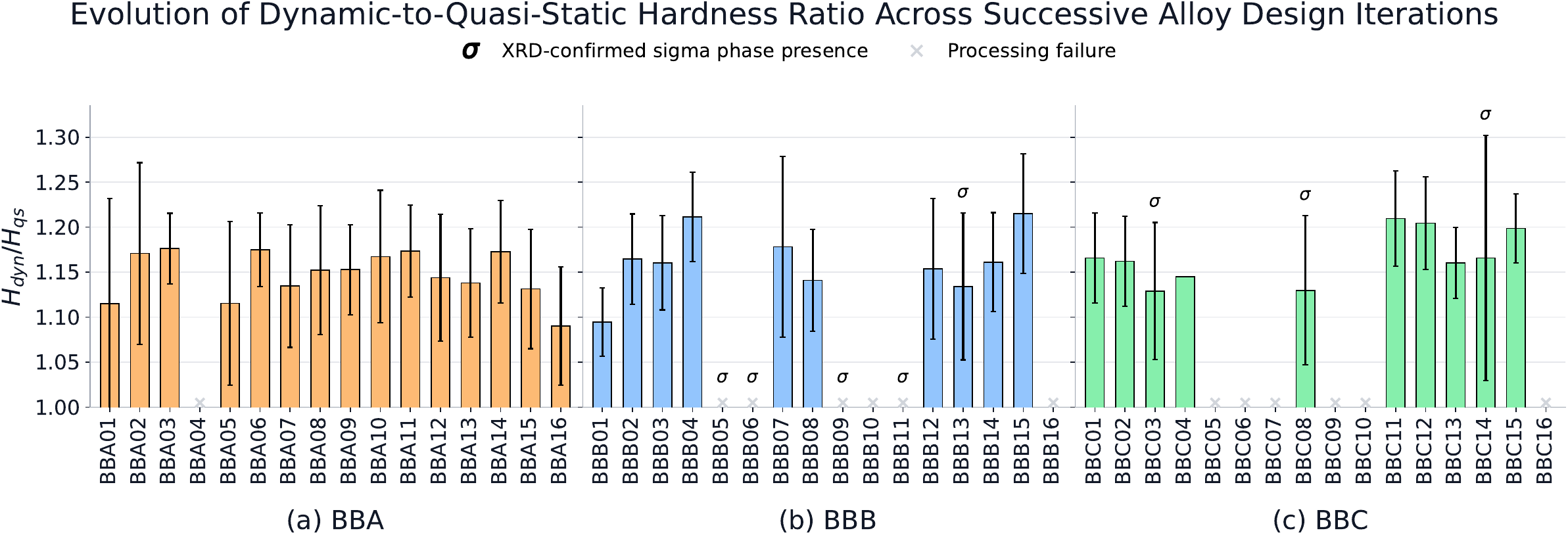}
    \caption{Dynamic-to-quasistatic hardness ratio ($H_{\mathrm{dyn}}/H_{\mathrm{qs}}$) for all candidate alloys across iterations (a) BBA, (b) BBB, and (c) BBC. Bar colors are iteration-specific within each panel. Gray \texttimes{} markers denote processing failures with no hardness-ratio data, and $\sigma$ markers indicate alloys with XRD-confirmed sigma phase presence.}
    \label{fig:HSRNI2}
\end{figure*}

The dynamic-to-quasi-static hardness ratio, $H_{\text{DYN}}/H_{\text{QS}}$ (\autoref{fig:HSRNI2}), ranged from 1.09 to 1.22 across all samples, with no clear compositional trend. Error bars were computed via propagation of uncertainty from the standard deviations of both dynamic and quasi-static measurements. None of the tested compositions showed error bars overlapping with unity, confirming a measurable rate-dependent strengthening effect in every alloy.

The contact-depth-to-indentation-depth ratio ($h_c/h$) was close to 1 for most samples, indicating minimal pile-up or sink-in. BBC03 was an exception ($h_c/h$ up to 1.03), consistent with its sigma-phase-induced deformation behavior. The near-unity $h_c/h$ for single-phase FCC alloys supports the use of nominal hardness calculations, reducing the need for the time-consuming imaging and tracing steps required for precise contact area estimation.

\subsubsection{Depth of Penetration}

Simulated DoP values across all iterations ranged from 2.22 to 3.35~mm (tabulated in \textbf{SI}). The five best-performing alloys (lowest DoP)---BBC03, BBB13, BBC02, BBB01, and BBC04---were concentrated in iterations BBB and BBC, while the five worst-performing alloys (BBA13, BBA16, BBC15, BBA03, and BBB07) were predominantly from iteration BBA, with one (BBC15) from iteration BBC. This shift confirms that the Bayesian optimization loop progressively identified compositions with improved impact resistance. The 1.13~mm spread in DoP across the full dataset (3.35 $-$ 2.22~mm), though narrow in absolute terms, provided sufficient differentiation for the EHVI acquisition function to distinguish candidates.

\subsubsection{Property Evolution}

\begin{figure*}[t]
    \centering
    \includegraphics[width=0.9\textwidth]{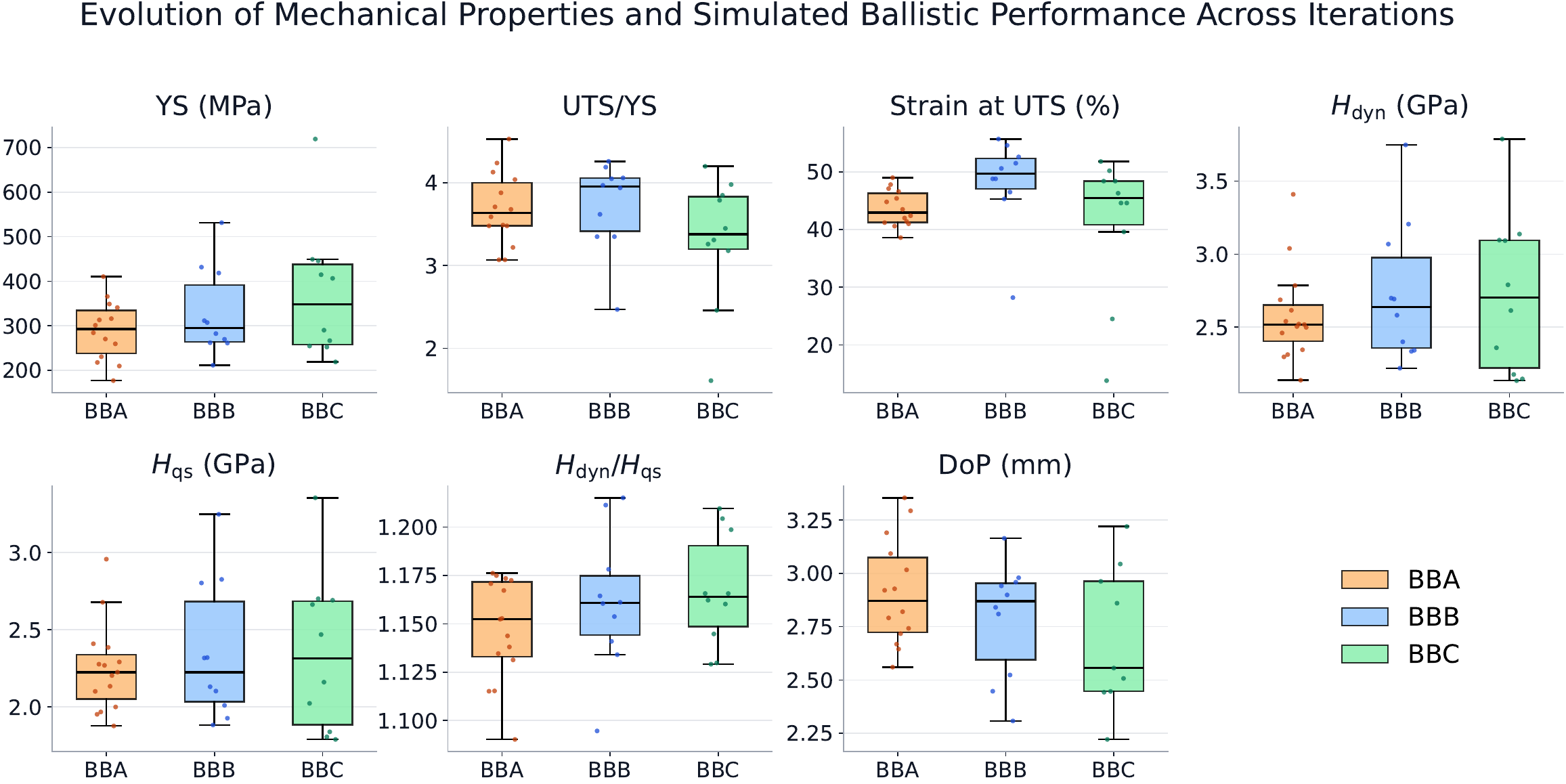}
    \caption{Evolution of mechanical properties and simulated ballistic performance across iterations. YS = yield strength, UTS = ultimate tensile strength, $H_{\mathrm{dyn}}$ = dynamic hardness, $H_{\mathrm{qs}}$ = quasistatic hardness, and DoP = simulated depth of penetration. Median strength metrics and $H_{\mathrm{dyn}}/H_{\mathrm{qs}}$ increase from BBA to BBC, while simulated depth of penetration decreases and strain at UTS is preserved.}
    \label{fig:boxplot_properties}
\end{figure*}

\autoref{fig:boxplot_properties} summarizes the statistical evolution of all measured properties across the three iterations. Median values for strength-related properties (YS, UTS, $H_\text{dyn}$, $H_\text{qs}$) increased from BBA to BBC while inter-quartile ranges narrowed, indicating convergence toward consistently high-performing compositions. Median DoP decreased across iterations, with the distribution tightening around lower penetration depths. Strain at UTS was maintained or improved despite the strength gains, confirming that the multi-objective framework avoided the classical strength--ductility trade-off.

These improvements in median properties should be interpreted alongside a processing caveat. The optimization campaign treats composition as the sole design variable, but the measured properties also depend on thermomechanical processing history---specifically, the degree of cold work, recrystallization temperature, and resulting grain size. Several alloys that passed phase verification (single-phase FCC by XRD) showed anomalous tensile behavior: low ductility, premature failure, or high scatter between replicate specimens. Post-mortem analysis traced these anomalies to incomplete recrystallization or non-uniform grain size distributions rather than compositional effects. The consequence for the optimization loop is twofold. First, noisy property measurements reduce the signal-to-noise ratio available to the GP surrogates, slowing convergence. Second, alloys whose measured properties reflect processing defects rather than intrinsic compositional effects can mislead the acquisition function, directing the search toward or away from regions for the wrong reasons. Grain sizes in this campaign ranged from 10 to 35~$\mu$m (\autoref{fig:BSE_SEM}), and the more compositionally complex alloys (7--8 elements) generally showed finer, more uniform grains---consistent with sluggish diffusion reducing grain growth---and correspondingly lower property scatter. Standardizing the thermomechanical route to produce consistent, fully recrystallized microstructures across all compositions is a prerequisite for isolating compositional effects in future campaigns.

\subsection{Composition--Property Relationships}

\subsubsection{Correlation Structure and Feature Importance}

\begin{figure*}[t]
    \centering
    \includegraphics[width=1.0\textwidth]{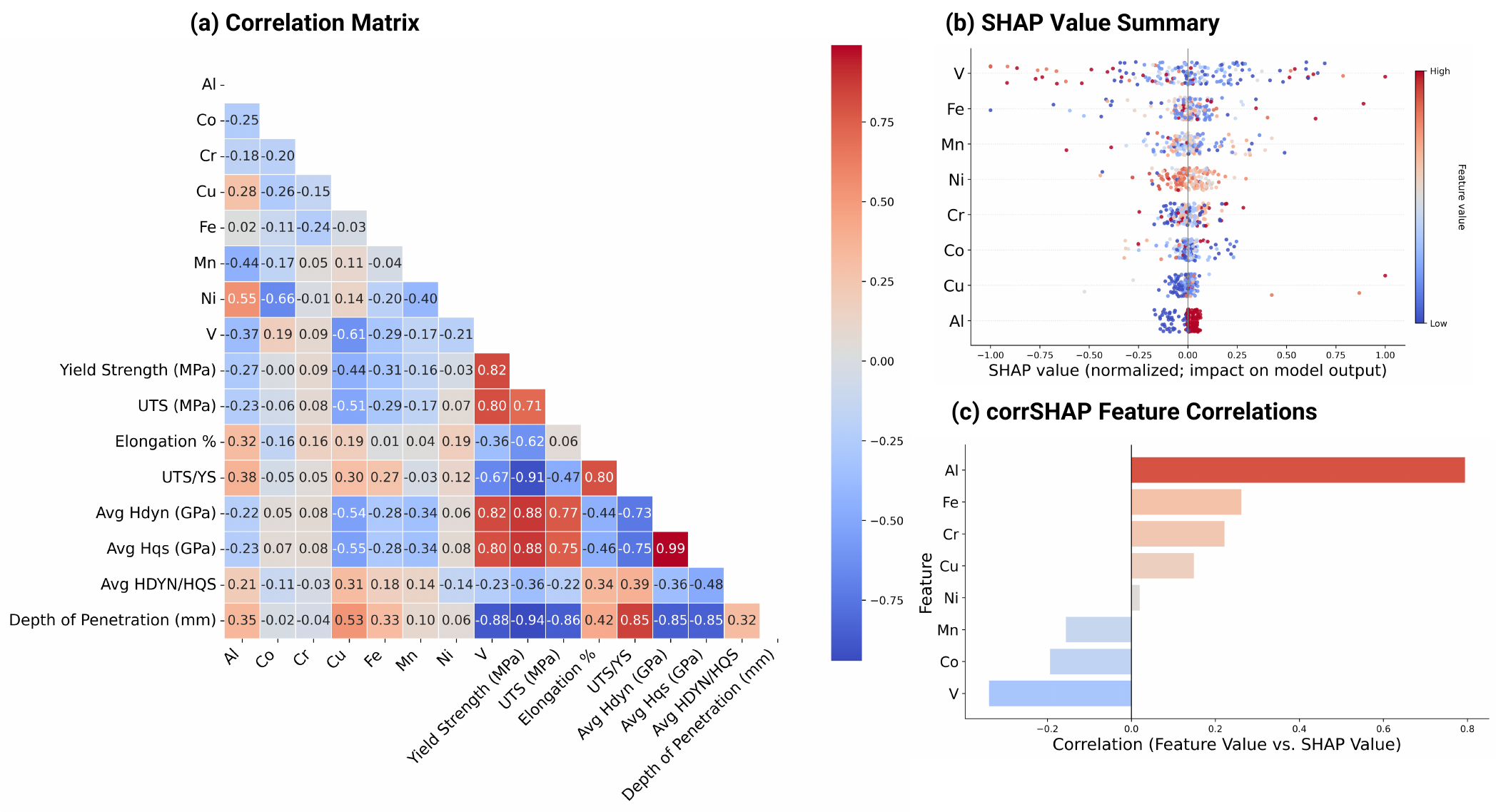}
    \caption{(a) Pearson correlation matrix among elemental concentrations, mechanical properties, and DoP. (b) SHAP beeswarm plot ranked by mean absolute SHAP value; dot color encodes feature value (red = high, blue = low). (c) corrSHAP bar chart showing the net directional influence of each element on the model output.}
    \label{fig:correlation_matrix}
\end{figure*}

\autoref{fig:correlation_matrix} combines three complementary analyses: (a)~a Pearson correlation matrix quantifying linear relationships among elemental concentrations and mechanical responses, (b)~a SHAP beeswarm plot showing each element's importance and directional behavior in the predictive model, and (c)~a corrSHAP bar chart measuring the correlation between each element's compositional value and its SHAP contribution. SHAP values were computed from the gradient-boosted regression models trained independently on each of the five target objectives normalized per target, and pooled to give a single multi-objective view. The multi-objective corrSHAP (\autoref{fig:correlation_matrix}c) is the per-target Pearson correlation between feature value and SHAP value, averaged over the five target models. Positive values mean the element systematically raises the composite response across targets; cancellation across targets reflects a multi-objective trade-off.

Vanadium and copper define the primary compositional axis governing strength and penetration resistance. Vanadium shows strong positive correlations with YS ($r = 0.82$, $n = 37$), UTS ($r = 0.80$), and both hardness measures ($r \approx 0.81$), and a strong negative correlation with DoP ($r = -0.88$). Copper shows the opposite pattern: negative correlations with all strength metrics ($r = -0.44$ to $-0.55$) and positive correlation with DoP ($r = 0.53$). The V--Cu anticorrelation ($r = -0.61$) means the design space naturally partitions along this axis.

Among property--property relationships, YS and hardness are tightly coupled ($r = 0.88$), and the two hardness metrics are nearly identical ($r = 0.99$). UTS/YS is strongly anticorrelated with YS ($r = -0.91$) and positively correlated with elongation ($r = 0.80$), confirming that strain-hardening capacity accompanies ductility. DoP correlates almost perfectly (inversely) with YS ($r = -0.94$) and strongly with UTS ($r = -0.86$), indicating that penetration resistance in this alloy family is governed predominantly by strength.

SHAP and corrSHAP reflect the chosen multi-objective target and not single-property linear correlations. The SHAP results align with the linear correlations. Vanadium dominates mean $|\text{SHAP}|$, with normalized contributions spanning the full $[-1, +1]$ range; high-V points (red) at the positive tail and low-V (blue) at the
negative tail, matching V's positive correlations with strength metrics. Iron, manganese, and nickel follow; aluminum has the smallest mean $|\text{SHAP}|$ and clusters near zero. Mn and Ni dots show mixed colors at both positive and negative SHAP values, indicating interaction-mediated (non-monotonic) effects whose sign depends on the concentrations of other elements.
 
Aluminum has the strongest positive corrSHAP ($r \approx +0.78$), significantly higher than iron ($+0.27$), chromium ($+0.22$), and copper ($+0.15$) --- higher Al consistently pushes the composite up despite Al's modest $|\text{SHAP}|$ magnitude. This is consistent with aluminum's role as a solid solution strengthener in Ni-stabilized HEAs, where small additions of Al improve YS and UTS/YS without strongly penalizing ductility. Nickel is near zero ($r \approx 0$), confirming the sign-flipping, interaction-mediated behavior seen in SHAP beeswarm plot. The negative tail is vanadium ($-0.32$), cobalt ($-0.22$), and manganese ($-0.17$). Vanadium's negative multi-objective corrSHAP coexists with its strong positive correlations with strength because the composite also includes elongation and UTS/YS, with which V correlates $-0.36$ and $-0.67$, respectively. The ductility penalty associated with V (likely tied to its role in promoting sigma phase formation and reducing strain hardening capacity) dominates the strength benefit when all five objectives are weighted together.

\subsubsection{Strength--Ductility Trade-off}

\begin{figure*}[t]
    \centering
    \includegraphics[width=\textwidth]{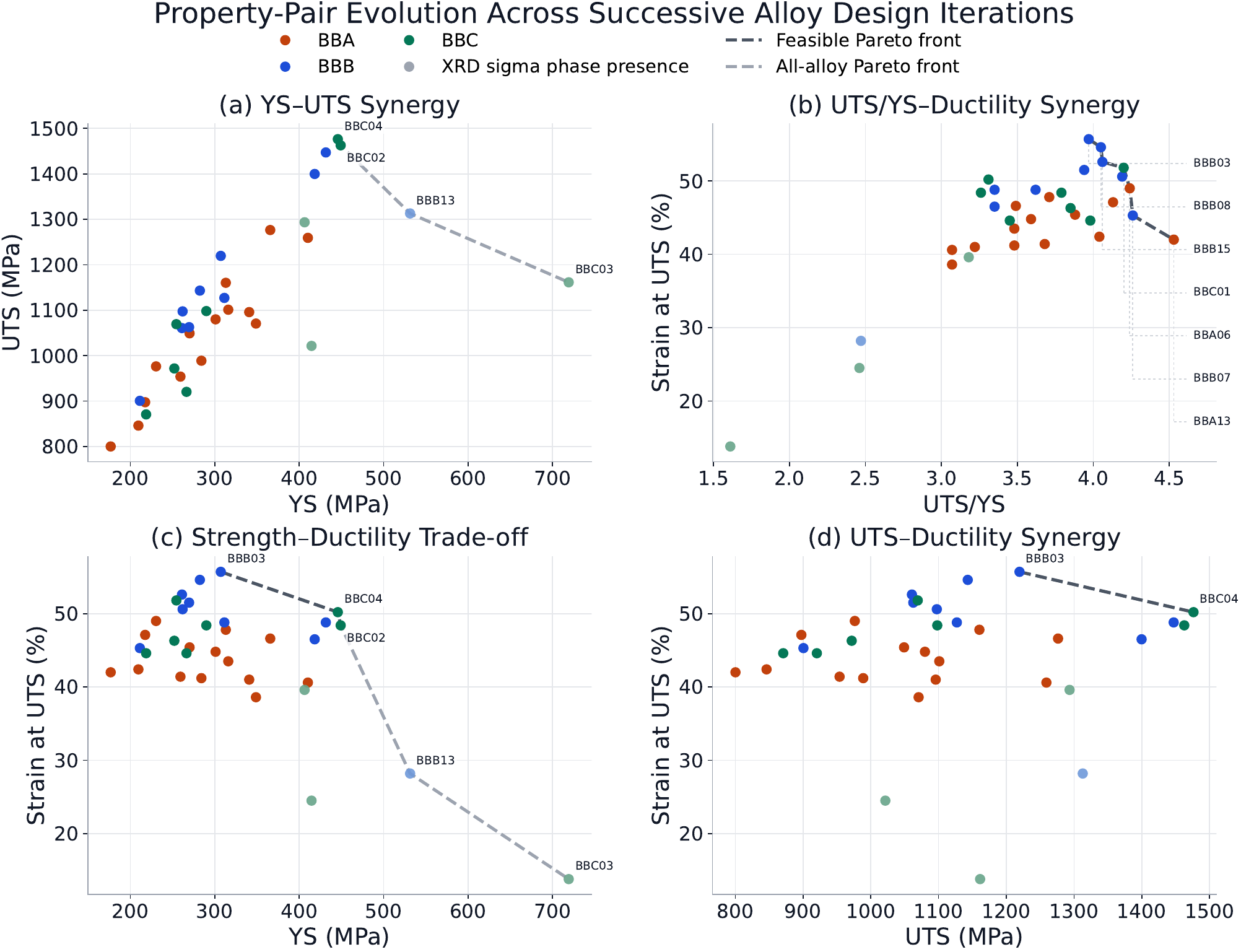}
    \caption{Pair plots of mechanical property combinations across three successive design iterations. Filled points denote single-phase FCC alloys and desaturated points indicate XRD-confirmed sigma phase presence. Dashed dark-grey lines trace the Pareto front considering only feasible alloys, while dashed light-grey lines trace the Pareto front considering all alloys.}
    \label{fig:pair_plots}
\end{figure*}

\autoref{fig:pair_plots} shows how the BBA (iteration~0), BBB (iteration~1), and BBC (iteration~2) alloys evolve through key strength--ductility property spaces, illustrating both improvement in feasible FCC alloys and the trade-off imposed by sigma phase formation.

In panel~(a), most BBA (iteration~0) alloys occupy the lower-strength region, while BBB (iteration~1) and BBC (iteration~2) extend further into the upper-right portion of the plot. BBC04 reaches ${\sim}450$~MPa YS at ${\sim}1480$~MPa UTS, while BBC03 achieves ${\sim}720$~MPa YS but at lower UTS (${\sim}1160$~MPa), consistent with sigma-phase embrittlement. In panel~(b), several of the highest-ductility, highest-work-hardening alloys across the campaign, including BBB03, BBB08, BBB15, BBC01, and BBA06, cluster above 45\% elongation with UTS/YS $>$ 3.5. In panel~(c), the population follows the expected negative trend between elongation and YS, but BBC04 and BBC02 (${\sim}450$--490~MPa YS and ${\sim}48$--50\% elongation) sit above it. In panel~(d), BBB (iteration~1) and BBC (iteration~2) increasingly populate the upper-right quadrant, with BBC04 combining ${\sim}1480$~MPa UTS with ${\sim}50\%$ elongation.

The sigma-bearing alloys BBC03 and BBB13 serve as counterexamples: elevated YS comes at steep ductility cost, reinforcing the importance of phase stability as a design constraint. A detailed analysis of the compositional and microstructural factors driving these improvements will be presented in a forthcoming publication.

\subsubsection{Strength--DoP Correlations and Rate Sensitivity}

\begin{figure*}[t]
    \centering
    \includegraphics[width=0.8\textwidth]{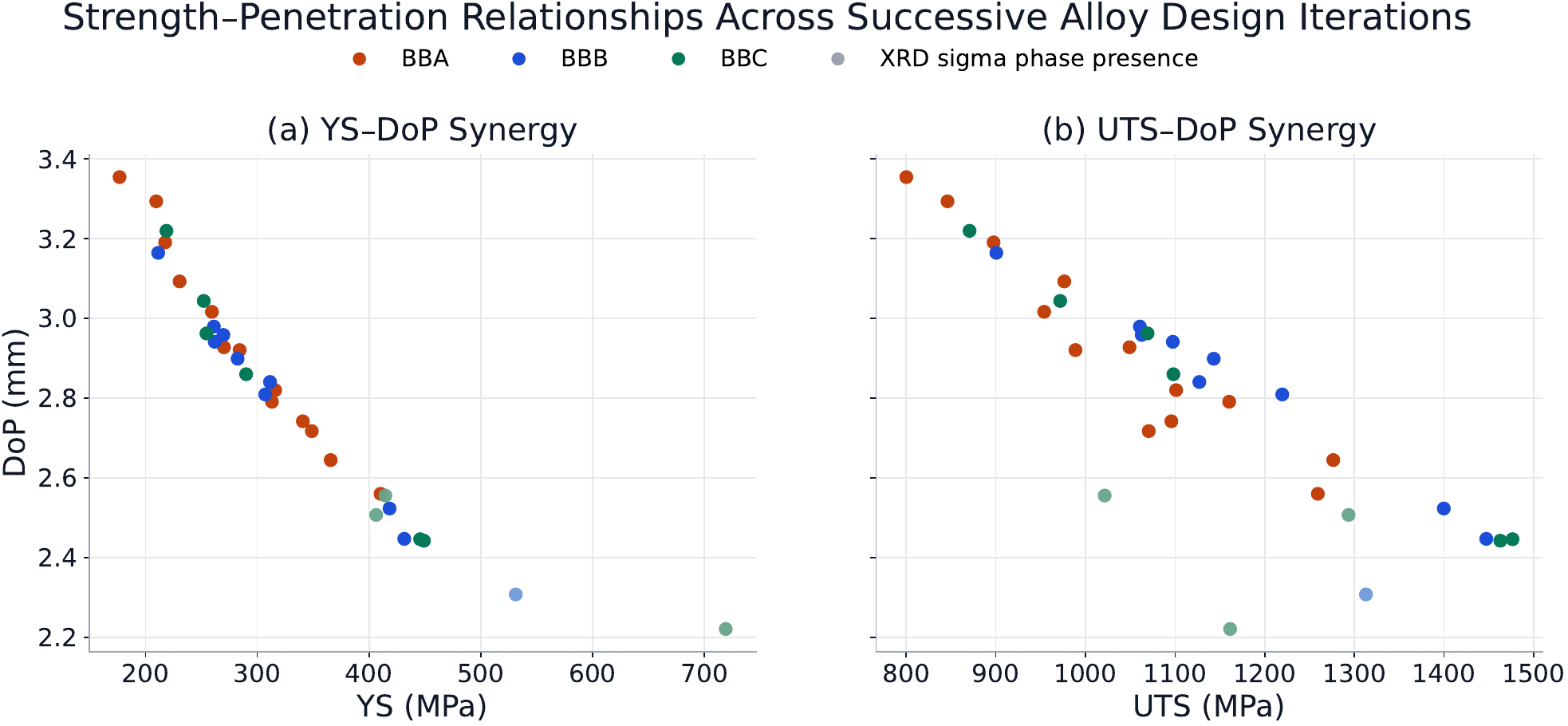}
    \caption{Simulated depth of penetration (DoP) versus quasi-static strength metrics across all tested alloys: (a) yield strength (YS) and (b) ultimate tensile strength (UTS). In both panels, higher strength is associated with lower DoP, indicating that the strongest alloys also show the greatest simulated impact resistance.}
    \label{fig:DOPvAllStrengths}
\end{figure*}

DoP shows the expected inverse correlation with strength metrics (\autoref{fig:DOPvAllStrengths}): higher YS and UTS correspond to lower penetration depth. The strong DoP--YS correlation ($r = -0.94$) raises the question of what DoP adds beyond YS alone. The residual variance captures the contribution of rate-dependent strengthening through the Cowper--Symonds constitutive model: alloys with the same YS but different rate-sensitivity exponents $m$ produce different DoP values, providing a simulation-derived discriminant that quasi-static testing alone cannot.

\begin{figure*}[t]
    \centering
    \includegraphics[width=0.9\textwidth]{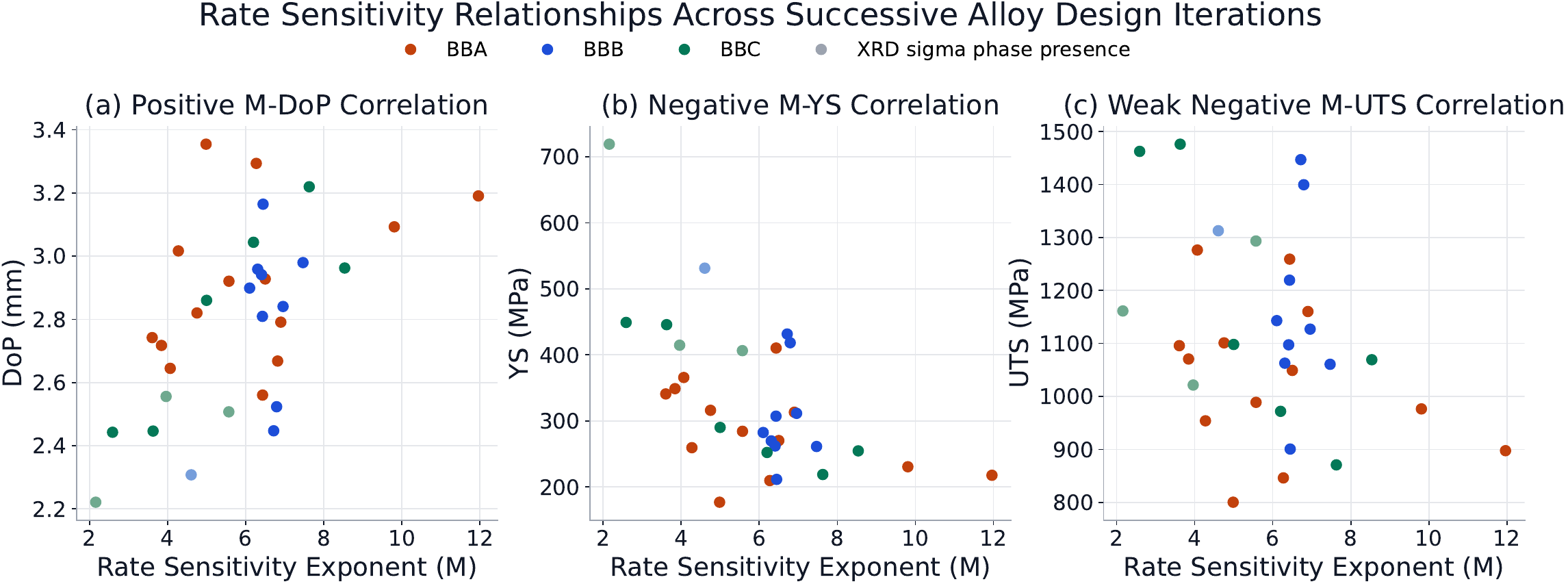}
    \caption{Rate sensitivity exponent ($m$) versus (a) simulated DoP, (b) yield strength, and (c) UTS for all candidate alloys across three iterations.}
    \label{fig:MRelationships}
\end{figure*}

The relationship between DoP and the calibrated rate-sensitivity exponent $m$ (\autoref{fig:MRelationships}(a)) is moderately positive, meaning that higher $m$ values are associated with larger penetration depths. YS and UTS both correlate negatively with $m$ (\autoref{fig:MRelationships}(b,c)), with the YS relationship stronger than the UTS relationship. Together, these trends indicate that alloys with higher quasi-static strength tend to exhibit lower calibrated rate-sensitivity exponents in this dataset, consistent with a trade-off between quasi-static strength and rate sensitivity in the present Cowper--Symonds fits.

\subsubsection{Emergent Compositional Design Rules}

The optimization campaign revealed two distinct compositional routes to high performance within the FCC-feasible space, which we describe below.

The first is a \emph{high-strength regime} defined by V = 20--24~at.\%, Cr $\leq$ 4~at.\%, Ni = 36--60~at.\%, and Cu = 0. The four strongest single-phase FCC alloys---BBC04 (${\sim}1480$~MPa UTS, 50\% elongation), BBC02 (${\sim}1460$~MPa, 48\%), BBB01 (${\sim}1450$~MPa, 49\%), and BBB12 (${\sim}1400$~MPa, 47\%)---all share this compositional signature. The strong V--YS correlation ($r = 0.82$, $n = 37$; Section~3.3.1) is consistent with lattice distortion strengthening from the large atomic-size and modulus mismatch of V in the FCC matrix~\cite{varvenne2016theory}. We note, however, that the Varvenne--Luque--Curtin model assumes a fully random solid solution; at V = 20--24~at.\%, deviations from randomness (short-range ordering or local V clustering) could affect quantitative agreement, and this has not been independently validated for the alloys in this campaign. Nickel stabilizes the FCC phase at high V concentrations (VEC effect), and chromium is minimized to avoid sigma formation. BBC04 is the most extreme example: 60~at.\% Ni and 24~at.\% V with no Co, no Cu, and only 4~at.\% Cr---an alloy at the outer edge of the design space that the BO identified precisely because the constraint-aware acquisition permitted boundary exploration.

The second is a \emph{high-ductility regime} with moderate V (8--12~at.\%), higher Mn (16~at.\%), and moderate Cr. BBB03 (${\sim}1220$~MPa UTS, 56\% elongation, UTS/YS = 3.97) and BBC01 (${\sim}1070$~MPa, 52\%, UTS/YS = 4.20) represent this compositional region. The elevated UTS/YS ratios and sustained elongation in these Mn-rich alloys are consistent with enhanced strain hardening, possibly through SFE reduction and twinning~\cite{li2016metastable, otto2013influences}, although direct microstructural evidence of the operating deformation mechanisms was not obtained in this campaign. These alloys trade peak strength for superior ductility and strain hardening capacity.

The two routes occupy different regions of the Pareto front and validate the 8-element expansion from Campaign~1. The high-strength route exploits V--Ni synergy that was inaccessible in the 6-element system (which lacked the compositional freedom to simultaneously maximize V and Ni while suppressing Cr). The high-ductility route exploits Mn, which was absent from Campaign~1 entirely. Cu, by contrast, showed negative Pearson correlations with all strength metrics ($r = -0.44$ to $-0.51$; Section~3.3.1) while its multi-objective corrSHAP is slightly positive ($r \approx +0.15$), reflecting partial cancellation across the five targets; it appeared in none of the top-performing alloys. The BO effectively learned to deprioritize Cu-containing compositions---a result that informs future campaign design in this alloy family.

\subsection{Bayesian Optimization Performance and Feasibility Analysis}

Sections 3.1--3.3 reported the properties and composition--property relationships across the three iterations. The optimization metrics below quantify how the BRAVE framework produced these outcomes, and what the pattern of successes and failures exposes about the structure of the constrained design problem.

Table~\ref{tab:bo_metrics} summarizes the optimization performance across the three iterations. The initial Pareto front (BBA) contained 10 non-dominated alloys with a hypervolume of 658.24. Iteration BBB increased the hypervolume to 817.77 (24.2\% gain) and expanded the Pareto front to 17 alloys, with 9 new non-dominated compositions. Iteration BBC reached a hypervolume of 856.30 (4.7\% gain) with 21 Pareto-optimal alloys.

The Bayesian advantage of the campaign is best measured by the concentration of experimental budget in high-value regions of the design space, rather than by Pareto front size or hypervolume trajectory---both of which are expected to increase under random sampling in correlated objective spaces. The campaign explored 0.12\% of the feasible space (33 feasible alloys out of 27,240 candidates) and concentrated four of its strongest alloys (BBC04, BBC02, BBB01, BBB12---all with V $\geq$ 20~at.\% and Ni $\geq$ 36~at.\%) in a compositional region comprising approximately 100 of the 27,240 feasible candidates ($<$0.4\% of the feasible space). Under random sampling, the probability of placing four or more of 33 draws in this region is $P \approx 6.5 \times 10^{-6}$ (hypergeometric distribution). The feasibility-weighted acquisition strategy is further validated by a controlled computational benchmark (Appendix), in which risk-weighted EHVI matched oracle performance (perfect feasibility knowledge) while reducing wasted evaluations approximately fivefold relative to feasibility-blind BO, across two bi-objective optimization problems on the 6-component Al--V--Cr--Fe--Co--Ni system.

\begin{table*}[t]
\centering
\caption{Bayesian Optimization Performance Across Iterations}
\label{tab:bo_metrics}
\begin{tabular}{lccccc}
\toprule
\textbf{Iteration} & \textbf{Avg EHVI} & \textbf{Hypervolume} & \textbf{HV Gain} & \textbf{Pareto Size} & \textbf{Infeasible} \\
\midrule
BBA & N/A & 658.24 & --- & 10 & 1/16 \\
BBB & 836.41 & 817.77 & 24.2\% & 17 (9 new) & 5/16 \\
BBC & 1059.2 & 856.30 & 4.7\% & 21 (5 new) & 9/16 \\
\bottomrule
\end{tabular}
\end{table*}

\subsubsection{Feasibility--Performance Trade-off}

The number of alloys infeasible for Bayesian optimization (sigma phase or manufacturing failure preventing XRD verification) increased across iterations: 1/16 in BBA (BBA04, manufacturing failure), 5/16 in BBB (BBB05, BBB06, BBB09, BBB11, BBB13---all sigma), and 9/16 in BBC (BBC03, BBC08, BBC14 sigma; BBC05, BBC06, BBC07, BBC09, BBC10, BBC16 failed cold rolling before XRD). Of the 33 FCC-confirmed alloys, most proceeded to full characterization. A few FCC-confirmed alloys (BBA09, BBB10, BBB16) showed mechanical anomalies (brittle fracture or processing-induced defects) that limited the quality of their objective data but did not change their phase classification.

\emph{Sigma-phase failures} decreased from 0 in BBA to 5 in BBB to 3 in BBC. The reduction from 5 to 3 between BBB and BBC reflects two mid-campaign interventions: the transition from TCHEA6 to TCHEA7 with dual-temperature screening, and the retraining of the GPC feasibility classifier on the 5 sigma outcomes from BBB. The GPC and updated CALPHAD screening reduced the sigma failure rate despite the BO continuing to target V-rich, high-strength compositions near the FCC/sigma boundary.

\emph{Manufacturing and processing failures} (alloys that never reached XRD verification or that failed during cold rolling) went from 1 in BBA to 0 in BBB to 6 in BBC. The six BBC processing failures all failed cold rolling. The current framework has no mechanism to predict or prevent these failures, as the GPC is trained only on phase outcomes. Post-hoc analysis suggests that V-containing nano-precipitates may be responsible for embrittlement in certain compositions, but this finding postdates the present study and will be reported separately.

This observation highlights a deeper tension in this alloy system. Table~\ref{tab:v_failure} quantifies the relationship between V content and outcome across all 48 alloys: no alloy with V = 0 failed, while failure rates exceed 50\% for V $\geq$ 12~at.\%. The top row of the performance landscape (V $\geq$ 20~at.\%) contains both the three strongest alloys in the campaign (BBC04, BBC02, BBB01---all at V = 24~at.\%) and four of the eight sigma-phase failures. At V = 24~at.\%, half the alloys fail and half are the best performers---the definition of a boundary-adjacent optimum. High V content simultaneously increases the risk of sigma-phase formation and processing-induced embrittlement. The highest-performing alloys therefore reside adjacent to multiple feasibility boundaries

\begin{table*}[t]
\centering
\caption{Failure rate and average yield strength by vanadium content across all 48 alloys.}
\label{tab:v_failure}
\small
\begin{tabular}{ccccp{0.48\textwidth}}
\toprule
\textbf{V (at.\%)} & \textbf{Total} & \textbf{Failed} & \textbf{Failure Rate} & \textbf{Notable Alloys} \\
\midrule
0  & 8  & 0 & 0\%   & --- \\
4  & 10 & 1 & 10\%  & BBA04 (manufacturing) \\
8  & 6  & 1 & 17\%  & BBC05 (cold rolling) \\
12 & 9  & 2 & 22\%  & BBC06, BBC07 (cold rolling) \\
16 & 4  & 4 & 100\% & BBB09, BBB11 (sigma); BBC10, BBC16 (cold rolling) \\
20 & 5  & 4 & 80\%  & BBB06, BBB13, BBC08 (sigma); BBC09 (cold rolling) \\
24 & 6  & 3 & 50\%  & BBB05, BBC03, BBC14 (sigma); BBC04, BBC02, BBB01 (FCC, top 3 UTS) \\
\bottomrule
\end{tabular}
\end{table*}

To quantify whether the relationship between V content and infeasibility is statistically significant, we applied two complementary tests~\cite{virtanen2020scipy}. First, treating V content as a continuous variable, the Pearson correlation between V content and infeasibility (coded as 0 or 1) is $r = 0.54$ ($P = 9.0 \times 10^{-5}$, $n = 48$): higher V is associated with higher probability of failure, and the probability that this association arises by chance is less than 0.01\%. This association is driven by V specifically, not by the V + Cr combination that defines the sigma stability threshold: V and Cr are uncorrelated across the 48 alloys ($r = 0.06$), confirming that V is a dominant predictor of infeasibility. Second, dividing alloys into high-V ($\geq 16$~at.\%) and low-V ($< 16$~at.\%) groups, 11 of 15 high-V alloys were infeasible compared to 4 of 33 low-V alloys---an odds ratio of 19.9 (Fisher's exact test, $P = 5.3 \times 10^{-5}$), meaning a high-V alloy is approximately 20 times more likely to fail than a low-V alloy. This association is robust to the choice of threshold: the odds ratio remains significant ($P < 0.05$) for all cutoffs from V $\geq$ 12 to V $\geq$ 20~at.\%.

Among the 30 feasible alloys with yield strength data, V content correlates with YS at $r = 0.84$ ($P = 8.7 \times 10^{-9}$). The same element that most strongly predicts mechanical performance also most strongly predicts failure---both at significance levels below $P < 10^{-4}$. The BO's preferential sampling of high-V alloys in iterations BBB and BBC (15 of 32 alloys with V $\geq$ 16~at.\%, compared to 0 of 16 in the diversity-sampled BBA) is itself evidence that the surrogates correctly identified V as the performance driver. The selection bias toward high V is not a confound---it is the optimizer doing its job, concentrating budget where the predicted payoff is highest. The consequence is that both the best alloys and the most failures accumulate in the same compositional region, which is the signature of a boundary-adjacent optimum.

This dual role is a pattern consistent with the general observation in constrained optimization that optimal solutions tend to lie on or near active constraint boundaries~\cite{ray2009infeasibility}. This coupling arises when the same compositional variable drives both performance and constraint violation: increasing V improves strength through lattice distortion effects (the objective gradient points toward high V) but simultaneously promotes sigma-phase formation (the constraint gradient points in the same direction). The unconstrained optimum would be a very high-V alloy that forms sigma. The constrained optimum is therefore the alloy with the highest V content that remains single-phase FCC---on the phase stability boundary by definition.

(Formally, the Karush--Kuhn--Tucker conditions for constrained optimality require $\nabla f + \lambda \nabla g = 0$ with multiplier $\lambda > 0$ at the solution, placing the optimum on the active constraint boundary $g(\mathbf{x}) = 0$~\cite{nocedal2006numerical}. The KKT conditions apply strictly to continuous, single-objective problems; the present campaign operates on a discrete compositional grid with multiple objectives, but the underlying intuition---that aligned objective and constraint gradients push optima toward boundaries---transfers to this setting.) If the optimum were far from the constraint, the constraint would not be restricting the search, and the problem would be effectively unconstrained. This has a practical implication for campaign design: in alloy systems where the primary strengthening element also destabilizes the target phase (e.g., V in the present system), a feasibility-aware acquisition strategy is a prerequisite for accessing the high-performing region of the design space. The present campaign provides an experimental realization of this principle: vanadium drives both the objective ($r = 0.84$ with YS) and the constraint ($r = 0.54$ with infeasibility), and the three strongest alloys in the campaign share the V = 24~at.\% boundary with three sigma failures. The boundary phenomenon has been observed implicitly in metastability-engineered HEAs, where optimal TRIP-mediated strength--ductility combinations arise at the edge of FCC phase stability~\cite{li2016metastable}, and has been proposed as an algorithmic strategy for constrained Bayesian optimization~\cite{tian2024boundary, maguire2025good}. However, the explicit connection between performance--constraint coupling through a shared compositional variable and the KKT prediction that constrained optima lie on active boundaries has not, to our knowledge, been demonstrated quantitatively in an experimental alloy campaign.

The sigma-phase trend (0 $\rightarrow$ 5 $\rightarrow$ 3) is the more informative metric for evaluating the feasibility-aware acquisition strategy. In iteration BBA, candidates were selected by diversity-aware k-medoids sampling across the interior of the FCC stability region, producing no sigma failures. In BBB, the BO targeted high-V, high-Cr regions where the GP surrogates predicted strong mechanical performance---the same regions where sigma is thermodynamically competitive. The 5 sigma failures provided the GPC with its first substantial training data and triggered the TCHEA7 transition. In BBC, the updated screening and retrained GPC reduced sigma failures to 3 despite continued exploration of the V-rich boundary---evidence that the feasibility-aware acquisition learned to avoid the most sigma-prone compositions while preserving access to the high-performing V-rich region.

The counterfactual remains instructive. A hard-filtered strategy would have excluded the high-V region entirely after the BBB sigma failures. BBC04 and BBC02---the two highest-UTS alloys in the campaign---would never have been synthesized. The BRAVE framework found them by accepting controlled risk near the phase boundary, with the GPC progressively sharpening the feasibility boundary between iterations.

The relevant performance metric is the performance gained per iteration, not the absolute failure rate. Despite 9/16 total failures in BBC (3 sigma + 6 processing), the 7 feasible alloys from that iteration included the strongest compositions in the campaign (BBC04, BBC02), and the hypervolume continued to increase (4.7\% gain over BBB). Extending the feasibility model to incorporate processing constraints---predicting which compositions are likely to fail cold rolling, independent of phase stability---is a priority for future campaigns.

The feasibility analysis also exposes CALPHAD database fidelity as a practical bottleneck for closed-loop alloy campaigns. The TCHEA6 database, used for initial screening, failed to predict sigma formation in several V- and Cr-rich compositions that were experimentally confirmed as multiphase. The mid-campaign transition to TCHEA7---with its expanded sigma, BCC, HCP, and liquid phase assessments across $>$100 ternary subsystems---reduced but did not eliminate these prediction errors (Section~3.1). The GPC feasibility classifier absorbs residual prediction errors by learning the empirical feasibility boundary from experimental phase outcomes, but its effectiveness depends on the quality of the thermodynamic screening that defines the initial candidate pool. A poorly calibrated CALPHAD database shifts the entire feasible space, potentially excluding high-performing regions from the outset or flooding the candidate pool with infeasible compositions that consume experimental budget before the GPC has sufficient training data to compensate. This dependency has practical implications: future campaigns should (i)~validate CALPHAD predictions against a small exploratory batch before committing to full-scale optimization, (ii)~incorporate dual-temperature screening (as we adopted here after iteration BBB) from the first iteration, and (iii)~prioritize thermodynamic databases with assessed coverage of the relevant ternary and quaternary subsystems.

\section{Conclusions}\label{sec:Conclusion}

In this work, we applied a risk-aware Bayesian optimization campaign (BRAVE) to the Al--V--Cr--Mn--Fe--Co--Ni--Cu FCC HEA system, synthesizing 48 alloys over three closed-loop iterations. We optimized five objectives simultaneously (YS, UTS/YS, strain at UTS, $H_\text{dyn}/H_\text{qs}$, DoP) while exploring 0.12\% of the 27,240-composition feasible space.

\begin{enumerate}

\item \textbf{The highest-performing compositions reside on the FCC phase stability boundary.} Vanadium drives both yield strength ($r = 0.84$, $n = 30$ feasible alloys) and sigma-phase formation ($r = 0.54$, $P = 9 \times 10^{-5}$, $n = 48$). At V = 24~at.\%, the three strongest alloys and three sigma failures share the same compositional point---the definition of a boundary-adjacent optimum. This outcome corresponds to the KKT prediction that constrained optima lie on active boundaries when objective and constraint gradients align~\cite{nocedal2006numerical, ray2009infeasibility}. Although such optima have been observed implicitly in metastability-engineered HEAs~\cite{li2016metastable} and proposed algorithmically~\cite{tian2024boundary, maguire2025good}, they have not, to our knowledge, been demonstrated with statistical significance in a closed-loop experimental campaign.

\item \textbf{BRAVE navigated the boundary where simpler strategies cannot.} Hard filtering would have excluded the V-rich region after the first sigma failures; BBC04 and BBC02 (the two highest-UTS alloys) would never have been synthesized. Conversely, feasibility-blind BO would have repeatedly queried infeasible compositions without learning to avoid them. BRAVE embedded a learned feasibility probability into EHVI, concentrating resources in the V-rich, Ni-rich corner ($<$0.4\% of the feasible space, $P \approx 6.5 \times 10^{-6}$ under random sampling) while reducing the sigma rate from 5/16 to 3/16. Processing failures (1 $\rightarrow$ 0 $\rightarrow$ 6) lie outside the framework's scope; however, despite 9/16 total infeasible alloys in BBC, the 7 feasible compositions included the strongest in the campaign. A computational benchmark (Appendix) confirms that risk-weighted EHVI matches oracle performance with ${\sim}5\times$ fewer wasted evaluations.

\item \textbf{Two compositional regimes emerged from the 8-element space.} A high-strength regime (V = 20--24~at.\%, Cr $\leq$ 4~at.\%, Ni = 36--60~at.\%) produced BBC04 (${\sim}1480$~MPa UTS, 50\% elongation), where Ni stabilizes FCC at extreme V concentrations through VEC effects. A high-ductility regime (V = 8--12~at.\%, Mn = 16~at.\%) achieved UTS/YS up to 4.20 at $>$50\% elongation, consistent with SFE reduction and enhanced strain hardening~\cite{li2016metastable, otto2013influences}. Neither regime was accessible in Campaign~1's 6-element system. SHAP analysis confirmed V as the most influential feature (highest mean $|\text{SHAP}|$), consistent with its strong linear correlation with strength ($r = 0.82$). Ni, ranked fourth by mean $|\text{SHAP}|$, shows near-zero linear correlation with strength ($r = -0.03$) but non-negligible SHAP contributions whose sign reverses with compositional context---consistent with interaction-mediated, non-monotonic influence. Cu correlated negatively with all strength metrics and positively with DoP.

\item \textbf{The campaign avoided the strength--ductility trade-off and compounded gains from Campaign~1.} Median YS increased across iterations while ductility was preserved, and inter-quartile ranges narrowed---indicating convergence. Every single-phase FCC alloy exceeded Campaign~1's best UTS/YS (2.93~\cite{hastings2025accelerated}): minimum 3.07, median 3.79, 9 above 4.0. We attribute this to three compounding factors: (i)~informative priors from Campaign~1 data and the Varvenne--Luque--Curtin model, (ii)~a design space expanded based on Campaign~1's lessons, and (iii)~objectives refined from the finding that hardness alone is an incomplete proxy for mechanical response.

\item \textbf{DoP operated as a simulation-in-the-loop objective.} Calibrated from tensile and SHPB data via a Cowper--Symonds model, DoP tracks YS ($r = -0.94$) but also captures rate-sensitivity effects invisible to quasi-static testing. The strongest alloys (high V) showed lower rate-dependent strengthening, consistent with dislocation-glide-dominated deformation---a trade-off that is invisible without a rate-sensitive objective.

\item \textbf{CALPHAD database fidelity limits the feasibility classifier.} The mid-campaign TCHEA6 $\rightarrow$ TCHEA7 transition reduced but did not eliminate false negatives---the errors that produce failed experiments. The GPC learns the empirical boundary from phase outcomes; however, a poorly calibrated database can exclude high-performing regions or admit too many infeasible compositions before the GPC acquires sufficient training data.

\end{enumerate}

We note several limitations. The campaign optimized composition only; microstructure and processing were not design variables. Variability in grain size (10--35~$\mu$m) and recrystallization response introduced scatter that compositional surrogates cannot capture---several phase-pure alloys showed anomalous tensile behavior traceable to incomplete recrystallization, reducing the signal-to-noise available to the GP models. The DoP model uses a Cowper--Symonds framework without damage criteria, limiting its scope to relative penetration ranking. Furthermore, the GPC (trained on 8 sigma failures) cannot predict the 7 processing failures (i.e., cold-rolling defects) that dominated BBC's budget---a distinct failure mode outside the classifier's design scope.

The three-iteration budget was set by experimental throughput rather than convergence criteria, and the diminishing HV gain (24.2\% in BBB, 4.7\% in BBC) leaves open whether additional iterations would yield substantial further improvement.

In future work, we plan to incorporate microstructural descriptors (e.g., grain size, texture, precipitate fraction) as design variables and extend the feasibility model to processing failures. Testing BRAVE on systems with more complex phase competition (e.g., refractory HEAs with BCC/HCP/sigma boundaries) would assess generalizability. The KKT observation predicts boundary-adjacent optima should recur whenever the performance-driving element also destabilizes the target phase.

\section*{Declaration of competing interest}

    The authors declare that they have no known competing financial interests or personal relationships that could have appeared to influence the work reported in this paper.

\section*{Data Availability}

    The data generated and analyzed during this study is included in \underline{Supplementary Information} and is also available in the github repository below, provided as an Excel database.

\section*{Code Availability}

    Code used in this project is available on https://github.com/mrinalinimulukutla/BRAVE.

\section{Acknowledgements}

The research was sponsored by the Army Research Laboratory and was accomplished under Cooperative Agreement Number W911NF-22-2-0106. The views and conclusions contained in this document are those of the authors. They should not be interpreted as representing the official policies, either expressed or implied, of the Army Research Laboratory or the US Government. The US Government is authorized to reproduce and distribute reprints for Government Purposes, notwithstanding any copyright notation herein. 
    
The authors would like to acknowledge Dr.\ Anup and his group at Texas A\&M University for access to their XRD Bruker D8 Discover instrument, and Texas A\&M's Materials Characterization Facility for access to their EBSD-capable FIB-SEM. The authors also acknowledge Dr.\ Ken Vecchio and his group at the University of California, San Diego for their support with Split-Hopkinson Pressure Bar (SHPB) testing, and Texas A\&M's High Performance Research Computing group for access to the computational infrastructure used in this work.

We also acknowledge the use of Claude Sonnet 4.6 for data analysis and some sentence structure formations in the manuscript. After using this tool/service, the author(s) reviewed and edited the content as needed and take(s) full responsibility for the content of the publication.

\bibliography{references.bib}

\section*{Appendix}
\label{sec:appendix}

\subsection*{A. Mathematical Details of the BRAVE Framework}
\label{sec:appendix_math}

\begin{figure*}[!ht]
    \centering
    \includegraphics[width=0.95\textwidth]{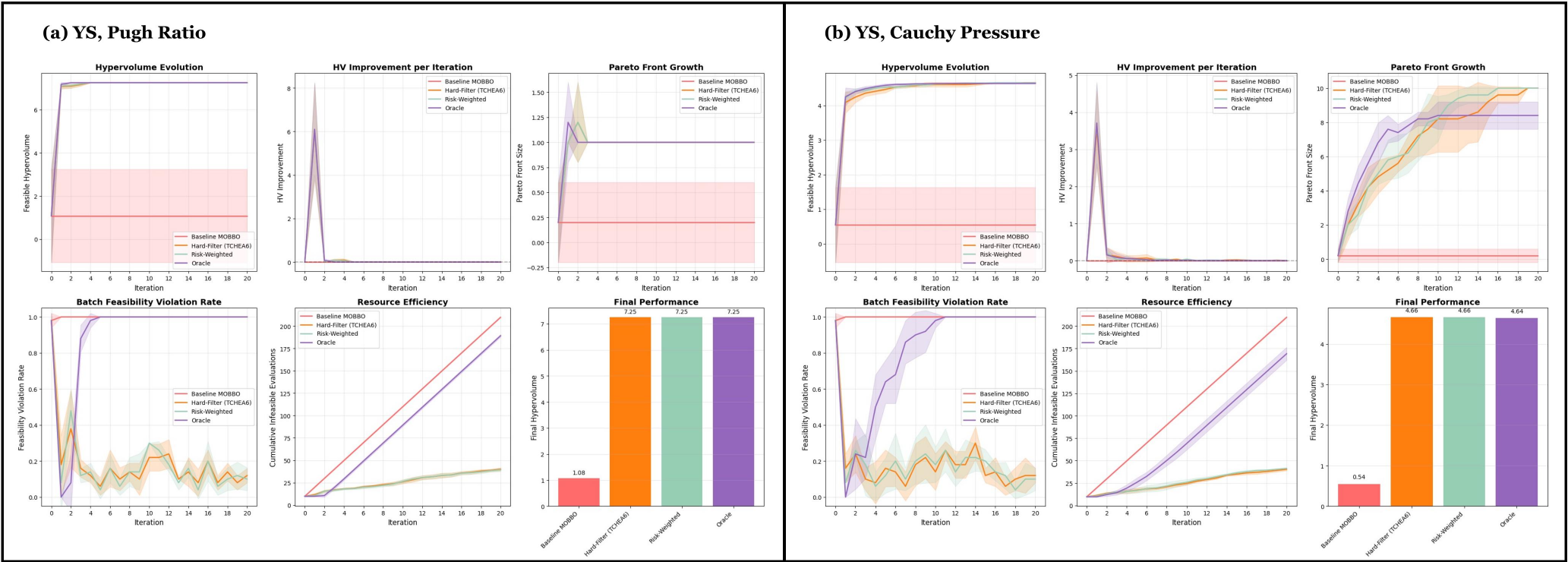}
    \caption{Benchmarking results for four MOBBO strategies on the six-component HEA system. Two bi-objective tasks are shown: (a) CV(YS)--Pugh ratio and (b) CV(YS)--Cauchy pressure. Metrics: feasible hypervolume (HV), Pareto front size, and feasibility-violation rate (FVR). Risk-weighted MOBBO matches oracle performance while maintaining low FVR.}
    \label{fig:benchmark_results}
\end{figure*}

This section collects the mathematical details of the surrogate modeling, multi-objective acquisition, multi-source fusion, and batch selection components summarized in Section~2.6.

\subsubsection*{A.1 Gaussian Process Regression}

In Bayesian optimization, probabilistic surrogate models are used to represent expensive-to-evaluate objective functions and guide candidate selection through uncertainty-aware predictions. We used Gaussian process regression (GPR) as the surrogate due to its low computational cost and flexible distance-based correlation structure~\cite{Rasmussen:2005:GPM:1162254}.

Assuming there are $N$ previously observed data denoted by $\{\mathbf{X}_{N}, \mathbf{y}_{N}\}$, where $\mathbf{X}_{N} = (\mathbf{x}_{1}, \ldots, \mathbf{x}_{N})$ and $\mathbf{y}_{N} = \left(f(\mathbf{x}_{1}), \ldots, f(\mathbf{x}_{N})\right)$, then the GP probabilistic prediction at unobserved location \textbf{x} is a normal distribution:
\begin{equation}
\label{GP11}
f_{\textrm{GP}}(\mathbf{x}) \mid \mathbf{X}_{N}, \mathbf{y}_{N} \sim \mathcal{N}\left(\mu(\mathbf{x}),\sigma_{\textrm{GP}}^2(\mathbf{x})\right)
\end{equation}
where
\begin{equation}
\begin{aligned}
\label{meancov}
\mu(\mathbf{x}) &= K(\mathbf{X}_{N},\mathbf{x})^\textrm{T}[K(\mathbf{X}_{N},\mathbf{X}_{N})+\sigma^2_{n}I]^{-1} \mathbf{y}_{N}\\
\sigma_{\textrm{GP}}^2 (\mathbf{x})  &=  k(\mathbf{x},\mathbf{x}) - K(\mathbf{X}_{N},\mathbf{x})^\textrm{T} [K(\mathbf{X}_{N},\mathbf{X}_{N})+\sigma^2_{n}I]^{-1}K(\mathbf{X}_{N},\mathbf{x})
\end{aligned}
\end{equation}
here, $k$ is a real-valued kernel function to calculate correlations, $K(\mathbf{X}_{N},\mathbf{X}_{N})$ as a $N \times N$ matrix with $m,n$ entry as $k(\mathbf{x}_{m}, \mathbf{x}_{n})$, and $K(\mathbf{X}_{N}, \mathbf{x})$ is a $N \times 1$ vector with $m^{th}$ entry as $k(\mathbf{x}_{m}, \mathbf{x})$.  The term $\sigma^2_{n}$ is used to model experiment errors. A common choice for kernel function is squared exponential:
\begin{equation}
\label{eq:5}
k(\mathbf{x},\mathbf{x'}) = \sigma_s^2 \exp\left(- \sum_{h = 1}^{d} \frac {(x_h-x'_h)^2}{2l_h^2}\right )
\end{equation}
where $d$ is the dimensionality of the input space, $\sigma_s^2$ is the signal variance, and $l_h$, where $h = 1,2,\ldots,d$, is the characteristic length-scale to determine correlation strength between data points within dimension $h$.

\subsubsection*{A.2 Informative Priors}

Although standard GPR sets the mean function to zero, this forfeits the opportunity to encode prior knowledge. An informative prior instead sets the mean function to a physics-based model prediction, $\mu(\mathbf{x})$, and trains the GPR on the discrepancy $f'(\mathbf{x}) = f_{\textrm{exp}}(\mathbf{x}) - \mu(\mathbf{x})$ between experimental observations and the prior. The posterior $f = f' + \mu$ defaults to the physics-based prediction where no experimental data exist, and overrides it where observations are available.

In this work, we used GPR with an informative prior that integrates physical insights from the Varvenne--Luque--Curtin solid solution strengthening model~\cite{varvenne2016theory} as the prior, $\mu$. This model captures the strengthening behavior of FCC HEAs by accounting for solute misfit volumes, modulus mismatch, and local lattice distortions in a fully random solution. We combined this physics-based prior with limited experimental data, including Vickers hardness (HV) and temperature-dependent yield strength (YS) measurements.

Let the true composition-temperature-dependent mechanical behavior of the HEAs be denoted by $f_{\textrm{exp}}(\mathbf{x})$, where $\mathbf{x}$ is a vector describing the alloy composition and test temperature. The GPR model is built over the residual between experimental observations and the prior prediction, $f_{\textrm{dis}}(\mathbf{x}) = f_{\textrm{exp}}(\mathbf{x}) - \mu(\mathbf{x})$. The final GPR prediction is then:
\[
f(\mathbf{x}) = f_{\textrm{dis}}(\mathbf{x}) + \mu(\mathbf{x}).
\]

The machine learning models were developed using data from~\cite{hastings2025accelerated} in conjunction with the {\tt CBFV}~\cite{Murdock2020IsProperties} and {\tt HEACalculator} packages~\cite{doguhan_sariturk_2022_7429046} for composition feature generation. These tools generated over 5000 features for each composition, which were subsequently optimized using the RecursiveByShapValues method~\cite{Prokhorenkova2018CatBoost:Features} to eliminate redundant features based on SHAP values. Given the tabular nature of the dataset, with a limited number of datapoints and a high dimensionality of features, the CatBoost gradient boosting algorithm~\cite{Prokhorenkova2018CatBoost:Features} was used.

\subsubsection*{A.3 Multi-Objective Optimization via EHVI}

Because EHVI is implemented in minimization form, the four desirable-to-maximize objectives (YS, UTS/YS, strain at UTS, and $H_\text{dyn}/H_\text{qs}$) were sign-inverted before surrogate modeling and hypervolume calculation, while DoP remained in its original minimization form.

A multi-objective optimization problem is defined as
\begin{equation}
\label{eq:problem_app}
    \mbox{minimize}\,\,\,  \{f_1(\textbf{x}), ...,f_n(\textbf{x})\}, \textbf{x} \in \mathcal{X}
\end{equation}
where $f_1(\textbf{x}),\ldots, f_n(\textbf{x})$ are the objectives and $\mathcal{X}$ is the feasible design space. Typically, no single solution simultaneously optimizes all objectives. The solution is instead a set of non-dominated designs with trade-offs, forming the Pareto front in the objective space. The optimal solutions $\mathbf{y}$ are denoted as $\mathbf{y}\prec \mathbf{y'}$ and expressed as
\begin{equation}
\begin{aligned}
   \{\mathbf{y}: \mathbf{y} & = (y_1, y_2, \ldots, y_n), \; y_i \leq y_i' \;\; \forall \; i \in \{1, 2, \ldots, n\}, \\
   &  \exists \; j \in \{1, 2, \ldots, n\} : y_j < y'_j\}
\end{aligned}
\end{equation}
where $\mathbf{y}' = (y_1', y_2', \ldots, y_n')$ denotes any possible objective output.

Several approaches exist for approximating the Pareto front, including weighted sum~\cite{marler2010weighted}, adaptive weighted sum~\cite{kim2005adaptive}, normal boundary intersection~\cite{das1998normal}, and hypervolume indicator methods~\cite{beume2009s,bradstreet2010fast,emmerich2011hypervolume,fonseca2006improved,russo2013quick,yang2007novel,zitzler1999multiobjective}. Hypervolume is the space bounded by a fixed reference point and the current Pareto front approximation; maximizing it pushes the approximation toward the true Pareto front. The acquisition function used here is Expected Hypervolume Improvement (EHVI), computed via the recursive decomposition method of Zhao et al.~\cite{zhao2018fast}. Following~\cite{feliot2017bayesian}, EHVI is defined as
\begin{equation}
    \mathbb{E}[\textrm{{HVI}}({\mathbf{y}})] =\int_U \mathbb{P}(\mathbf{y} \prec{\mathbf{y}'})d{{\mathbf{y}'}}
    \label{Expected_app}
\end{equation}
where $\mathbb{P}(\mathbf{y} \prec \mathbf{y}')$ is the probability that $\mathbf{y}'$ dominates $\mathbf{y}$ and $U$ is the dominated hypervolume. Given an independent GP for each objective, the posterior predictive is $y_i \sim \mathcal{N}(\mu_i , \sigma_i^2)$ where $i\leq m$ and $\mu_i,\sigma_i^2$ are the mean and variance of the $i^{th}$ objective. For a new candidate:
\begin{equation}
    \mathbb{P}(\mathbf{y} \prec \mathbf{y}') = \prod_{i=1}^m \Phi \left ( \frac{y_i' - \mu_i}{\sigma_i} \right )
    \label{prob_app}
\end{equation}
where $\Phi$ is the CDF of the standard normal. Details regarding the closed-form expression of Eq.~(\ref{Expected_app}) can be found in Refs.~\cite{zhao2018fast,feliot2017bayesian,while2011fast,jaszkiewicz2018improved,khatamsaz2020efficient}.

\subsubsection*{A.4 Multi-Source Modeling via Reification}

Four objectives (YS, UTS/YS, strain at UTS, $H_\text{dyn}/H_\text{qs}$) were measured experimentally; the fifth (DoP) was a simulation output derived from FEM models calibrated to the tensile and SHPB data acquired in the same iteration. DoP therefore passed through a two-level model chain---experiment $\rightarrow$ constitutive calibration $\rightarrow$ FEM $\rightarrow$ GP---and its uncertainty compounded surrogate approximation error with upstream calibration error. All five objectives entered the EHVI computation as GP posterior distributions regardless of their origin.

To avoid assumptions about the hierarchy of information sources, we estimated inter-source correlations using the reification technique~\cite{allaire2012fusing,thomison2017model}. Following~\cite{Winkler_1981}, the fused mean and variance are:
\begin{equation}
\label{e:WinklerMeanGen_app}
\mathbb{E} [\hat{f}(\mathbf{x})]=\frac{\mathbf{e}^\textrm{T} \tilde{\Sigma}(\mathbf{x})^{-1}  \boldsymbol{\mu}(\mathbf{x})}{\mathbf{e}^\textrm{T} \tilde{\Sigma}(\mathbf{x})^{-1} \mathbf{e}}
\end{equation}
\begin{equation}
\label{e:WinklerVarianceGen_app}
\textrm{Var}\left(\hat{f}(\mathbf{x})\right) = \frac{1}{\mathbf{e}^\textrm{T} \tilde{\Sigma}(\mathbf{x})^{-1} \mathbf{e}}
\end{equation}
where $\mathbf{e} = [1,\ldots,1]^\textrm{T}$, $\boldsymbol{\mu}(\mathbf{x}) = [\mu_1(\mathbf{x}),\ldots,\mu_m(\mathbf{x})]^\textrm{T}$ given $m$ information sources, and $\tilde{\Sigma}(\mathbf{x})$ is the covariance matrix with diagonal elements $\sigma_i^2$ and off-diagonal elements $\bar{\rho}_{ij} \sigma_i \sigma_j$. The correlation coefficient $\bar{\rho}_{ij}$ is calculated via reification:
\begin{equation}
\bar{\rho}_{ij}(\mathbf{x}) = \frac{\sigma_j^2(\mathbf{x})}{\sigma_i^2(\mathbf{x}) + \sigma_j^2(\mathbf{x})} \rho_{ij}(\mathbf{x}) + \frac{\sigma_i^2(\mathbf{x})}{\sigma_i^2(\mathbf{x})+\sigma_j^2(\mathbf{x})} \rho_{ji}(\mathbf{x})
\label{e:VWrho_app}
\end{equation}
where
\begin{equation}
\begin{aligned}
\rho_{ij}(\mathbf{x}) &= \frac{\sigma_i(\mathbf{x})}{\sqrt{\left(\mu_i(\mathbf{x})-\mu_j(\mathbf{x})\right)^2 + \sigma_i^2(\mathbf{x})}}, \\
\rho_{ji}(\mathbf{x}) &= \frac{\sigma_j(\mathbf{x})}{\sqrt{\left(\mu_i(\mathbf{x})-\mu_j(\mathbf{x})\right)^2 + \sigma_j^2(\mathbf{x})}}.
\end{aligned}
\label{e:rho_app}
\end{equation}

\subsubsection*{A.5 Batch Bayesian Optimization}

GP surrogates depend on kernel length-scale hyperparameters that control correlation strength across the input space. Although standard tuning methods (e.g., maximum likelihood~\cite{Rasmussen:2005:GPM:1162254}, cross-validation) are effective with sufficient data, they can become trapped in local optima when training data are sparse~\cite{joy2020batch,couperthwaite2020materials}.

To address this, we constructed a large ensemble of GPs with randomly sampled length-scales, each encoding a different smoothness assumption~\cite{joy2020batch}. Each GP made a different inference about the underlying objective function, yielding different candidate suggestions:
\begin{equation}
    \textbf{x}_{1:n} = \arg\max_{\mathbf{x} \in \chi}  \textrm{EHVI}^{N}(\textbf{x} | \textrm{GP}(\mathbb{D}_N,\theta_{1:n}))
\end{equation}
where $n$ Gaussian processes were built with different hyperparameter sets $\theta$ given data $\mathbb{D}$. From the $n$ candidate suggestions, k-medoids clustering selected a batch of compositions that maximized diversity while preserving aggregate hypervolume gain.

\subsection*{B. Benchmarking Risk-Aware Multi-Objective Bayesian Optimization}
\label{sec:appendix_benchmark}

To quantitatively assess the effectiveness of the proposed risk-aware multi-objective Bayesian optimization (MOBBO) strategy, we constructed a controlled computational benchmark based on the six-component Al--V--Cr--Fe--Co--Ni HEA space. The full design space was discretized at 4~at.\% resolution, generating a pool of 23{,}694 distinct compositions with precomputed objective properties and phase stability metrics.

Two bi-objective optimization problems were defined:
(i) maximize yield strength proxy (CV\_YS) and Pugh ratio ($B/G$), and
(ii) maximize CV\_YS and Cauchy pressure ($C_{12} - C_{44}$).
CV\_YS was computed using the Curtin--Varvenne solid solution strengthening model~\cite{varvenne2016theory}; Pugh ratio and Cauchy pressure were approximated via rule-of-mixtures aggregation of elemental elastic constants.

Feasibility was introduced through a dual-database configuration that mimics realistic model discrepancy: screening used TCHEA6 (FCC $\geq 0.99$ at 700$^\circ$C), while ground-truth feasibility labels were assigned from TCHEA7. Any composition with TCHEA7 FCC fraction $<0.99$ was treated as infeasible.

Four acquisition strategies were compared:

\begin{enumerate}
    \item \textbf{Baseline MOBBO}: EHVI applied across the full design space without feasibility awareness.
    \item \textbf{Hard-Filter MOBBO}: EHVI restricted to TCHEA6-predicted FCC-stable compositions.
    \item \textbf{Risk-Weighted MOBBO}: EHVI multiplied by $p_{\text{feas}} = (\text{FCC}_{\text{TCHEA7}})^{\alpha}$ with $\alpha = 2$.
    \item \textbf{Oracle MOBBO}: EHVI weighted by the true binary feasibility label from TCHEA7 (upper bound).
\end{enumerate}

Each campaign selected 10 initial compositions via $k$-medoids clustering, then ran 20 sequential iterations of batch size 10, with GP regressors for objectives and a Random Forest classifier for feasibility. Five random seeds yielded 100 total campaigns per objective pair. Performance was evaluated using feasible hypervolume (HV), Pareto front size, and feasibility violation rate (FVR).

\autoref{fig:benchmark_results} summarizes the results. The differences between strategies are consistent across both objective pairs:

\begin{itemize}

\item \textbf{Baseline MOBBO} fails completely: near-zero feasible HV, trivial Pareto sets, FVR $\approx 1.0$, with $>$90\% of evaluations wasted on infeasible candidates.

\item \textbf{Hard-Filter MOBBO} improves feasibility (FVR $\approx 0.15$) but plateaus early when TCHEA6 mispredictions over-restrict exploration. It achieves HV = 7.25 (Pugh) and 4.66 (Cauchy) but lacks robustness near misclassified boundaries.

\item \textbf{Risk-Weighted MOBBO} matches the oracle in final HV (7.25 and 4.66), maintains low FVR ($\approx 0.15$--0.20), and converges within 3--5 iterations. By penalizing rather than excluding uncertain candidates, it preserves exploration flexibility near critical boundaries.

\item \textbf{Oracle MOBBO} achieves identical final HV but wastes $\approx$170--190 evaluations on infeasible candidates early on, demonstrating that feasibility knowledge alone is insufficient---it must be incorporated into the acquisition function.

\end{itemize}

The oracle result is instructive: perfect feasibility knowledge does not prevent waste. The oracle still aggressively queries high-objective infeasible regions early on ($\approx$170--190 wasted evaluations), because feasibility labels alone do not suppress the acquisition function's attraction to unexplored regions with high predicted performance. Risk-weighted MOBBO avoids this by embedding feasibility directly into the acquisition score, achieving identical final HV with ${\sim}5\times$ fewer wasted evaluations. The lesson parallels the experimental campaign: identifying the feasibility boundary is necessary but not sufficient---the acquisition function must be designed to navigate it.



\section*{Supplementary material}

## 1. Summary of Prior Work

In our previous work, the framework was demonstrated on the Al–V–Cr–Fe–Co–Ni FCC high entropy alloy (HEA) system, targeting three key performance objectives: ultimate tensile strength/yield strength ratio, hardness, and strain rate sensitivity. The study combines several key innovations: (i) initial design space filtering using CALPHAD-based phase stability predictions and property constraints to reduce ∼53,000 candidate compositions to 2,437 feasible alloys; (ii) division of the design space based on predicted stacking fault energy (SFE) into low ($<$50 mJ/m$^2$) and high ($>$50 mJ/m$^2$) SFE regimes to explore different deformation mechanisms; (iii) batch Bayesian optimization using an ensemble of Gaussian processes with varying length scales to enable parallel experimentation while avoiding premature convergence; and (iv) Expected Hypervolume Improvement (EHVI) as the acquisition function to guide multi-objective optimization.

Through five iterative design-make-test-learn loops, each evaluating 8 alloys in parallel, we successfully identified three-objective Pareto sets by exploring only 0.15% of the feasible design space. The experimental campaign used vacuum arc melting for synthesis, comprehensive characterization including SEM and XRD for phase verification, quasi-static tensile testing for mechanical properties, and nanoindentation for hardness and strain rate sensitivity measurements. The framework demonstrated significant acceleration over traditional methods, with statistical analysis showing that the observed Pareto set sizes (14 and 15 points for low and high SFE regimes, respectively) had less than 0.5% probability of occurring through random sampling, validating the effectiveness of the Bayesian approach.

This prior work established the foundational principles of the BIRDSHOT framework within the Al–V–Cr–Fe–Co–Ni system, integrating physics-informed modeling, high-throughput experimentation, and CALPHAD-based phase prediction. It provided insights into the trade-offs between composition, mechanical properties, and phase stability that informed the design of the present campaign.

## 2. Integrated Workflow

The BIRDSHOT workflow is a tightly coupled, multi-stage closed loop that candidate suggestion, synthesis, sample preparation, multi-modal characterization, physics-based simulation, and Bayesian optimization within a single closed-loop architecture (Figure 1).

The mathematical details of the BRAVE computational framework (GP surrogates, informative priors, EHVI acquisition, multi-source fusion, and batch selection) are provided in Appendix A of the main text.

## 3. Detailed Sampling and Candidate Selection Strategy for Iteration 0

### *3.1. Overview*

Iteration 0 required selecting a small, manufacturable set of alloys that (i) lie within a thermodynamically feasible single-phase FCC region, (ii) span the chemically diverse subspaces

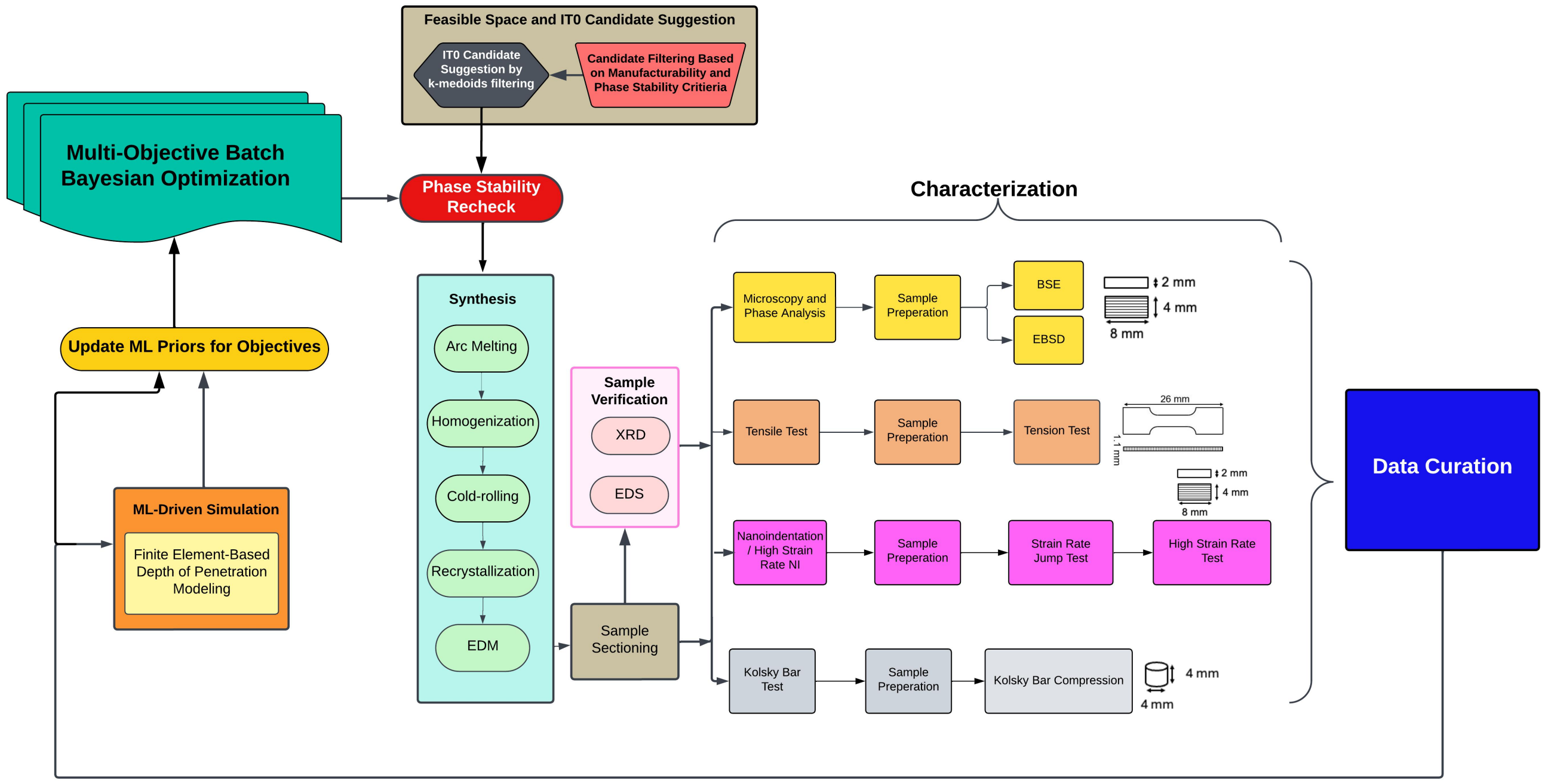


Figure 1: Integrated BIRDSHOT workflow from candidate suggestion through synthesis, sample preparation, characterization, simulation, and Bayesian optimization for selection of the next batch of candidates.

of the multi-component system, and (iii) are compatible with synthesis and characterization constraints. Because the full combinatorial space is extremely large, the selection strategy combined CALPHAD-based thermodynamic screening, manifold learning for exploratory analysis, diversity-aware clustering, and subsystem-balanced down-selection. Figure 2 provides a schematic overview of the complete workflow, with each stage annotated by the number of surviving candidates.

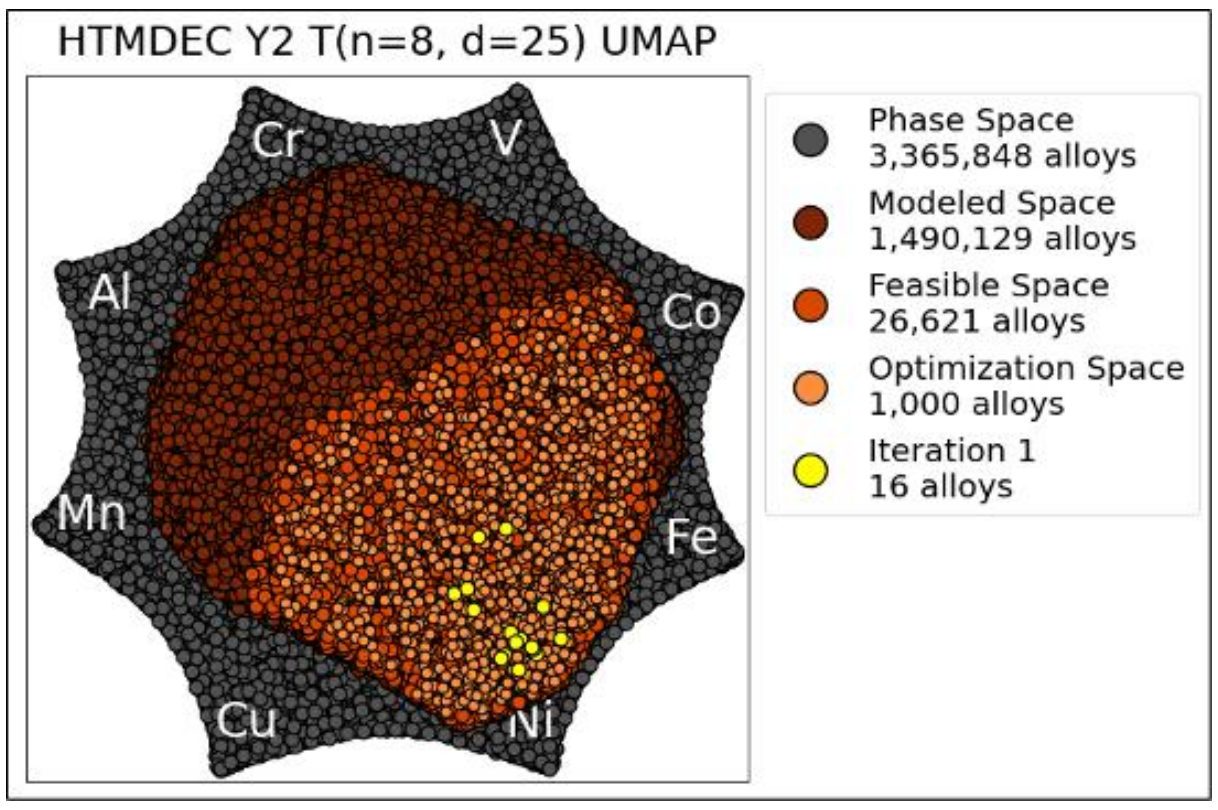


Figure 2: Schematic overview of the Iteration 0 candidate selection pipeline, showing the progressive reduction from 3.37 million unconstrained compositions to the final 16-alloy synthesis batch. Each stage is annotated with the number of surviving candidates.

### *3.2. Compositional Space Enumeration and Subsystem Partitioning*

The design space which is the eight element system Al–V–Cr–Mn–Fe–Co–Ni–Cu, is discretized on a 4 at.% compositional grid and subject to the simplex constraint $\sum_i x_i = 1$, $x_i \in \{0, 0.04, 0.08, \dots\}$. The number of unique compositions is given by the stars-and-bars formula:

$$T(n, d) = \binom{n + d - 1}{d}, \qquad d = \frac{100}{\Delta}, \tag{1}$$

where $n = 8$ and $\Delta = 4$ at.% yields $d = 25$, giving $T(8, 25) =$ 3,365,848 unique compositions in the unconstrained simplex. For reference, 2 at.% and 1 at.% resolutions yield 264,385,828 and 26,075,972,538 compositions, respectively. Figure 3 illustrates the exponential growth of the combinatorial space with increasing elements and finer discretization.

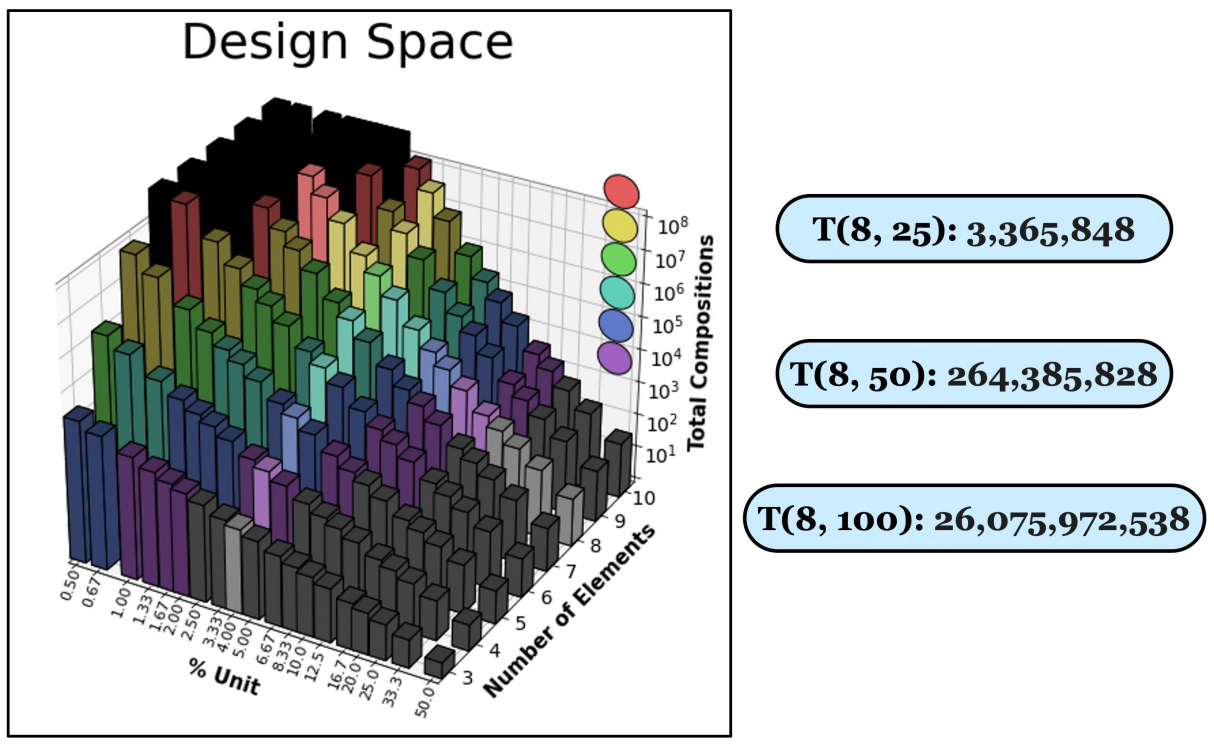


Figure 3: Number of unique compositions as a function of system size ($n$ elements) and discretization resolution. The eight-element, 4 at.% system investigated in this work (red star) contains 3.37 million compositions before applying compositional bounds or phase-stability constraints.

Element specific upper bounds were imposed to ensure manufacturability and avoid extreme chemistries: Al $\leq 0.24$, V $\leq 0.24$,

$Cr \leq 0.50$, $Mn \leq 0.40$, $Fe \leq 0.50$, $Co \leq 0.50$, $Ni \leq 0.60$, and $Cu \leq 0.24$. Applying these constraints reduced the space to $1{,}504{,}938$ candidate compositions.

Each bounded composition was assigned to a chemical subsystem defined by the set of elements present at nonzero atomic fraction on the discretized grid. This partitioning yielded 37 distinct subsystems spanning six- to eight-component alloys (Table 1). The full eight element subsystem (type 1) contains $240{,}000$ compositions after bounding, while seven element subsystems range from $75{,}993$ to $92{,}595$ depending on the excluded element. Six element subsystems show even greater variation, with populations between $15{,}788$ and $29{,}506$. Subsystems of five or fewer elements account for an additional $252{,}629$ compositions. This partitioning was used throughout the pipeline to track chemical diversity and prevent over-sampling of dominant subsystems during candidate selection.

Table 1: Chemical subsystem partitioning of the bounded design space. Elements marked with ** are absent from the subsystem. "Compositions" gives the number of grid points in each subsystem after applying element specific upper bounds. Subsystems of size 5 and below are reported as aggregated totals.

| Size | Type | Al | V | Cr | Mn | Fe | Co | Ni | Cu | Compositions |
|---|---|---|---|---|---|---|---|---|---|---|
| 8 | 1 | Al | V | Cr | Mn | Fe | Co | Ni | Cu | 240,000 |
| 7 | 2 | Al | V | Cr | Mn** | Fe | Co | Ni | Cu | 78,828 |
| 7 | 3 | Al | V | Cr | Mn | Fe** | Co | Ni | Cu | 76,830 |
| 7 | 4 | Al | V | Cr | Mn | Fe | Co | Ni** | Cu | 75,993 |
| 7 | 5 | Al | V | Cr** | Mn | Fe | Co | Ni | Cu | 76,830 |
| 7 | 6 | Al | V** | Cr | Mn | Fe | Co | Ni | Cu | 92,595 |
| 7 | 7 | Al** | V | Cr | Mn | Fe | Co | Ni | Cu | 92,595 |
| 7 | 8 | Al | V | Cr | Mn | Fe | Co | Ni | Cu** | 92,595 |
| 7 | 9 | Al | V | Cr | Mn | Fe | Co** | Ni | Cu | 76,830 |
| 6 | 10 | Al | V** | Cr | Mn | Fe | Co | Ni | Cu** | 29,506 |
| 6 | 11 | Al | V | Cr** | Mn | Fe | Co | Ni | Cu** | 22,584 |
| 6 | 12 | Al** | V | Cr | Mn | Fe | Co | Ni | Cu** | 29,506 |
| 6 | 13 | Al | V | Cr | Mn** | Fe | Co | Ni | Cu** | 23,694 |
| 6 | 14 | Al | V | Cr | Mn | Fe** | Co | Ni | Cu** | 22,584 |
| 6 | 15 | Al | V | Cr | Mn | Fe | Co** | Ni | Cu** | 22,584 |
| 6 | 16 | Al** | V | Cr** | Mn | Fe | Co | Ni | Cu | 22,584 |
| 6 | 17 | Al | V** | Cr | Mn | Fe | Co** | Ni | Cu | 22,584 |
| 6 | 18 | Al | V | Cr** | Mn | Fe | Co** | Ni | Cu | 16,436 |
| 6 | 19 | Al** | V | Cr | Mn | Fe | Co** | Ni | Cu | 22,584 |
| 6 | 20 | Al | V | Cr | Mn | Fe** | Co** | Ni | Cu | 16,436 |
| 6 | 21 | Al | V | Cr | Mn** | Fe | Co** | Ni | Cu | 17,496 |
| 6 | 22 | Al** | V | Cr | Mn** | Fe | Co | Ni | Cu | 23,694 |
| 6 | 23 | Al | V** | Cr | Mn** | Fe | Co | Ni | Cu | 23,694 |
| 6 | 24 | Al** | V** | Cr | Mn | Fe | Co | Ni | Cu | 29,506 |
| 6 | 25 | Al | V | Cr | Mn** | Fe** | Co | Ni | Cu | 17,496 |
| 6 | 26 | Al | V | Cr** | Mn** | Fe | Co | Ni | Cu | 17,496 |
| 6 | 27 | Al | V** | Cr | Mn | Fe** | Co | Ni | Cu | 22,584 |
| 6 | 28 | Al** | V | Cr | Mn | Fe** | Co | Ni | Cu | 22,584 |
| 6 | 29 | Al | V | Cr** | Mn | Fe** | Co | Ni | Cu | 16,436 |
| 6 | 30 | Al | V** | Cr** | Mn | Fe | Co | Ni | Cu | 22,584 |
| 6 | 31 | Al | V** | Cr | Mn | Fe | Co | Ni** | Cu | 21,930 |
| 6 | 32 | Al | V | Cr | Mn** | Fe | Co | Ni** | Cu | 16,848 |
| 6 | 33 | Al | V | Cr** | Mn | Fe | Co | Ni** | Cu | 15,788 |
| 6 | 34 | Al** | V | Cr | Mn | Fe | Co | Ni** | Cu | 21,930 |
| 6 | 35 | Al | V | Cr | Mn | Fe** | Co | Ni** | Cu | 15,788 |
| 6 | 36 | Al | V | Cr | Mn | Fe | Co** | Ni** | Cu | 15,788 |
| 6 | 37 | Al | V | Cr | Mn | Fe | Co | Ni** | Cu** | 21,930 |
| 5 | all | | | | | | | | | 218,786 |
| 4 | all | | | | | | | | | 32,285 |
| 3 | all | | | | | | | | | 1,548 |
| 2 | all | | | | | | | | | 10 |

### 3.3. CALPHAD Equilibrium Screening

Equilibrium phase fractions were computed at 700 °C for all 1,504,938 bounded compositions using Thermo-Calc 2022.2 with the TCHEA6 database, interfaced through TC-Python. Calculations were parallelized across a high-performance computing cluster and partitioned by chemical subsystem to improve throughput and reduce licensing conflicts. Of the attempted compositions, 1,490,129 (99.02%) converged successfully, with peak throughput reaching ∼72,000 equilibrium calculations per core per day and an average of ∼44,000 per core per day. Compositions with predicted FCC phase fraction $\geq 0.99$ were retained as feasible, yielding 27,074 candidates – 1.54% of the evaluated space (Figure 4).

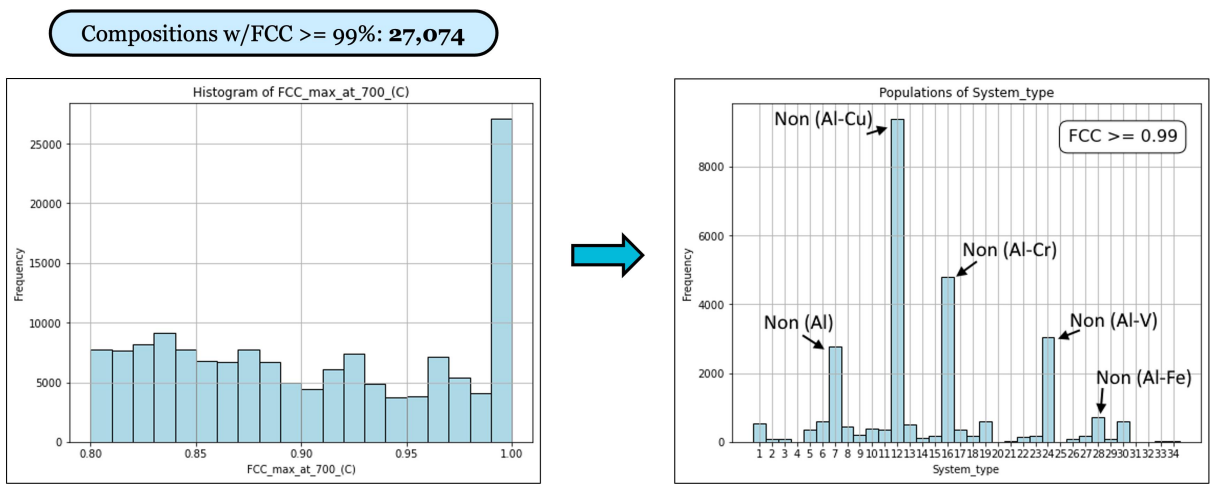


Figure 4: Progressive filtering of the design space. Application of the FCC $\geq 0.99$ phase-fraction threshold reduces the modeled space from 1,490,129 to 27,074 feasible compositions.

### 3.4. Local FCC Stability Perturbation Analysis

To assess how robust CALPHAD-predicted FCC stability is to small local composition changes, we performed a perturbation analysis on candidate alloys that satisfied FCC $\geq 0.99$ (Figure 5. For each candidate composition on the 4 at.% design grid, we generated nearby compositions by varying each elemental fraction by $\pm 2$ at.% and recalculated equilibrium phase stability using Thermo-Calc, subject to the composition simplex constraint. This probes whether a composition predicted to be FCC remains FCC under small local composition shifts.

Across 1,154,207 perturbed compositions evaluated, 903,326 retained FCC $\geq$ 0.99, corresponding to 78.3% local phase-stability retention. We therefore interpret a feasibility probability near 0.8 as corresponding to a composition that is locally robust to small perturbations. This provides empirical support for the acquisition threshold $P_{\text{feas}}(\mathbf{x}) \geq 0.8$ used in the main text.

### 3.5. Observations from FCC Feasibility Screening

The FCC screening results show several structural features of the feasible design space. Nickel is essential for FCC stability in this system: all subsystems lacking nickel, including types 4, 31–37, and all five-element and smaller subsystems without nickel returned zero viable candidates at the $\geq 0.99$ FCC threshold. The maximum FCC phase fraction achieved by any nickel-free composition remained below 0.90 (type 4), with some subsystems not exceeding 0.87–0.88 (types 31, 32). In contrast, cobalt-free subsystems (types 9, 15, 17–21) retained thousands of feasible compositions.

Effective solubility limits imposed by the FCC stability constraint differ markedly from the imposed compositional bounds.

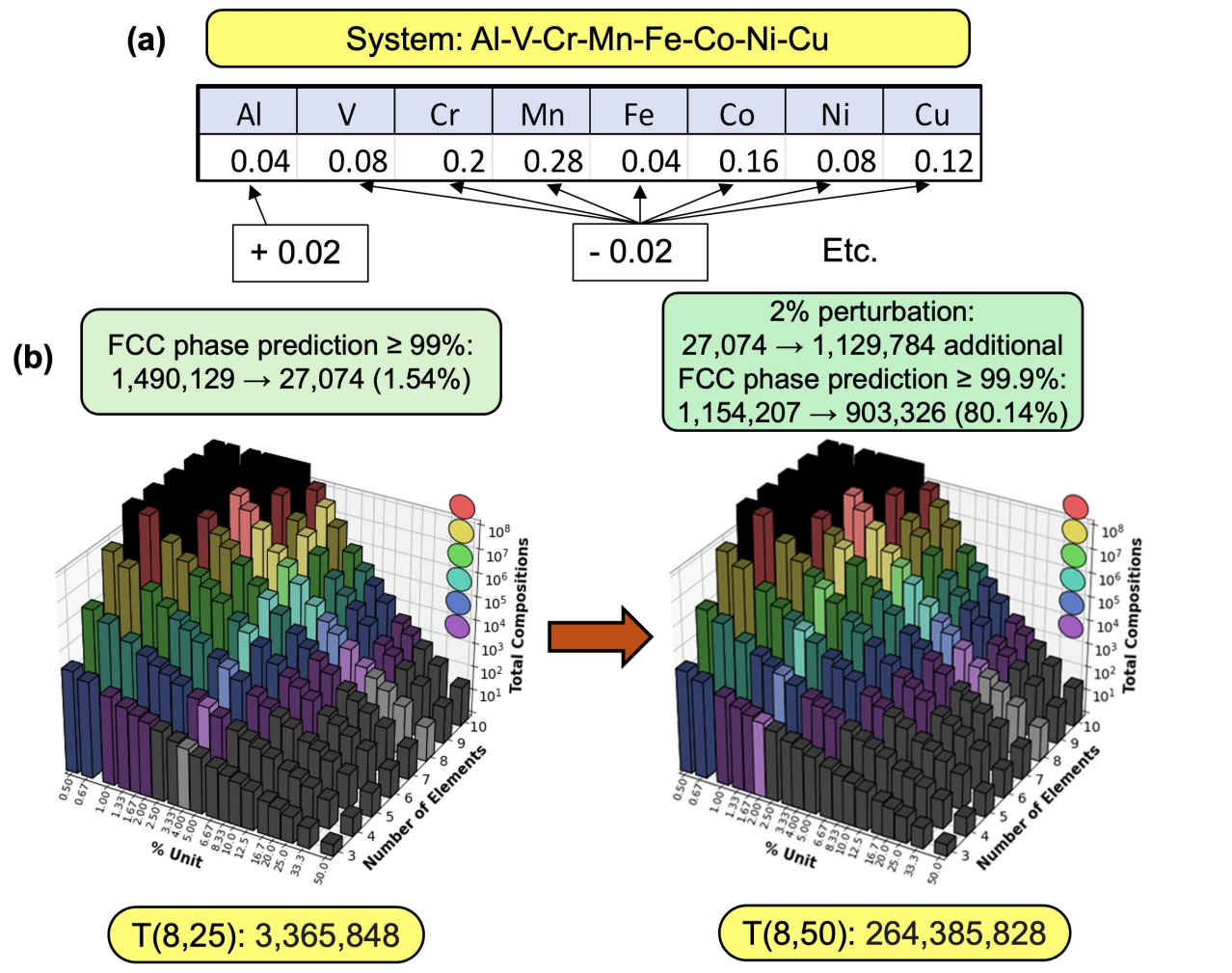


Figure 5: Local FCC stability perturbation analysis. (a) Illustration of $\pm 2$ at.% perturbations in 8D composition space. (b) Distribution of phase-stability outcomes: 78.3% of perturbed alloys remain fully FCC (FCC $\geq 0.99$), while the remainder show secondary phases such as BCC or SIGMA.

Copper, although bounded at 0.24, is effectively confined to much lower concentrations in FCC feasible compositions, reflecting limited solid solubility in the FCC matrix. Chromium is similarly constrained, with feasible compositions not exceeding ∼28 at.% despite a bound of 0.50. Nickel, by contrast, occupies the highest atomic fractions among the feasible candidates, consistent with its role as the primary FCC stabilizer.

The strong population imbalance across subsystems (Figure 6) has direct implications for sampling: naive random selection from the 27,074 feasible compositions would over-represent high-population subsystems and miss chemically distinct but sparsely populated regions. Subsystems with fewer than 100 FCC-feasible candidates were excluded, retaining 26,621 compositions across the feasible subsystems for subsequent clustering.

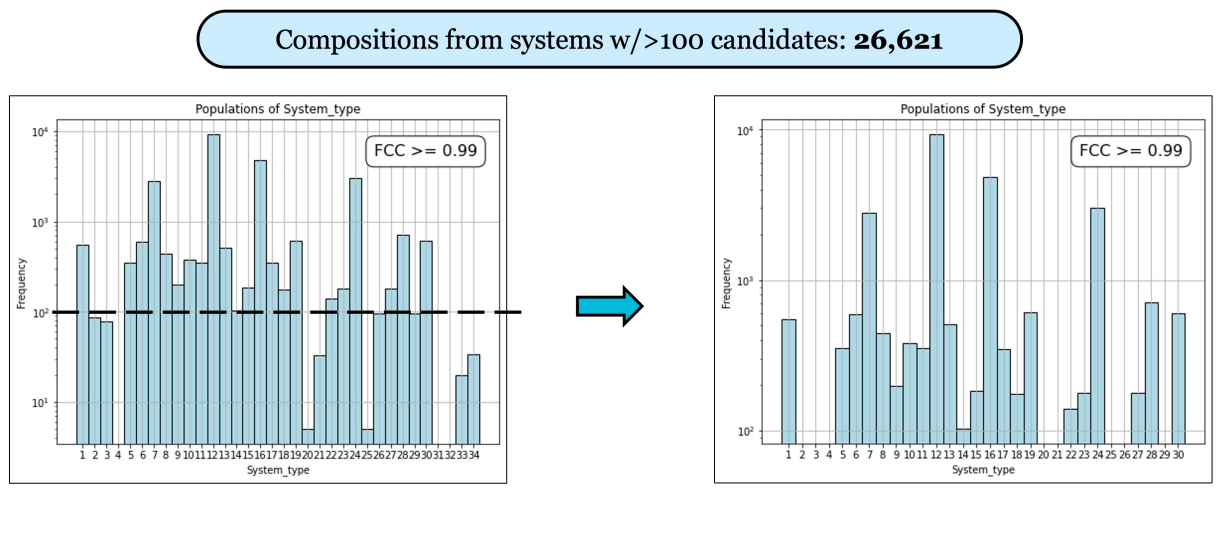


Figure 6: Number of FCC-feasible compositions per chemical subsystem, showing strong population imbalance. Subsystems with fewer than 100 candidates were excluded from subsequent sampling, retaining 26,621 compositions.

### 3.6. Diversity Sampling via $k$-Medoids Clustering

To construct a representative and computationally manageable subset, $k$-medoids clustering was applied to the atomic-fraction vectors of the 26,621 viable compositions using Euclidean distance, with $k = 1{,}000$. Because medoids are actual data points, all selected representatives are physically realizable compositions. The 1,000 medoids span both high density core regions and peripheral areas of the feasible space. Standard initialization

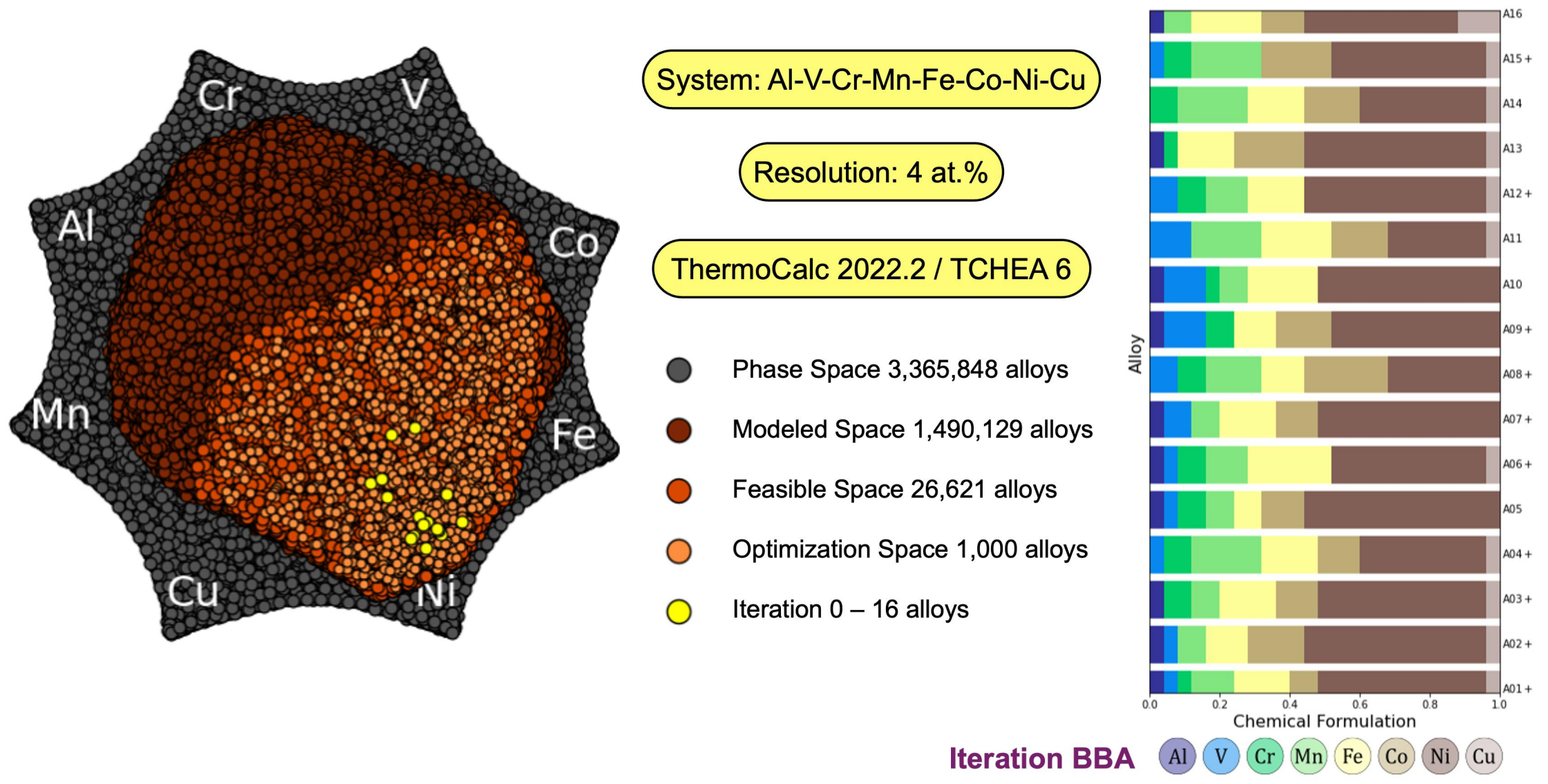


Figure 7: Candidate alloy suggestions for Iteration 0 (BBA), selected by diversity-aware $k$-medoids sampling across 37 chemical subsystems.

using only atomic fractions as dimensions tends to favor high population subsystems; explicit subsystem aware down selection (described below) was therefore required to ensure chemical diversity in the final batch.

### *3.7. Down Selection to the Iteration 0 Batch*

The final Iteration 0 (BBA) batch consisted of 16 alloys (Figure 7, constrained by experimental throughput. Down selection from the 1,000 medoids used a quota-based diversity algorithm to ensure coverage across chemical subsystems. Each subsystem $i$ was assigned a quota

$$q_i = \max\left(1,\ \mathrm{round}\left(16 \cdot \frac{n_i}{\sum_j n_j}\right)\right),$$

where $n_i$ is the number of feasible candidates in subsystem $i$, with greedy correction to enforce $\sum_i q_i = 16$. Within each subsystem, medoids were ranked by a combined criterion balancing proximity to the subsystem centroid (representativeness) and distance from the global centroid (chemical diversity). A final greedy refinement step maximized the minimum pairwise distance among the 16 selected candidates.

## 4. High-Throughput Sample Preparation, Characterization Automation, and Data Management

### *4.1. Sample Preparation Workflow Evolution*

Microstructural characterization and nanoindentation samples were initially prepared by mechanical polishing using a Buehler AutoMet 250 autopolisher with diamond suspensions down to 1 $\mu$m. For electron backscatter diffraction (EBSD) analysis, the surfaces were subsequently polished to a mirror finish using a Buehler VibroMet 2 vibratory polisher for 12 h in a 0.04 $\mu$m colloidal silica suspension.

In iterations BBA and BBB, samples were individually mounted using graphite mounting powder. Although effective, this method was labor-intensive and caused misalignment during indentation testing in a few cases. In addition, it limited batch sizes to eight samples, thereby reducing throughput. To improve efficiency and sample quality, a new preparation technique was developed in iteration BBC. This revised protocol used large aluminum polishing platens that allowed simultaneous processing of 16 or more samples. Samples were fixed to these platens using crystalbond adhesive, ensuring uniform mounting height and eliminating misalignment issues that had previously impacted mechanical characterization.

To prevent sample mix-ups during high-throughput processing, a laser marking system was introduced. Each sample is lightly marked on its reverse side with a unique identifier indicating the iteration and sample number. Although this step adds to the workflow, it has been optimized for rapid batch processing and ensures reliable traceability throughout testing.

Future plans aim to further increase efficiency by adapting techniques that enable mounting of eight samples per aluminum puck. This adjustment will allow up to 48 samples to be ground and polished concurrently, while maintaining the precise alignment required for subsequent mechanical testing. Overall, these improvements are projected to result in a sixfold increase in sample preparation throughput compared to the initial methods.

### 4.2. *Nanoindentation and High Strain Rate Nanoindentation Automation*

Custom Python and ImageJ scripts were implemented to automate image acquisition and projected contact area analysis for nanoindentation, addressing one of the primary challenges in establishing an automated system for measuring contact area. Ensuring consistent reliability across different materials and strain rates was a critical hurdle that was overcome through iterative refinements. These improvements enabled direct comparisons between nanoindentation data and results from high strain rate methods, such as the split Hopkinson pressure bar. The strain rate sensitivities obtained from these different techniques were found to be in close agreement, typically within one standard deviation.

Integrating the high strain rate nanoindentation system into the workflow presented additional technical challenges. The instrument outputs data in encrypted binary files, which required development of Python scripts for reliable data extraction. Further automation scripts were developed to handle complex data analysis tasks, reducing processing time. Nanoindentation data was cross-validated with tensile and Hopkinson bar test results to identify correlations in rate sensitivity across the three measurement techniques. The rate sensitivities from all methods were found to be within one standard deviation of each other.

### 4.3. *Data Management Architecture*

Data management in the prior BIRDSHOT study was organized around a **sample-centric framework**, designed to meet the distinct requirements of the VAM and DED synthesis workflows [1]. This approach enabled end-to-end traceability from synthesis through characterization and analysis, supporting distributed collaboration and ensuring data integrity via natural version control. However, the path-dependent folder navigation imposed significant challenges for new users and complicated retrieval of specific results even for existing users. This experience motivated the adoption of a more discoverable, database-driven framework in subsequent studies.

In the current BIRDSHOT study, we implemented a **method-centric data architecture**, as shown in Figure 8, designed to maximize traceability, scalability, and efficiency while reducing redundancy. In this system, sample identifiers remain linked to synthesis origin, but folder organization treats design, processing and characterization methods as first-class entities beneath synthesis-level groupings. Sample folders use short, descriptive labels (e.g., SEM, Tensile, NI) to enable rapid identification, and measurement identifiers follow a hierarchical and extensible convention <SampleID>_<Synthesis>_<Technique>[_<Subsample>][_...]. Metadata associated with each method, such as sem-details.json or tensile-details.json, is stored in structured JSON files within the corresponding method folders, allowing automated parsing, integration with downstream analysis pipelines, and programmatic traversal. For high-throughput synthesis, folders labeled by production batch capture shared processing metadata to minimize duplication while maintaining logical clarity and supporting future extensibility for new characterization techniques or processing steps.

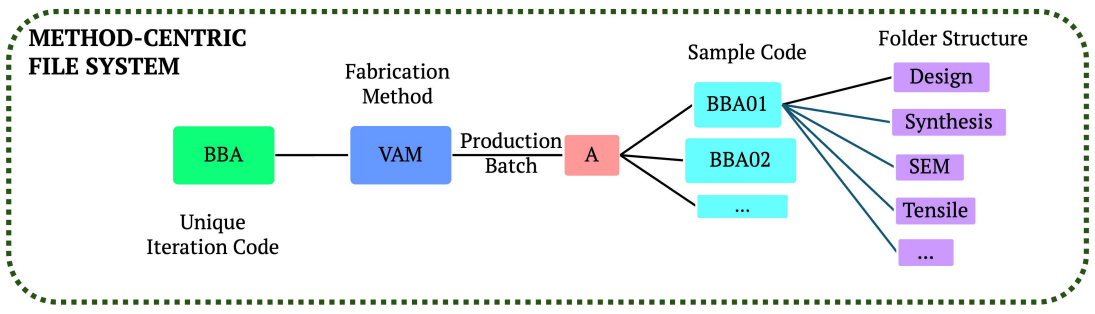


Figure 8: Folder structure illustrating synthesis-level grouping and method-focused organization.

### 4.4. *IGSN Sample Registration*

Each candidate alloy in this study is assigned an International Generic Sample Number (IGSN), a globally unique, resolvable identifier for physical samples, thereby linking the physical specimens, their metadata, and downstream analytical results in a persistent, cross-platform framework. IGSNs are issued and managed under the IGSN–DataCite partnership, ensuring sustainability and interoperability across research data systems [2].

By design, IGSN identifiers behave like DOIs: they are registered in a dedicated IGSN catalog within the DataCite system and resolved using standard DOI infrastructure, but are distinguished semantically as "physical object" identifiers. The metadata for each IGSN follows the DataCite Metadata Schema, including mandatory fields such as identifier, creator, title, publisher, publication year, and resource type, along with recommended fields like related identifier, contributors, dates, descriptions, and geolocation when applicable [3]. In particular, our sample hierarchy is reflected in the related Identifier fields so that parent–child relationships (like subsamples) are explicitly recorded in the IGSN metadata. Because IGSNs are globally registered and resolvable, they support FAIR principles—especially findability and interoperability—by enabling external users (or data infrastructures) to unambiguously reference and access sample information, even beyond our internal data ecosystem.

## 5. IGSN METADATA AND COMPOSITIONS FOR ALL ALLOYS

All alloys synthesized in this study are registered with International Generic Sample Numbers (IGSNs) through the DataCite system. Each IGSN is resolvable via the IGSN ID service (https://support.datacite.org/docs/igsn-ids). Table 2, Table 3, and Table 4 list the alloy identifier, target composition (at.%), and IGSN for every alloy across the three design iterations.

### 5.1. ***Supplementary Table 2*** | *Iteration 0 (BBA) alloys*

Table 2

| Alloy Name | Target Atomic % | | | | | | | | IGSN |
|---|---|---|---|---|---|---|---|---|---|
| | Al | Co | Cr | Cu | Fe | Mn | Ni | V | |
| BBA01 | 4 | 8 | 4 | 4 | 16 | 12 | 48 | 4 | TMAMAK00097 |
| BBA02 | 4 | 16 | 0 | 4 | 12 | 8 | 52 | 4 | TMAMAK00098 |
| BBA03 | 4 | 12 | 8 | 4 | 16 | 8 | 48 | 0 | TMAMAK00099 |
| BBA04 | 0 | 12 | 8 | 4 | 16 | 20 | 36 | 4 | TMAMAK00100 |
| BBA05 | 4 | 12 | 8 | 0 | 8 | 8 | 56 | 4 | TMAMAK00101 |
| BBA06 | 4 | 0 | 8 | 4 | 24 | 12 | 44 | 4 | TMAMAK00102 |
| BBA07 | 4 | 12 | 0 | 0 | 16 | 8 | 52 | 8 | TMAMAK00103 |
| BBA08 | 0 | 24 | 8 | 0 | 12 | 16 | 32 | 8 | TMAMAK00104 |
| BBA09 | 4 | 16 | 8 | 0 | 12 | 0 | 48 | 12 | TMAMAK00105 |
| BBA10 | 4 | 0 | 4 | 0 | 20 | 8 | 52 | 12 | TMAMAK00106 |
| BBA11 | 0 | 16 | 0 | 4 | 20 | 20 | 28 | 12 | TMAMAK00107 |
| BBA12 | 0 | 0 | 8 | 4 | 16 | 12 | 52 | 8 | TMAMAK00108 |
| BBA13 | 4 | 20 | 4 | 4 | 16 | 0 | 52 | 0 | TMAMAK00109 |
| BBA14 | 0 | 16 | 8 | 4 | 16 | 20 | 36 | 0 | TMAMAK00110 |
| BBA15 | 0 | 20 | 8 | 4 | 0 | 20 | 44 | 4 | TMAMAK00111 |
| BBA16 | 4 | 12 | 0 | 12 | 20 | 8 | 44 | 0 | TMAMAK00112 |

### 5.2. ***Supplementary Table 3*** | *Iteration 1 (BBB) alloys*

Table 3

| Alloy Name | Target Atomic % | | | | | | | | IGSN |
|---|---|---|---|---|---|---|---|---|---|
| | Al | Co | Cr | Cu | Fe | Mn | Ni | V | |
| BBB01 | 0 | 24 | 4 | 0 | 8 | 4 | 36 | 24 | TMAMAK00113 |
| BBB02 | 4 | 32 | 8 | 4 | 4 | 20 | 24 | 4 | TMAMAK00114 |
| BBB03 | 4 | 0 | 12 | 4 | 4 | 16 | 52 | 8 | TMAMAK00115 |
| BBB04 | 4 | 12 | 4 | 0 | 28 | 0 | 40 | 12 | TMAMAK00116 |
| BBB05 | 0 | 44 | 12 | 0 | 4 | 4 | 12 | 24 | TMAMAK00117 |
| BBB06 | 0 | 16 | 20 | 0 | 4 | 4 | 36 | 20 | TMAMAK00118 |
| BBB07 | 4 | 40 | 0 | 0 | 12 | 4 | 36 | 4 | TMAMAK00119 |
| BBB08 | 4 | 8 | 12 | 4 | 4 | 24 | 40 | 4 | TMAMAK00120 |
| BBB09 | 0 | 4 | 8 | 0 | 4 | 28 | 40 | 16 | TMAMAK00121 |
| BBB10 | 4 | 0 | 4 | 4 | 8 | 16 | 52 | 12 | TMAMAK00122 |
| BBB11 | 0 | 24 | 4 | 0 | 4 | 32 | 20 | 16 | TMAMAK00123 |
| BBB12 | 0 | 20 | 4 | 0 | 4 | 4 | 48 | 20 | TMAMAK00124 |
| BBB13 | 0 | 40 | 4 | 0 | 28 | 4 | 4 | 20 | TMAMAK00125 |
| BBB14 | 4 | 0 | 12 | 16 | 4 | 12 | 52 | 0 | TMAMAK00126 |
| BBB15 | 4 | 0 | 8 | 24 | 4 | 16 | 44 | 0 | TMAMAK00127 |
| BBB16 | 4 | 36 | 0 | 0 | 4 | 4 | 40 | 12 | TMAMAK00128 |

5.3. ***Supplementary Table 4*** | *Iteration 2 (BBC) alloys*

Table 4

| Alloy Name | Target Atomic % | | | | | | | | IGSN |
|---|---|---|---|---|---|---|---|---|---|
| | Al | Co | Cr | Cu | Fe | Mn | Ni | V | |
| BBC01 | 4 | 12 | 4 | 4 | 20 | 16 | 32 | 8 | TMAMAK00129 |
| BBC02 | 4 | 8 | 0 | 0 | 4 | 8 | 52 | 24 | TMAMAK00130 |
| BBC03 | 4 | 4 | 12 | 0 | 0 | 4 | 52 | 24 | TMAMAK00131 |
| BBC04 | 4 | 0 | 4 | 0 | 4 | 4 | 60 | 24 | TMAMAK00132 |
| BBC05 | 0 | 24 | 8 | 0 | 16 | 40 | 4 | 8 | TMAMAK00133 |
| BBC06 | 4 | 28 | 4 | 0 | 4 | 4 | 44 | 12 | TMAMAK00134 |
| BBC07 | 4 | 32 | 4 | 0 | 0 | 4 | 44 | 12 | TMAMAK00135 |
| BBC08 | 0 | 8 | 0 | 4 | 4 | 28 | 36 | 20 | TMAMAK00136 |
| BBC09 | 0 | 36 | 4 | 0 | 16 | 16 | 8 | 20 | TMAMAK00137 |
| BBC10 | 4 | 4 | 16 | 0 | 8 | 0 | 52 | 16 | TMAMAK00138 |
| BBC11 | 4 | 12 | 4 | 4 | 12 | 16 | 36 | 12 | TMAMAK00139 |
| BBC12 | 4 | 4 | 4 | 8 | 28 | 16 | 32 | 4 | TMAMAK00140 |
| BBC13 | 4 | 28 | 0 | 20 | 4 | 8 | 36 | 0 | TMAMAK00141 |
| BBC14 | 0 | 12 | 4 | 0 | 4 | 16 | 40 | 24 | TMAMAK00142 |
| BBC15 | 4 | 16 | 0 | 20 | 4 | 12 | 44 | 0 | TMAMAK00143 |
| BBC16 | 4 | 28 | 4 | 0 | 8 | 0 | 40 | 16 | TMAMAK00144 |

## 6. TCHEA6 TO TCHEA7 DATABASE UPDATE: TERNARY SUBSYSTEM COVERAGE

During this study, Thermo-Calc released the updated TCHEA7 database with refined phase stability predictions across more than 100 ternary and higher order systems. To assess the impact on the Al–V–Cr–Mn–Fe–Co–Ni–Cu design space, all 56 ternary subsystems were evaluated for assessment availability in both TCHEA6 and TCHEA7, along with any newly predicted secondary phases (sigma, BCC, FCC, HCP). Table 5 summarizes the results. A "Yes" in only the TCHEA7 column indicates a newly added assessment; "Yes" in both columns indicates an assessment that was already available in TCHEA6 and carried over; blank or "No" entries indicate no assessment is available.

Key changes from TCHEA6 to TCHEA7 relevant to this campaign include the addition of sigma phase predictions in several subsystems (e.g., V–Cr–Fe, V–Cr–Co, V–Fe–Co), new BCC predictions (V–Co–Ni, Fe–Ni–Cu, Co–Ni–Cu), and expanded HCP phase assessments in multiple subsystems (Al–Cr–Ni, Al–Mn–Fe, Al–Co–Ni, Al–Ni–Cu, V–Fe–Ni, Cr–Fe–Ni, Mn–Fe–Cu, Mn–Ni–Cu, Fe–Co–Cu, Fe–Ni–Cu, Co–Ni–Cu). Of particular relevance to this campaign, the V–Cr–Fe and V–Fe–Ni subsystems (highlighted in Table 5) received updated assessments that introduced sigma, FCC, and HCP phase predictions not present in TCHEA6. Several alloys containing these ternary combinations showed secondary phases experimentally, consistent with the refined TCHEA7 predictions. These updates motivated the rescreening of the full feasible space using TCHEA7 with the additional constraint of FCC $\geq 0.99$ at $950\,^{\circ}$C, as described in the main text.

Table 5: Ternary subsystem assessment availability in TCHEA6 and TCHEA7 for the Al–V–Cr–Mn–Fe–Co–Ni–Cu system. "Yes" in both columns indicates the assessment was available in TCHEA6 and carried over; "Yes" in only the TCHEA7 column indicates a newly added assessment; blank or "No" indicates no assessment is available. Additional columns indicate whether sigma, BCC, FCC, or HCP phases are newly predicted in TCHEA7 for that subsystem. Bold rows indicate subsystems where experimentally observed secondary phases align with the updated TCHEA7 predictions.

| Subsystem | TCHEA6 | TCHEA7 | SIGMA | BCC | FCC | HCP |
|---|---|---|---|---|---|---|
| *Al–V–X* | | | | | | |
| Al–V–Cr | Yes | Yes | SIGMA | | | |
| Al–V–Mn | | No | | | | |
| Al–V–Fe | No | No | SIGMA | | | |
| Al–V–Co | No | No | SIGMA | | | |
| Al–V–Ni | Yes | Yes | SIGMA | | | |
| Al–V–Cu | | No | | | | |
| *Al–Cr–X* | | | | | | |
| Al–Cr–Mn | | No | | | | |
| Al–Cr–Fe | Yes | Yes | | | | |
| Al–Cr–Co | Yes | Yes | SIGMA | | | |
| Al–Cr–Ni | Yes | Yes | SIGMA | | | HCP |
| Al–Cr–Cu | | No | | | | |
| *Al–Mn–X* | | | | | | |
| Al–Mn–Fe | | Yes | | | | HCP |
| Al–Mn–Co | | No | | | | |
| Al–Mn–Ni | | Yes | | | | |
| Al–Mn–Cu | | Yes | | | | |
| *Al–Fe–X* | | | | | | |
| Al–Fe–Co | No | No | | | | |
| Al–Fe–Ni | Yes | Yes | SIGMA | | | HCP |
| Al–Fe–Cu | | Yes | | | | |
| *Al–Co–X* | | | | | | |
| Al–Co–Ni | Yes | Yes | | | | HCP |
| Al–Co–Cu | | No | | | | |
| *Al–Ni–X* | | | | | | |
| Al–Ni–Cu | | Yes | | | | HCP |
| *V–Cr–X* | | | | | | |
| V–Cr–Mn | | No | | | | |
| **V–Cr–Fe** | **Yes** | **Yes** | **SIGMA** | | **FCC** | **HCP** |
| V–Cr–Co | Yes | Yes | SIGMA | | | |
| V–Cr–Ni | Yes | Yes | | | | |
| V–Cr–Cu | | No | | | | |
| *V–Mn–X* | | | | | | |
| V–Mn–Fe | | No | | | | |
| V–Mn–Co | | No | | | | |
| V–Mn–Ni | | Yes | | | | |
| V–Mn–Cu | | No | | | | |
| *V–Fe–X* | | | | | | |
| V–Fe–Co | No | Yes | SIGMA | | | |
| **V–Fe–Ni** | **Yes** | **Yes** | | | **FCC** | **HCP** |
| V–Fe–Cu | | Yes | | | | HCP |
| *V–Co–X* | | | | | | |
| V–Co–Ni | Yes | Yes | | BCC | | HCP |
| V–Co–Cu | | No | | | | |
| *V–Ni–X* | | | | | | |
| V–Ni–Cu | | No | | | | |
| *Cr–Mn–X* | | | | | | |
| Cr–Mn–Fe | | Yes | | | | |
| Cr–Mn–Co | | Yes | | | | |
| Cr–Mn–Ni | | Yes | | | | |
| Cr–Mn–Cu | | No | | | | |
| *Cr–Fe–X* | | | | | | |
| Cr–Fe–Co | Yes | Yes | | | | |
| Cr–Fe–Ni | Yes | Yes | | | | HCP |

*Continued on next page*

| **Subsystem** | **TCHEA6** | **TCHEA7** | **SIGMA** | **BCC** | **FCC** | **HCP** |
|---|---|---|---|---|---|---|
| Cr–Fe–Cu | | Yes | | | | |
| *Cr–Co–X* | | | | | | |
| Cr–Co–Ni | Yes | Yes | | | | |
| Cr–Co–Cu | | Yes | | | | |
| *Cr–Ni–X* | | | | | | |
| Cr–Ni–Cu | | Yes | | | | |
| *Mn–Fe–X* | | | | | | |
| Mn–Fe–Co | | Yes | | | | |
| Mn–Fe–Ni | | Yes | | | | |
| Mn–Fe–Cu | | Yes | | | | HCP |
| *Mn–Co–X* | | | | | | |
| Mn–Co–Ni | | Yes | | | | |
| Mn–Co–Cu | | Yes | | | | |
| *Mn–Ni–X* | | | | | | |
| Mn–Ni–Cu | | Yes | | | | HCP |
| *Fe–Co–X* | | | | | | |
| Fe–Co–Ni | Yes | Yes | | | | |
| Fe–Co–Cu | | Yes | | | | HCP |
| *Fe–Ni–X* | | | | | | |
| Fe–Ni–Cu | | Yes | | BCC | | HCP |
| *Co–Ni–X* | | | | | | |
| Co–Ni–Cu | | Yes | | BCC | | HCP |

## 7. COMPOSITIONAL AND PHASE ANALYSIS OF ALL ALLOYS

Some alloys which violated project constraints, highlighted in red, had manufacturing issues or secondary phases and were not characterized with EDS.

### *7.1.* ***Supplementary Table 6*** *| Iteration 0 (BBA) alloys*

Table 6

| Alloy Name | Target Atomic % | | | | | | | | EDS Measured Atomic % | | | | | | | | XRD Phase |
|---|---|---|---|---|---|---|---|---|---|---|---|---|---|---|---|---|---|
| | Al | Co | Cr | Cu | Fe | Mn | Ni | V | Al | Co | Cr | Cu | Fe | Mn | Ni | V | |
| BBA01 | 4 | 8 | 4 | 4 | 16 | 12 | 48 | 4 | 4.1 | 8.2 | 4.2 | 4.2 | 16.1 | 12.2 | 47.0 | 4.1 | FCC |
| BBA02 | 4 | 16 | 0 | 4 | 12 | 8 | 52 | 4 | 4.2 | 15.0 | 0.0 | 4.2 | 12.2 | 8.2 | 52.2 | 4.1 | FCC |
| BBA03 | 4 | 12 | 8 | 4 | 16 | 8 | 48 | 0 | 4.3 | 12.1 | 8.2 | 4.2 | 16.1 | 8.2 | 47.0 | 0.0 | FCC |
| **BBA04** | 0 | 12 | 8 | 4 | 16 | 20 | 36 | 4 | · | · | · | · | · | · | · | · | — |
| BBA05 | 4 | 12 | 8 | 0 | 8 | 8 | 56 | 4 | 4.1 | 12.2 | 8.2 | 0.0 | 8.2 | 8.0 | 55.2 | 4.1 | FCC |
| BBA06 | 4 | 0 | 8 | 4 | 24 | 12 | 44 | 4 | 4.2 | 0.0 | 8.3 | 4.1 | 24.0 | 11.9 | 43.4 | 4.1 | FCC |
| BBA07 | 4 | 12 | 0 | 0 | 16 | 8 | 52 | 8 | 5.0 | 12.3 | 0.0 | 0.0 | 16.2 | 7.9 | 50.5 | 8.1 | FCC |
| BBA08 | 0 | 24 | 8 | 0 | 12 | 16 | 32 | 8 | 0.0 | 22.9 | 8.1 | 0.0 | 12.2 | 16.0 | 32.7 | 8.1 | FCC |
| BBA09 | 4 | 16 | 8 | 0 | 12 | 0 | 48 | 12 | 4.9 | 16.1 | 8.2 | 0.0 | 12.0 | 0.0 | 46.6 | 12.1 | FCC |
| BBA10 | 4 | 0 | 4 | 0 | 20 | 8 | 52 | 12 | 4.4 | 0.0 | 4.2 | 0.0 | 20.2 | 8.0 | 51.0 | 12.2 | FCC |
| BBA11 | 0 | 16 | 0 | 4 | 20 | 20 | 28 | 12 | 0.0 | 16.1 | 0.0 | 4.0 | 20.1 | 20.3 | 27.4 | 12.0 | FCC |
| BBA12 | 0 | 0 | 8 | 4 | 16 | 12 | 52 | 8 | 0.0 | 0.0 | 8.3 | 4.1 | 16.1 | 12.3 | 51.2 | 8.0 | FCC |
| BBA13 | 4 | 20 | 4 | 4 | 16 | 0 | 52 | 0 | 4.0 | 20.4 | 4.1 | 4.2 | 16.3 | 0.0 | 51.0 | 0.0 | FCC |
| BBA14 | 0 | 16 | 8 | 4 | 16 | 20 | 36 | 0 | 0.0 | 16.1 | 8.2 | 4.0 | 16.0 | 20.3 | 35.3 | 0.0 | FCC |
| BBA15 | 0 | 20 | 8 | 4 | 0 | 20 | 44 | 4 | 0.0 | 19.9 | 8.2 | 4.2 | 0.0 | 20.5 | 43.2 | 4.0 | FCC |
| BBA16 | 4 | 12 | 0 | 12 | 20 | 8 | 44 | 0 | 4.1 | 12.3 | 0.0 | 11.8 | 20.4 | 8.1 | 43.2 | 0.0 | FCC |

### *7.2.* ***Supplementary Table 7*** *| Iteration 1 (BBB) alloys*

Table 7

| Alloy Name | Target Atomic % | | | | | | | | EDS Measured Atomic % | | | | | | | | XRD Phase |
|---|---|---|---|---|---|---|---|---|---|---|---|---|---|---|---|---|---|
| | Al | Co | Cr | Cu | Fe | Mn | Ni | V | Al | Co | Cr | Cu | Fe | Mn | Ni | V | |
| BBB01 | 0 | 24 | 4 | 0 | 8 | 4 | 36 | 24 | 0.0 | 20.5 | 4.2 | 0.0 | 8.1 | 4.0 | 38.9 | 24.4 | FCC |
| BBB02 | 4 | 32 | 8 | 4 | 4 | 20 | 24 | 4 | 4.2 | 30.4 | 8.2 | 4.1 | 4.1 | 20.1 | 24.7 | 4.1 | FCC |
| BBB03 | 4 | 0 | 12 | 4 | 4 | 16 | 52 | 8 | 4.1 | · | 12.3 | 4.1 | 4.1 | 16.1 | 51.2 | 8.2 | FCC |
| BBB04 | 4 | 12 | 4 | 0 | 28 | 0 | 40 | 12 | 4.0 | 11.1 | 4.1 | 0.0 | 28.0 | 0.0 | 40.8 | 12.0 | FCC |
| **BBB05** | 0 | 44 | 12 | 0 | 4 | 4 | 12 | 24 | · | · | · | · | · | · | · | · | FCC + $\sigma$ |
| **BBB06** | 0 | 16 | 20 | 0 | 4 | 4 | 36 | 20 | · | · | · | · | · | · | · | · | FCC + $\sigma$ |
| BBB07 | 4 | 40 | 0 | 0 | 12 | 4 | 36 | 4 | 4.0 | 39.1 | 0.0 | 0.0 | 12.3 | 3.9 | 36.7 | 4.0 | FCC |
| BBB08 | 4 | 8 | 12 | 4 | 4 | 24 | 40 | 4 | 3.8 | 8.0 | 12.3 | 4.2 | 4.1 | 24.0 | 39.4 | 4.1 | FCC |
| **BBB09** | 0 | 4 | 8 | 0 | 4 | 28 | 40 | 16 | · | · | · | · | · | · | · | · | FCC + $\sigma$ |
| BBB10 | 4 | 0 | 4 | 4 | 8 | 16 | 52 | 12 | · | · | · | · | · | · | · | · | FCC |
| **BBB11** | 0 | 24 | 4 | 0 | 4 | 32 | 20 | 16 | · | · | · | · | · | · | · | · | FCC + $\sigma$ |
| BBB12 | 0 | 20 | 4 | 0 | 4 | 4 | 48 | 20 | 0.0 | 20.2 | 4.2 | 0.0 | 4.1 | 4.0 | 47.1 | 20.4 | FCC |
| **BBB13** | 0 | 40 | 4 | 0 | 28 | 4 | 4 | 20 | 0.0 | 39.7 | 4.2 | 0.0 | 27.9 | 3.9 | 4.2 | 20.1 | FCC + $\sigma$ |
| BBB14 | 4 | 0 | 12 | 16 | 4 | 12 | 52 | 0 | 4.4 | 0.0 | 12.1 | 15.7 | 4.1 | 12.4 | 51.3 | 0.0 | FCC |
| BBB15 | 4 | 0 | 8 | 24 | 4 | 16 | 44 | 0 | 4.3 | 0.0 | 8.3 | 23.5 | 4.1 | 16.2 | 43.7 | 0.0 | FCC |
| BBB16 | 4 | 36 | 0 | 0 | 4 | 4 | 40 | 12 | · | · | · | · | · | · | · | · | FCC |

### 7.3. ***Supplementary Table 8*** *| Iteration 2 (BBC) alloys*

Table 8

| Alloy Name | Target Atomic % | | | | | | | | EDS Measured Atomic % | | | | | | | | XRD Phase |
|---|---|---|---|---|---|---|---|---|---|---|---|---|---|---|---|---|---|
| | Al | Co | Cr | Cu | Fe | Mn | Ni | V | Al | Co | Cr | Cu | Fe | Mn | Ni | V | |
| BBC01 | 4 | 12 | 4 | 4 | 20 | 16 | 32 | 8 | 4.0 | 12.2 | 4.2 | 4.0 | 20.1 | 16.0 | 31.4 | 8.2 | FCC |
| BBC02 | 4 | 8 | 0 | 0 | 4 | 8 | 52 | 24 | 4.0 | 8.2 | 0.0 | 0.0 | 4.1 | 8.0 | 51.2 | 24.4 | FCC |
| **BBC03** | 4 | 4 | 12 | 0 | 0 | 4 | 52 | 24 | 4.0 | 4.1 | 12.2 | 0.0 | 0.0 | 3.9 | 51.3 | 24.4 | **FCC** $+ \sigma$ |
| BBC04 | 4 | 0 | 4 | 0 | 4 | 4 | 60 | 24 | 4.0 | 0.0 | 4.2 | 0.0 | 4.1 | 4.0 | 59.3 | 24.5 | FCC |
| **BBC05** | 0 | 24 | 8 | 0 | 16 | 40 | 4 | 8 | · | · | · | · | · | · | · | · | **—** |
| **BBC06** | 4 | 28 | 4 | 0 | 4 | 4 | 44 | 12 | · | · | · | · | · | · | · | · | **—** |
| **BBC07** | 4 | 32 | 4 | 0 | 0 | 4 | 44 | 12 | · | · | · | · | · | · | · | · | **—** |
| **BBC08** | 0 | 8 | 0 | 4 | 4 | 28 | 36 | 20 | · | 8.1 | · | 4.1 | 4.2 | 28.1 | 35.7 | 19.9 | **FCC** $+ \sigma$ |
| **BBC09** | 0 | 36 | 4 | 0 | 16 | 16 | 8 | 20 | · | · | · | · | · | · | · | · | **—** |
| **BBC10** | 4 | 4 | 16 | 0 | 8 | 0 | 52 | 16 | · | · | · | · | · | · | · | · | **—** |
| BBC11 | 4 | 12 | 4 | 4 | 12 | 16 | 36 | 12 | 4.3 | 12.2 | 4.2 | 3.9 | 12.2 | 15.9 | 35.1 | 12.3 | FCC |
| BBC12 | 4 | 4 | 4 | 8 | 28 | 16 | 32 | 4 | 4.1 | 4.3 | 4.2 | 7.8 | 28.0 | 16.1 | 31.3 | 4.1 | FCC |
| BBC13 | 4 | 28 | 0 | 20 | 4 | 8 | 36 | 0 | 4.3 | 28.0 | · | 19.9 | 4.1 | 8.4 | 35.4 | · | FCC |
| **BBC14** | 0 | 12 | 4 | 0 | 4 | 16 | 40 | 24 | · | 12.1 | 4.1 | · | 4.3 | 16.1 | 39.0 | 24.5 | **FCC** $+ \sigma$ |
| BBC15 | 4 | 16 | 0 | 20 | 4 | 12 | 44 | 0 | 4.4 | 16.2 | · | 19.7 | 4.2 | 12.3 | 43.3 | · | FCC |
| **BBC16** | 4 | 28 | 4 | 0 | 8 | 0 | 40 | 16 | · | · | · | · | · | · | · | · | **—** |

## 8. PROPERTIES OF ALL ALLOYS

### 8.1. ***Supplementary Table 9*** | *Tensile properties: Iteration 0 (BBA)*

Table 9

| Alloy Name | Tensile Properties | | | |
|---|---|---|---|---|
| | YS (MPa) | UTS (MPa) | UTS/YS | Elongation (%) |
| BBA01 | 284.20 | 988.80 | 3.48 | 41.2 |
| BBA02 | 301.00 | 1079.90 | 3.59 | 44.8 |
| BBA03 | 217.60 | 897.50 | 4.13 | 47.1 |
| **BBA04** | — | — | — | — |
| BBA05 | 313.10 | 1160.30 | 3.71 | 47.8 |
| BBA06 | 230.50 | 976.40 | 4.24 | 49.0 |
| BBA07 | 410.30 | 1259.20 | 3.07 | 40.6 |
| BBA08 | 270.20 | 1049.10 | 3.88 | 45.4 |
| **BBA09** | — | — | — | — |
| BBA10 | 365.70 | 1276.40 | 3.49 | 46.6 |
| BBA11 | 348.80 | 1070.60 | 3.07 | 38.6 |
| BBA12 | 316.00 | 1101.10 | 3.48 | 43.5 |
| BBA13 | 176.80 | 800.20 | 4.53 | 42.0 |
| BBA14 | 259.40 | 953.90 | 3.68 | 41.4 |
| BBA15 | 340.80 | 1095.80 | 3.22 | 41.0 |
| BBA16 | 209.60 | 846.10 | 4.04 | 42.4 |
| BBA17 | 310.20 | 1057.50 | 3.41 | 42.0 |
| BBA18 | 267.00 | 1019.40 | 3.82 | 46.7 |
| BBA19 | 232.60 | 911.20 | 3.92 | 42.1 |

*8.2. **Supplementary Table 10** | Tensile properties: Iteration 1 (BBB)*

Table 10

| Alloy Name | Tensile Properties | | | |
|---|---|---|---|---|
| | YS (MPa) | UTS (MPa) | UTS/YS | Elongation (%) |
| BBB01 | 431.54 | 1447.09 | 3.35 | 48.8 |
| BBB02 | 311.40 | 1127.03 | 3.62 | 48.8 |
| BBB03 | 307.14 | 1219.51 | 3.97 | 55.7 |
| BBB04 | 261.94 | 1097.43 | 4.19 | 50.6 |
| **BBB05** | — | — | — | — |
| **BBB06** | — | — | — | — |
| BBB07 | 211.35 | 900.56 | 4.26 | 45.3 |
| BBB08 | 282.39 | 1143.09 | 4.05 | 54.6 |
| **BBB09** | — | — | — | — |
| **BBB10** | — | — | — | — |
| **BBB11** | — | — | — | — |
| BBB12 | 418.34 | 1399.73 | 3.35 | 46.5 |
| **BBB13** | 531.30 | 1312.87 | 2.47 | 28.2 |
| BBB14 | 269.65 | 1062.73 | 3.94 | 51.5 |
| BBB15 | 261.17 | 1060.66 | 4.06 | 52.6 |
| **BBB16** | — | — | — | — |

### 8.3. ***Supplementary Table 11*** | *Tensile properties: Iteration 2 (BBC)*

Table 11

| Alloy Name | Tensile Properties | | | |
|---|---|---|---|---|
| | YS (MPa) | UTS (MPa) | UTS/YS | Elongation (%) |
| BBC01 | 254.55 | 1069.26 | 4.20 | 51.8 |
| BBC02 | 449.10 | 1462.73 | 3.26 | 48.4 |
| **BBC03** | 719.13 | 1161.35 | 1.61 | 13.8 |
| BBC04 | 445.73 | 1476.34 | 3.31 | 50.2 |
| **BBC05** | — | — | — | — |
| **BBC06** | — | — | — | — |
| **BBC07** | — | — | — | — |
| **BBC08** | 414.60 | 1021.45 | 2.46 | 24.5 |
| **BBC09** | — | — | — | — |
| **BBC10** | — | — | — | — |
| BBC11 | 290.05 | 1098.09 | 3.79 | 48.4 |
| BBC12 | 252.13 | 971.81 | 3.85 | 46.3 |
| BBC13 | 266.57 | 920.36 | 3.45 | 44.6 |
| **BBC14** | 406.38 | 1293.44 | 3.18 | 39.6 |
| BBC15 | 218.75 | 870.70 | 3.98 | 44.6 |
| **BBC16** | — | — | — | — |

8.4. ***Supplementary Table 12*** | *Dynamic and ballistic properties: Iteration 0 (BBA)*

Table 12

| **Alloy Name** | **Dynamic and Ballistic Properties** | | | |
|---|---|---|---|---|
| | **Avg $H_{dyn}$ (GPa)** | **Avg $H_{qs}$ (GPa)** | **Avg $H_{dyn}/H_{qs}$** | **Depth of Penetration (mm)** |
| BBA01 | 2.69 | 2.41 | 1.12 | 2.92 |
| BBA02 | 2.46 | 2.10 | 1.17 | — |
| BBA03 | 2.30 | 1.95 | 1.18 | 3.19 |
| **BBA04** | **—** | **—** | **—** | **—** |
| BBA05 | 2.54 | 2.28 | 1.12 | 2.79 |
| BBA06 | 2.31 | 1.97 | 1.17 | 3.09 |
| BBA07 | 3.04 | 2.68 | 1.13 | 2.56 |
| BBA08 | 2.62 | 2.27 | 1.15 | 2.93 |
| BBA09 | 3.41 | 2.96 | 1.15 | 2.67 |
| BBA10 | 2.78 | 2.39 | 1.17 | 2.64 |
| BBA11 | 2.50 | 2.13 | 1.17 | 2.72 |
| BBA12 | 2.52 | 2.20 | 1.14 | 2.82 |
| BBA13 | 2.14 | 1.88 | 1.14 | 3.35 |
| BBA14 | 2.34 | 2.00 | 1.17 | 3.02 |
| BBA15 | 2.52 | 2.23 | 1.13 | 2.74 |
| BBA16 | 2.50 | 2.29 | 1.09 | 3.29 |
| BBA17 | 2.51 | 2.20 | 1.14 | 2.85 |
| BBA18 | 2.16 | 1.84 | 1.17 | 2.94 |
| BBA19 | 2.36 | 2.01 | 1.17 | 3.11 |

8.5. ***Supplementary Table 13*** | *Dynamic and ballistic properties: Iteration 1 (BBB)*

Table 13

| **Alloy Name** | **Dynamic and Ballistic Properties** | | | |
|---|---|---|---|---|
| | **Avg $H_{dyn}$ (GPa)** | **Avg $H_{qs}$ (GPa)** | **Avg $H_{dyn}/H_{qs}$** | **Depth of Penetration (mm)** |
| BBB01 | 3.07 | 2.80 | 1.09 | 2.45 |
| BBB02 | 2.70 | 2.32 | 1.16 | 2.84 |
| BBB03 | 2.69 | 2.32 | 1.16 | 2.81 |
| BBB04 | 2.58 | 2.13 | 1.21 | 2.94 |
| **BBB05** | — | — | — | — |
| **BBB06** | — | — | — | — |
| BBB07 | 2.22 | 1.88 | 1.18 | 3.16 |
| BBB08 | 2.40 | 2.10 | 1.14 | 2.90 |
| **BBB09** | — | — | — | — |
| **BBB10** | — | — | — | — |
| **BBB11** | — | — | — | — |
| BBB12 | 3.75 | 3.25 | 1.15 | 2.52 |
| **BBB13** | 3.21 | 2.83 | 1.13 | 2.31 |
| BBB14 | 2.33 | 2.01 | 1.16 | 2.96 |
| BBB15 | 2.34 | 1.93 | 1.22 | 2.98 |
| **BBB16** | — | — | — | — |

8.6. ***Supplementary Table 14*** | *Dynamic and ballistic properties: Iteration 2 (BBC)*

Table 14

| Alloy Name | Dynamic and Ballistic Properties | | | |
|---|---|---|---|---|
| | Avg $H_{dyn}$ (GPa) | Avg $H_{qs}$ (GPa) | Avg $H_{dyn}/H_{qs}$ | Depth of Penetration (mm) |
| BBC01 | 2.36 | 2.02 | 1.17 | 2.96 |
| BBC02 | 3.10 | 2.66 | 1.16 | 2.44 |
| **BBC03** | 3.79 | 3.36 | 1.13 | 2.22 |
| BBC04 | 3.09 | 2.70 | 1.14 | 2.45 |
| **BBC05** | — | — | — | — |
| **BBC06** | — | — | — | — |
| **BBC07** | — | — | — | — |
| **BBC08** | 2.79 | 2.47 | 1.13 | 2.56 |
| **BBC09** | — | — | — | — |
| **BBC10** | — | — | — | — |
| BBC11 | 2.61 | 2.16 | 1.21 | 2.86 |
| BBC12 | 2.18 | 1.81 | 1.20 | 3.04 |
| BBC13 | 2.13 | 1.84 | 1.16 | — |
| **BBC14** | 3.14 | 2.69 | 1.17 | 2.51 |
| BBC15 | 2.14 | 1.79 | 1.20 | 3.22 |
| **BBC16** | — | — | — | — |

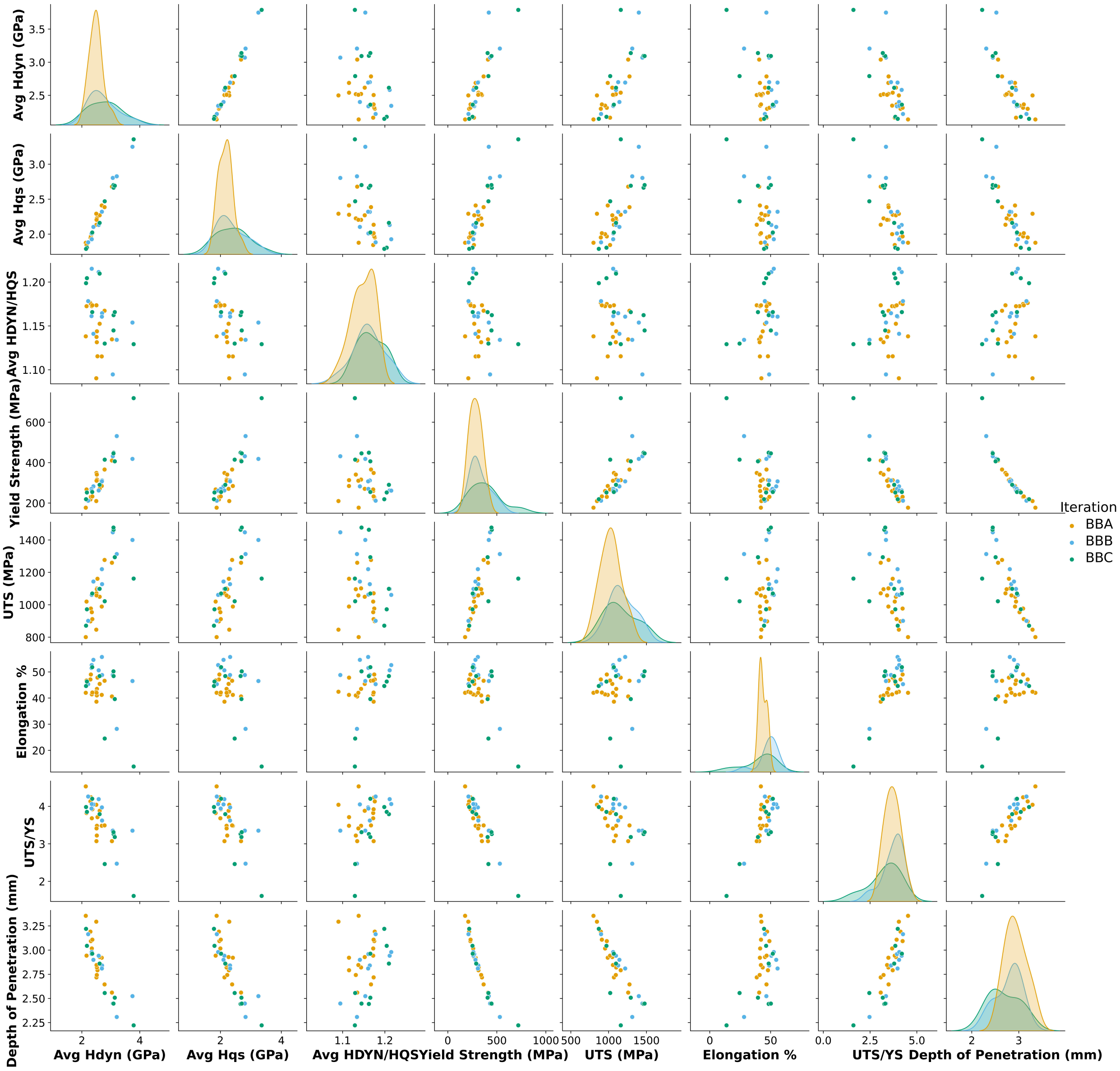


Figure 9: Comprehensive pairwise analysis of mechanical and ballistic properties for HEAs across all iterations within the study. Data points are color-coded by iteration: BBA (orange), BBB (blue), and BBC (green)

## 9. Property Correlation Analysis

The pairplot analysis in Figure 9 reveals relationships between the performance metrics: dynamic hardness (Hdyn), quasi-static hardness (Hqs), hardness ratio (Hdyn/Hqs), yield strength (YS), ultimate tensile strength (UTS), strain at UTS, strength ratio (UTS/YS), and penetration depth. The diagonal elements of the pairplot matrix show the distribution characteristics of each property, while the off-diagonal scatter plots illuminate pairwise correlations that are critical for understanding multi-objective optimization.

Strong positive correlations are observed among the quasi-static strength metrics: YS and hardness are tightly coupled (r = 0.88), and the two hardness measures are nearly identical (r = 0.99). The strength–ductility trade-off characteristic of FCC HEAs is reflected in the UTS/YS ratio, which is strongly anticorrelated with YS (r = -0.91) and positively correlated with elongation (r = 0.80).

Depth of penetration shows a near-perfect inverse correlation with YS (r = -0.94) and a strong inverse correlation with UTS (r = -0.86), indicating that penetration resistance in this alloy family is governed predominantly by quasi-static strength. The four highest-YS single-phase FCC alloys identified by the campaign (BBC04, BBC02, BBB01, BBB12) are accordingly also the best simulated ballistic performers.

The Hdyn/Hqs ratio quantifies rate-dependent strengthening at the indentation scale. Across the campaign it spans a narrow range (1.09–1.22) with no clear single-element compositional trend. At the per-alloy level, the rate-dependent component of ballistic response is captured more directly by the calibrated Cowper–Symonds exponent M, which correlates positively with

DoP and negatively with YS; the highest-strength alloys exhibit lower rate-sensitivity exponents, a trade-off that is invisible to quasi-static testing alone. The Hdyn/Hqs ratio is therefore retained as a complementary rate-sensitive objective within the Bayesian optimization loop rather than as a stand-alone predictor of ballistic performance.



\end{document}